\documentclass[usenatbib]{emulateapj}
\usepackage{epsf}
\usepackage{apjfonts}
\usepackage{xspace}
\usepackage{amsmath}
\def\Msun{\, M_{\odot}}
\def\Mcl{\, M_{\rm cl}}
\def\Mstcl{\, M_{\ast,\rm cl}}
\def\Mstclmax{\, M_{\ast,\rm cl,max}}
\def\Mclmax{\, M_{\rm cl,max}}
\def\Mclmin{\, M_{\rm cl,min}}
\def\Rcl{\, R_{\rm cl}}
\def\Scl{\, \Sigma_{\rm cl}}
\def\ecl{\, \varepsilon_{\rm cl}}
\def\tcl{\, t_{\rm cl}}

\def\MH2c{\, M_{\rm H_2,cell}}
\def\Msunpc2{\,\rm M_{\odot}\,pc^{-2}}

\newcommand{\Msol}{{\,M}_\odot} 
\newcommand{\Msolyr}{{\,M}_\odot\,{\rm yr}^{-1}} 

\newcommand{\pc} {{\,\rm pc}} 
\newcommand{\K} {{\,\rm K}} 
\newcommand{\cc}{{\,\rm {cm^{-3}}}}

\renewcommand{\vec}[1]{{\mathbf #1}} 
\newcommand{\kmsec}{{\,\rm {km\,s^{-1}} }}

\def\Gyr{\,{\rm Gyr}}
\def\Myr{\,{\rm Myr}}
\newcommand{\kpc} {{\,\rm kpc}}

\newcommand{\vel}{\mbox{\boldmath$v$}}

\newcommand{\nab}{\mbox{\boldmath$\nabla$}}

\begin{document}
\shorttitle{Stellar feedback models for galaxy formation simulations}
\slugcomment{{\em submitted to the Astrophysical Journal}} 
\shortauthors{Agertz et al.}

\title{Towards a complete accounting of energy and momentum from stellar feedback in galaxy formation simulations}

\author{
Oscar Agertz\altaffilmark{1,2}, Andrey
  V. Kravtsov\altaffilmark{1,2,3}, Samuel N. Leitner\altaffilmark{1,2,4}, Nickolay Y.\
  Gnedin\altaffilmark{1,2,5}}  
\altaffiltext{1}{Kavli Institute for Cosmological Physics and Enrico
  Fermi Institute, The University of Chicago, Chicago, IL 60637 USA} 
\altaffiltext{2}{Department of Astronomy \& Astrophysics, The
  University of Chicago, Chicago, IL 60637 USA} 
\altaffiltext{3}{Enrico Fermi Institute, The University of Chicago,
Chicago, IL 60637}
\altaffiltext{4}{Department of Astronomy, University of Maryland
College Park, MD 20742-2421}
\altaffiltext{5}{Particle Astrophysics Center, 
Fermi National Accelerator Laboratory, Batavia, IL 60510, USA}



\begin{abstract}
Stellar feedback plays a key role in galaxy formation by regulating star formation, driving interstellar turbulence and generating galactic scale outflows. Although modern simulations of galaxy formation can resolve scales of $\sim10-100\pc$, star formation and feedback operate on smaller, ``subgrid'' scales. Great care should therefore be taken in order to properly account for the effect of feedback on global galaxy evolution. We investigate the momentum and energy budget of feedback during different stages of stellar evolution, and study its impact on the interstellar medium using simulations of local star forming regions and galactic disks at the resolution affordable in modern cosmological zoom-in simulations. In particular, we present a novel subgrid model for the momentum injection due to radiation pressure and stellar winds from massive stars during early, pre-supernova evolutionary stages of young star clusters. This model is local and straightforward to implement in existing hydro codes without the need for radiative transfer. Early injection of momentum acts to clear out dense gas in star forming regions, hence limiting star formation. The reduced gas density mitigates radiative losses of thermal feedback energy from subsequent supernova explosions, leading to an increased overall efficiency of stellar feedback. The detailed impact of stellar feedback depends sensitively on the implementation and choice of parameters. Somewhat encouragingly, we find that implementations in which feedback is efficient lead to approximate self-regulation of global star formation efficiency. We compare simulation results using our feedback implementation to other phenomenological feedback methods, where thermal feedback energy is allowed to dissipate over time scales longer than the formal gas cooling time. We find that simulations with maximal momentum injection suppress star formation to a similar degree as is found in simulations adopting adiabatic thermal feedback. However, different feedback schemes are found to produce significant differences in the density and thermodynamic structure of the interstellar medium, and are hence expected to have a qualitatively different impact on galaxy evolution.  
\end{abstract}

\setcounter{figure}{0}
\section{Introduction}
\label{sect:intro}
Galaxy formation remains one of the most important, unsolved problems in modern astrophysics. In large part, this is because galaxy evolution depends on small-scale star formation and feedback processes, which are still poorly understood. For instance, it is now well established that stellar feedback from young massive stars can significantly affect the ISM by regulating star formation \citep[][and references therein]{maclow:review04,mckeeostriker07}, driving turbulence \citep{kim01,deavillez04,joungmaclow06,Agertz09,Tamburro2009} and generating galactic scale outflows \citep{Martin1999,Martin2005,OppenheimerDave2006}. 

One of the most salient and long-standing problems of galaxy formation modelling is the overprediction of baryon masses and concentrations of galaxies compared to expected values derived using dark matter halo abundance matching \citep{conroy_wechsler09,Guo2010}, satellite kinematics \citep{klypin_prada09,more2010}, and weak lensing \citep{Mandelbaum2006}, see \cite{Behroozi2010} for a comprehensive discussion. It is widely thought that the low baryon masses of galaxies are due to galactic winds driven by stellar feedback at the faint end of the stellar mass function \citep{DekelSilk86,Efstathiou00} and by the active galactic nuclei (AGN) and the bright end \citep{SilkRees1998,Benson2003}.

Pioneering numerical galaxy formation studies by \cite{Katz92}, \cite{NavarroWhite93} and \cite{Katz1996} demonstrated that thermal energy from SNe inefficiently coupled to the simulated ISM due to efficient radiative cooling in dense star forming regions. To avoid such radiative losses, an ad hoc delay of gas cooling in regions of recent star formation is often adopted \citep{Gerritsen1997PhDT, ThackerCouchman2000, ThackerCouchman2001, Stinson06, Governato07,Agertz09b, Agertz2011,Guedes2011,Aquila}, which is justified by the fact that the multiphase structure of star forming regions with pockets of low-density hot gas is not resolved. While excessive radiative losses may indeed be partially due to resolution effects, one may also argue that such losses should increase with increasing resolution, as star forming regions can collapse to higher densities at higher resolution. Indeed, simulations of supernova-driven blast waves with sub-parsec resolution show that most of thermal energy of hot gas is radiated away during the blast wave expansion \citep{Thornton1998,ChoKang2008}. It is thus necessary to consider other mechanisms of stellar feedback in addition to the energy-driven expansion of supernova (SN) bubbles. 

Observations of the giant molecular clouds (GMCs) hosting young star clusters indicate that gas is often dispersed well {\it before} the first SNe explode ($t\lesssim 4\Myr$). 
This can be partly due to the fact that natal GMCs are not gravitationally bound \citep{maclow:review04,Padoan1997,Li2004,Kritsuk2007,dobbs_etal11}. However, there is also plenty of evidence that dense star forming regions are destroyed by their HII regions, driven by ionization at low cluster masses \citep[][]{Walch2012} and radiation pressure, i.e. momentum transfer from radiation emitted by young massive stars to gas and dust \citep{Murray2005}, and stellar winds at high \citep{Matzner2002, KrumholzMatzner2009}. In the Milky Way, the majority of star formation is concentrated in a few hundred massive GMCs ($M_{\rm cl}\sim10^6\Msol$), containing the majority of the Galaxy's molecular gas reservoir \citep{Murray2011b}. \cite{Fall2010} presented arguments favoring radiation pressure as a gas removal mechanism
to explain the similarity in the shapes of the molecular clump and stellar
cluster mass functions. \citet{murray_etal10} argued that 
radiation pressure is the dominant force in driving the expansion of Galactic
HII regions and showed this explicitly in the case of several observed star forming regions. Multi-wavelength data of the evolved HII region around the dense R136 star cluster in the 30 Doradus region of LMC is also consistent with these arguments and implies that radiation pressure was the main driver of the HII region 
dynamics at radii $\lesssim 75$~pc \citep[][see however \citealt{pellegrini_etal11}]{lopez_etal11}. 

\begin{figure}[t]
\begin{center}
\includegraphics[scale=0.4]{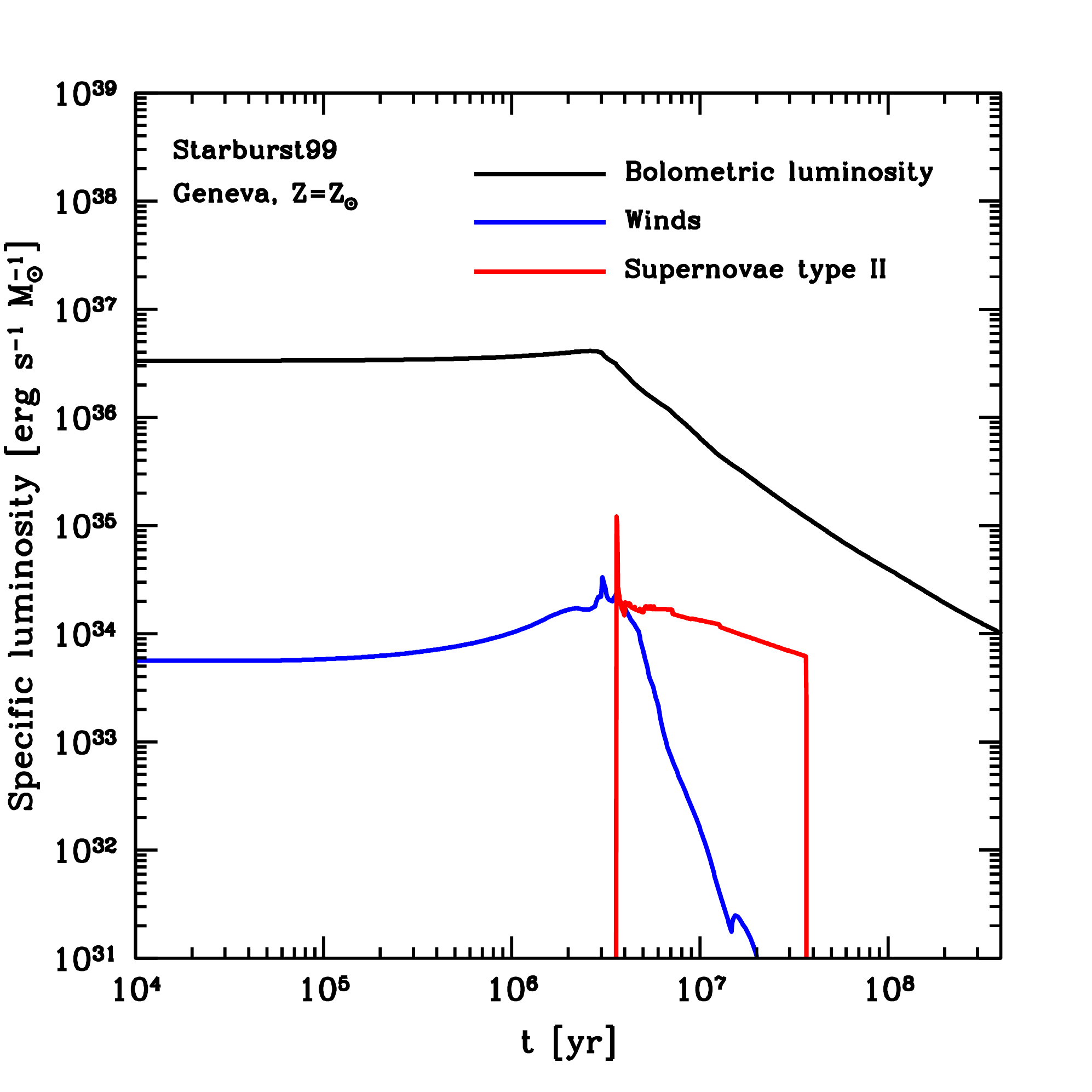}
\caption{Specific luminosity from radiation (black solid line), stellar winds (solid blue) and supernovae type II (solid red). The data is generated using {\small STARBURST99} assuming the Geneva high mass loss stellar tracks and solar metallicity.}
\label{fig:sb99}
\end{center}
\end{figure}

The radiation from a young stellar population indeed carries a large amount of energy and momentum, as illustrated in Figure \ref{fig:sb99}, which shows the specific luminosity and mechanical power from stellar winds and SNII from a stellar population assuming the \cite{Kroupa01} initial mass function (IMF) calculated using the {\small STARBURST99} code \citep{Leitherer1999}. A SNII event typically ejects $10\Msol$ at $\vel\sim 3000\kmsec$, and the stellar winds from young massive stars have a similar  velocity  \citep{Leitherer1999}. Although the mechanical luminosity of winds and SNII ejecta are two orders of magnitude smaller than the luminosity of emitted radition, their velocity is also two orders of magnitude smaller than the speed of light. This makes the actual momentum injection rate of all sources comparable,
\begin{equation}
\dot{p}_{\rm rad}\sim \frac{L_{\rm bol}}{c}\sim\dot{p}_{\rm SNII}\sim\dot{p}_{\rm winds}\sim\frac{L_{\rm mech}}{\vel},
\end{equation}
where $L_{\rm bol}$ is the bolometric luminosity of a stellar population. As can be seen in Figure~\ref{fig:sb99}, the first SNIIe occur  $\sim4\Myr$ after the birth of stellar population, while radiation pressure and stellar winds operate immediately after birth. Furthermore, the effect of radiation pressure can be significantly enhanced in dense, dusty regions as UV photons absorbed by dust re-radiate in the infrared, increasing the momentum injection rate in proportion to the infrared optical depth $\tau_{\rm IR}$, i.e. $\dot{p}_{\rm rad}\sim\tau_{\rm IR} L/c$ \citep[see, e.g.,][for discussion of momentum deposition by trapped radiation]{gayley_etal95}. In the environments of massive star clusters and central regions of starbursts, values of $\tau_{\rm IR}\sim10-100$ are plausible \citep{murray_etal10}, making momentum imparted by radiation pressure to the surrounding gas the dominant feedback source at early times ($t \lesssim 4\Myr$) in a stellar population. 

Early dispersal of gas can facilitate the survival and breakout of hot gas heated by SN blastwaves and dramatically increase the overall efficiency of stellar feedback. The momentum injection due to stellar winds and radiation pressure may also be an important feedback mechanism in its own right. Indeed, radiation pressure has been suggested to play a significant role in regulating global star formation in galaxies and launching galactic-scale winds \citep{Haehnelt1995, scoville_etal01,Murray2005}. Analytical work by \cite[][see also \citealt{NathSilk2009}]{Murray2011} demonstrated how massive star clusters can radiatively launch large scale outflows, provided star formation is vigorous enough ($\Sigma_{\rm SFR}\gtrsim0.05\,\Msol {\rm yr}^{-1} {\rm kpc}^{-2}$); an attractive property to capture in simulations of galaxy formation. 

Radiation pressure feedback has just recently been considered in numerical work studying isolated galactic disks \citep{Hopkins2011Prad,Hopkins2012structure,Chattopadhyay2012}. In the suite of papers by Hopkins et al., an implementation of radiation pressure feedback was explored using smoothed-particle-hydrodynamics (SPH), relying on high mass ($m_{\rm SPH}\sim 10^3\Msol$) and force resolution ($\sim {\rm few}\pc$). It was argued that radiation feedback has a significant effect on galaxies from dwarfs to extreme starbursts, where the contribution was most significant in the high surface density systems. \cite{Wise2012} demonstrated, using an adaptive-mesh-refinement (AMR) radiative transfer technique in a fully cosmological context, how radiation pressure in the single scattering regime could affect star formation rates and metal distributions in a dwarf galaxies in dark matter halos of $2.0\times 10^8\Msol$. \cite{Brook2012} and \cite{stinson_etal12} discussed the importance of ``early feedback'' in their SPH galaxy formation simulations. These authors assume that $10\%$ of the \emph{bolometric} luminosity radiated by young stars is converted into thermal energy of star forming gas over a 0.8 Myr time period, which significantly affects simulated galaxy properties. However, it is not clear how such a scheme relates to the actual processes of early feedback, which are thought to be momentum- rather than energy-driven. 

Cosmological zoom-in simulations of individual galaxies adopting a force resolution of $\lesssim 50-100\pc$, while reaching $z=0$, are becoming increasingly common \citep[][]{Agertz09b,Governato2010, Guedes2011}. At such resolution, the largest sites of star formation can be identified in simulations directly, although their internal structure would not be resolved. It is hence crucial to understand how well we can capture the \emph{global} effect of stellar feedback from star forming regions taking into account {\it all} plausible sources and mechanisms of stellar feedback at the resolution level affordable in modern cosmological simulations.

In this paper we discuss the available energy and momentum budget from stellar winds, SNe and radiation pressure. The latter is implemented using a novel empirically-based subgrid model. Using the adaptive-mesh-refinement (AMR) code {\small RAMSES} \citep{teyssier02}, we study the impact of these feedback sources in idealized simulations of star forming clouds and isolated disk galaxies. The simulations are performed at spatial resolution $\sim 50-100$~pc, comparable to that of modern state-of-the-art  cosmological simulations. We investigate how the detailed impact of stellar feedback depends on the implementation and choice of parameters of feedback schemes. We also compare a ``straight-injection'' approach, where energy and momentum is deposited directly onto the grid, compares to widely used phenomenological methods where thermal feedback energy is allowed to dissipate over longer time scales than expected by radiative cooling.

The paper is organized as follows. In \S~\ref{sect:SFFB} we discuss the feedback budget from SNe and stellar winds and radiation pressure. \S~\ref{sect:implementation} outlines the numerical implementation of stellar feedback in the AMR code {\small RAMSES}. In \S~\ref{sect:simulations} we present idealized cloud and galactic disk simulations, and discuss how the different sources of feedback affect global properties of star formation. 
We conclude by summarizing our results and conclusions in \S~\ref{sec:conclusions}.
We detail the empirically-based subgrid model used to compute momentum due to radiation pressure in Appendix A, and implementation of the second energy variable in Appendix B. 

\section{Stellar feedback and Star formation}
\label{sect:SFFB}

\subsection{Stellar feedback}
\label{sect:feedback}
Several processes are contributing to stellar feedback, as stars inject energy, momentum, mass and heavy elements over time via SNII, SNIa, stellar winds from massive stars, radiation pressure, and secular mass loss into surrounding interstellar gas. The feedback terms we aim to quantify in this section are: 
\begin{eqnarray}
\mbox{\emph{Energy:}}&\quad E_{\rm tot} &=  E_{\rm SNII}+E_{\rm SNIa}+E_{\rm wind} \nonumber \\
\mbox{\emph{Momentum:}}&\quad p_{\rm tot} &=  p_{\rm SNII}+p_{\rm wind}+p_{\rm rad}\label{eq:FBarray} \\
\mbox{\emph{Mass loss:}}&\quad m_{\rm tot} &=  {m}_{\rm SNII}+{m}_{\rm SNIa}+{m}_{\rm wind}+m_{\rm loss} \nonumber\\ 
\mbox{\emph{Metals:}}&\quad m_{\rm Z,tot} &=  {m}_{\rm Z,SNII}+m_{\rm Z,SNIa}+m_{\rm Z,wind}+m_{\rm Z,loss}. \nonumber
\end{eqnarray}
We choose to calculate and include the contribution of all feedback processes at every simulation timestep $\Delta t$ for every star particle formed by our star formation recipe (see \S~\ref{sect:SFlaw}). Feedback is thus not done instantaneously, but continuously in specific time periods when a given feedback process operates, taking into account the lifetime of stars of different masses in a stellar population. We assume that each star particle formed in our numerical simulations represents an ensemble of stars with a given initial mass function (IMF). For stellar masses $M\in [0.1-100]\Msol$, we assume the IMF form of \cite{Kroupa01}\footnote{The IMF suggested by \cite{Kroupa01} extends to $M=0.01\Msol$ with a slope of $\approx 0.3\pm0.7$ below $M=0.08$. For the purpose of stellar feedback, accounting for the low-mass range has a negligible effect for the feedback energy budget presented in this paper; the total number of available SNII events are reduced by only $\sim6\%$.},

\begin{equation} 
\label{eq:IMFK01}
\Phi(M) =  A\left\{
\begin{array}{rl} 
2\,M^{-1.3} & \text{for } 0.1\leq M < 0.5\,\Msol \\
M^{-2.3} & \text{for }  0.5\leq M < 100\,\Msol,
\end{array} \right.
\end{equation}
where $A$ normalizes $\Phi(M)$ such that total mass of stars is equal to the initial mass of a star particle, $m_{*,{\rm ini}}$. Note that the choice of IMF can significantly affect the amount of stellar feedback, especially the total energy and momentum output from massive stars. For example, the IMF of Equation~\ref{eq:IMFK01} has more than twice as many massive stars exploding as type II supernovae (assuming SNII mass range of $8-40\Msol$), and a Chabrier IMF \citep{chabrier03} three times as many, compared to the more bottom heavy IMF of \cite{Kroupa1993}.

\subsubsection{Stellar winds from massive stars}
\begin{figure}[t]
\begin{center}
\includegraphics[scale=0.43]{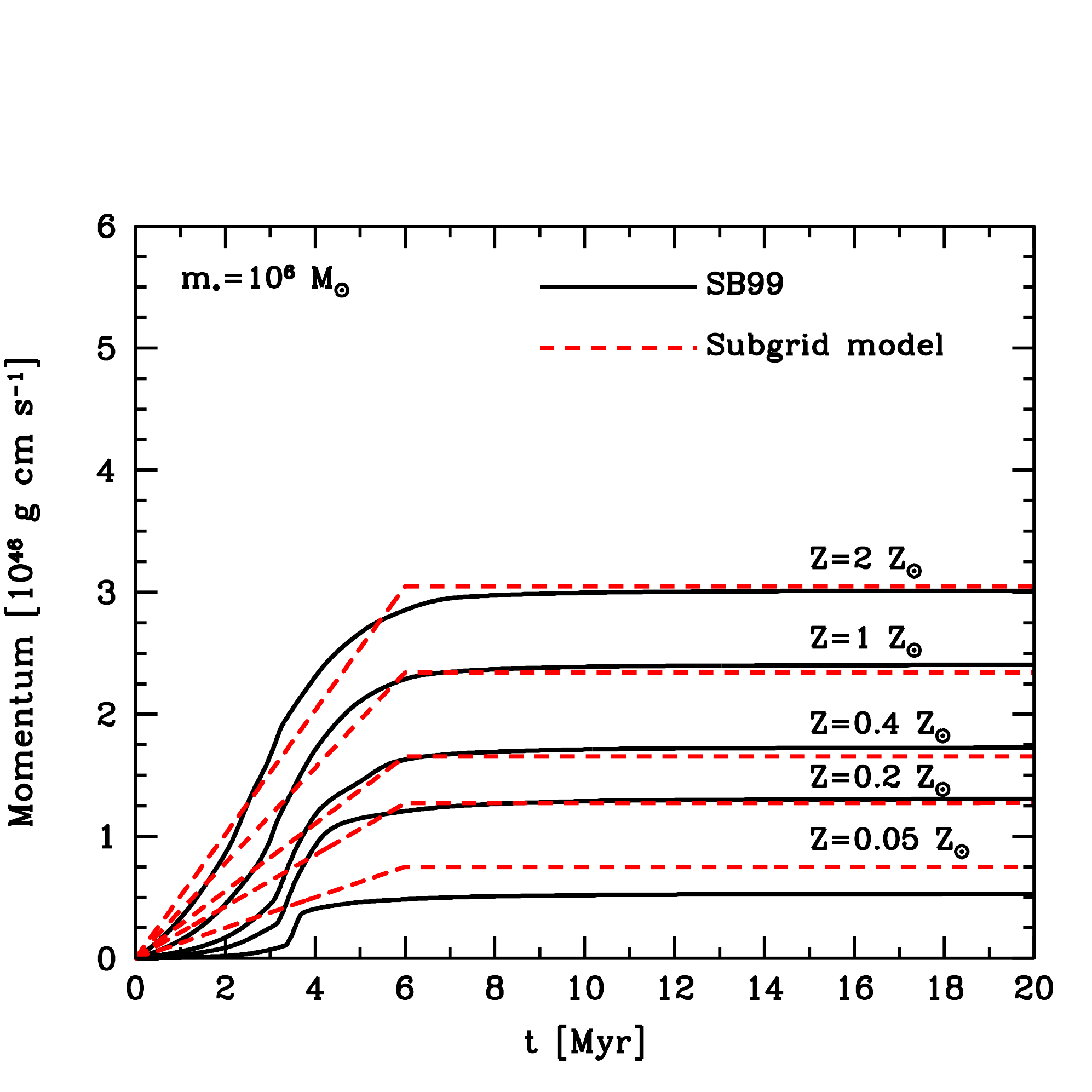}
\caption{Cumulative momentum from stellar winds from STARBURST99 (black lines) compared to the subgrid approximation in Equation\ref{eq:array2} (dashed red line).
}
\label{fig:pwind}
\end{center}
\end{figure}
Massive stars ($M\gtrsim 5\Msol$) can radiatively drive strong stellar winds from their envelopes during the first $6\Myr$ of stellar evolution, reaching terminal velocities of $1000-3000\kmsec$ \citep{LamersCassinelli1999}. The kinetic energy of these winds is expected to thermalize via shocks. To account for the energy, momentum, mass, and metal injection by such winds, we use calculations done with the {\small STARBURST99} code. We find that the dependence of energy and momentum injection on metallicity can be approximated by a simple function\footnote{We adopt the Geneva high mass loss stellar tracks, and fit for the provided metallicities $Z=2,1,0.4,0.2$ and $0.05\,Z_\odot$, assuming the IMF in Equation~\ref{eq:IMFK01}. We assume the feedback behavior at higher and lower metallicities to follow the extrapolation of our fits.} and we use such functional form in our simulations. Although the fit is approximate, its accuracy is sufficient given the uncertainties in the underlying wind models \citep[see discussion in][]{Leitherer92}. 

Specifically, we approximate the cumulative energy, momentum and mass injection, in CGS units, for a stellar population of age $t_*$ (in Myr), birth mass $m_{*,{\rm ini}}$ (in $M_\odot$) and stellar metallicity $Z_*$ (in units of solar metallicity $Z_\odot=0.02$), as
\begin{eqnarray}
\label{eq:array1}
E_{\rm wind} & = & m_{*,{\rm ini}}e_1\left(\frac{Z_*}{e_2}\right)^{e_3}\frac{t_*}{t_w}\,{\rm ergs} \quad{\rm for} \quad{t_* \leq t_w} \nonumber \\
\label{eq:array2}
p_{\rm wind} & = & m_{*,{\rm ini}}p_1\left(\frac{Z_*}{p_2}\right)^{p_3}\frac{t_*}{t_w}\,{\rm g\,cm\,s^{-1}} \quad{\rm for} \quad{t_* \leq t_w} \\
\label{eq:array3}
m_{\rm wind} & = & m_{*,{\rm ini}}m_1\ln\left(\frac{Z_*}{m_2}+1\right) \frac{t_*}{t_w}\,{\Msol}\quad{\rm for}\quad{t_* \leq t_w} \nonumber\\
m_{\rm Z,wind} & = & Z_*\,m_{\rm wind}\,\Msol \quad{\rm for}\quad{t_* \leq t_w} \nonumber,
\label{eq:array4}
\end{eqnarray}
where $e_{1,2,3}=[1.9\times10^{48}\,{\rm ergs}\,\Msol^{-1}, 0.50,0.38]$, $p_{1,2,3}=[1.8\times10^{40}\,{\rm g\,cm\,s^{-1}\,\Msol^{-1}}, 0.50,0.38]$ and $m_{1,2}=(2.4\times10^{-2}, 4.6\times10^{-4})$. The wind duration is $t_w=6.5\Myr$. 

In Figure~\ref{fig:pwind} we show an example of how the momentum injection for a $10^6\Msol$ star cluster, calculated using this approximation, compares to the {\small STARBURST99} calculation for different metallicities. The momentum injection agrees quite well for $Z\gtrsim0.1Z_{\odot}$, although we do oversimplify the time evolution, especially at early time ($t_*\lesssim 3\,$Myr). A similar conclusion holds for the wind energy injection and mass loss.

\subsubsection{Radiation pressure}
\label{sect:radpressure}
\begin{figure}[t]
\begin{center}
\includegraphics[scale=0.44]{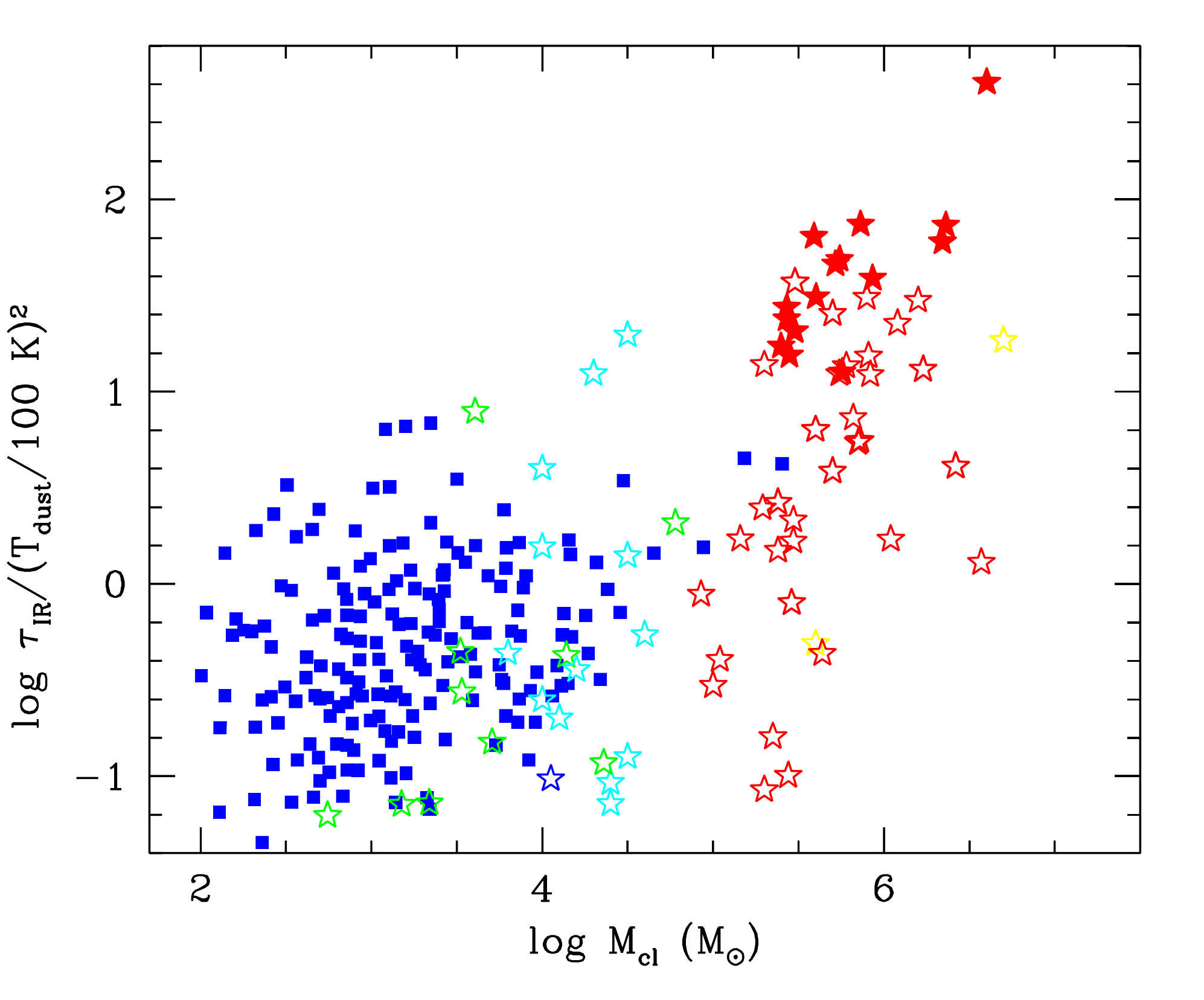}
\caption{Estimate of the infrared optical depth $\tau_{\rm IR}=\kappa_{\rm IR}\Sigma_{\rm cl}$ as a function of clump/cluster mass. We adopt $\kappa_{\rm IR}\approx 3{\rm\ cm^2\,g^{-1}}(T_{\rm d}/100 K)^2$ \citep{semenov_etal03} and the data are observations of molecular clumps in the Milky Way \protect\citep[blue squares][]{fall_etal10} and young star clusters (stars) (see Appendix~\ref{sect:Pradmodel} for further details). The green stars show clusters in the LMC \citep{mackey_gilmore03a} and SMC, while cyan stars show star clusters within Milky Way. The remaining star symbols show additional extragalactic star clusters, including starbursts such as M82 (solid red) and Antennae (red). The data for star clusters is from \protect\citet{portegies_etal10} \citep[M82 data from][]{krumholz_matzner09}. Although significant scatter exists,  massive star cluster tend to have high surface densities, giving rise to infrared optical depths $\tau_{\rm IR}\gtrsim 10$ for $M_{\rm cl}\gtrsim10^5\Msol$.
}
\label{fig:tauobs}
\end{center}
\end{figure}
The momentum injection rate from radiation can be written as
\begin{equation}
\label{eq:radpressure}
\dot{p}_{\rm rad}=(\eta_1+\eta_2\tau_{\rm IR})\frac{L(t)}{c},
\end{equation}
where $\tau_{\rm IR}$ is the infrared optical depth and $L(t)$ is the luminosity of the stellar population. The first term describes the direct radiation absorption/scattering, and should in principle be $\propto [1-\exp{(-\tau_{\rm UV})}]$. However, given the very large dust and HI opacities in the UV present in dense star forming regions, $\eta_1\approx 1$. The second term describes momentum transferred by infrared photons re-radiated by dust particles, and scattered multiple times by dust grains before they escape, where $\eta_2$ is added to scale the fiducial value of $\tau_{\rm IR}$ (i.e., in fiducial case $\eta_2=1$).

A simple, but crude, approach to account for radiation pressure feedback would be to assume that each star particle of mass $m_*$ is a single star cluster with luminosity $L(t)=L_1(t)m_*$, where the specific luminosity $L_1(t)$ is shown in Figure~\ref{fig:sb99}, and that the infrared optical depth $\tau_{\rm IR}$ is a constant on the order of $\sim 1-10$. The total momentum injected into the ISM at every time step is then simply $p_{\rm rad}=\dot{p}\Delta t$. However, this over-simplifies the impact of radiation pressure, as the effect is not expected to be of uniform strength in star clusters of different masses \citep[e.g.,][]{krumholz_matzner09}. This fact is illustrated in Figure~\ref{fig:tauobs} where we estimate $\tau_{\rm IR}=\kappa_{\rm IR}\Sigma_{\rm cl}$ using observational data for cluster/clump masses and radii, assuming $\kappa_{\rm IR}\approx 3{\rm\ cm^2\,g^{-1}}(T_{\rm d}/100 K)^2$ \citep{semenov_etal03} at solar dust-to-gas ratios (for dust temperatures of $T_{\rm d}\gtrsim 200$~K, $\kappa_{\rm IR}\gtrsim 5\rm\ cm^2\,g^{-1}$). Although the scatter is significant, this rough estimate illustrates that very large values of the infrared optical depth are plausible in massive star clusters; e.g., the observed densities of the star clusters in M82 allows for $\tau_{\rm IR}\sim 10-100$. In less massive star clusters ($\Mcl\sim10^2-10^4$), $\tau_{\rm IR}$ is of order unity and photoionization is the dominant source of radiative feedback \citep[see e.g. recent numerical work by][]{Walch2012}, although radiation pressure may be important source of momentum even the single scattering ($\tau_{\rm IR}=0$) regime \citep[e.g.,][]{Murray2011,Wise2012}. Note that these estimates assume a homogeneous and static distribution of dense gas around the young star clusters, and the effective values of $\tau_{\rm IR}$ around young clusters are quite uncertain \citep[e.g.,][]{Hopkins2011,Kuiper2012,Krumholztau2012}.

In our fiducial simulations  we use a subgrid model of radiation pressure, based on conservative empirical estimates of $\tau_{\rm IR}$. This approach differs from recent work by \cite{Hopkins2011} where attempt is made to calculate the optical depth directly from the density structure of the numerical simulations. The resolution of our simulations is matched to the typical resolution of modern state-of-the-art cosmological simulations and at such resolution the density field on the scale of star clusters is not resolved. 

In essence, a star particle formed via the adopted star formation prescription is assumed to consist of an ensemble of star clusters situated in an ensemble of natal molecular clumps. Via the time evolution of the bolometric luminosity of each star cluster, calculated using {\small STARBURST99}, we obtain the momentum injection rate exerted onto each molecular clump. By adopting a cluster/clump mass-size relation and mass function compatible with observations, we then compute the total momentum injection rate $\dot{p}_{\rm rad}$ as the integral over all star cluster masses represented by the star particle at each simulation time step. The full description of the subgrid model, and the adopted fiducial parameters, is presented in Appendix~\ref{sect:Pradmodel}.

\subsubsection{Supernovae type II}
\label{sect:SNII}
We calculate the time at which a star of mass $M$ ends its H and He burning phases, and leaves the main sequence, using the stellar age-mass-metallicity fit given by equation 3 in \cite{Raiteri1996}. By inverting this equation, we obtain the stellar masses exiting the main sequence at a given age and metallicity. At each simulation time step, $\Delta t$, we calculate the stellar masses $M_{t_*}$ and  $M_{t_*+\Delta t}$ that bracket the stellar masses exiting the main sequence over the current $\Delta t$. If the masses are in range of $8-40\Msol$, we assume they undergo core-collapse and end up as SNII events. The number of SNII events is hence given by 
\begin{equation}
N_{\rm SNII}=\int_{M_{t_*} }^{M_{t_*+\Delta t}}{\Phi}(M){\rm d}M.
\end{equation}
Initially, the SNII explosion energy is in the form of kinetic energy of ejecta, with a typical average value of $\bar{E}_{\rm SNII}=10^{51}\,{\rm ergs}$, which is thermalized via shocks. The total thermal energy injected by SNII is thus
\begin{equation}
E_{\rm SNII}=N_{\rm SNII}\bar{E}_{\rm SNII}.
\end{equation}
The SNII ejecta also carry momentum initial momentum, which should be accounted for explicitly. We assume each supernova event imparts momentum equivalent to an ejecta mass $m_{\rm ej}=12\Msol$ ejected at $\vel_{\rm ej}=3000\kmsec$,  amounting to a total release of 
\begin{equation}
p_{\rm SNII}=N_{\rm SNII}m_{\rm ej}v_{\rm ej}
\end{equation}
per time step. We find that the values for the amount of energy and momentum injected over $\approx 40\Myr$, as computed above, are in good agreement with the total momentum  and energy injection computed using the {\small STARBURST99} code.

Following \cite{Raiteri1996}, we adopt the following fits to the results of \citet{woosleyweaver1995} for the total ejected mass ($m_{\rm ej}$), as well as the ejected mass in iron and oxygen ($m_{\rm Fe}$ and $m_{\rm O}$), as a function of stellar mass $M$ (in $\Msol$):
\begin{eqnarray}
m_{\rm ej}&=&0.77\,M^{1.06}\nonumber \\
m_{\rm Fe}&=&2.8\times10^{-4}\,M^{1.86} \\
m_{\rm O}&=&4.6\times10^{-4}\,M^{2.72}\nonumber 
\end{eqnarray}
The total and enriched amount of ejecta released at a given time step becomes
\begin{equation}
\left.  m_{\Delta t} \right|_{\rm ej,Fe,O}=\int_{M_{t_*} }^{M_{t_*+\Delta t}}\left.  m \right|_{\rm ej,Fe,O}{\Phi}(M){\rm d}M.
\end{equation}
In the {\small RAMSES} implementation, we do not track separate variables of metal species, but simply one averaged metal density variable.
The total \emph{mass} of metals returned to the ISM, accounting for the pre-existing metallicity $Z_*$ of the stellar population, is
\begin{equation}
m_{Z,{\rm SNII}}=(m_{\rm Fe}+m_{\rm O})(1-Z_*)+m_{\rm ej}Z_*.
\end{equation}
After each feedback step the ejecta and metal mass is returned to the ISM, and the star particle mass is updated accordingly. A more sophisticated numerical treatment of chemical enrichment must ultimately include contributions from all relevant species, e.g. C, N, Ne, Mg, Si, Ca and S \citep[see e.g.][]{Wiersma2009}, which we leave for a future investigation. Note that oxygen dominates the ejected heavy elements by mass. 

For the IMF given in Equation~\ref{eq:IMFK01}, a stellar population of birth mass $m_{*,{\rm ini}}=10^4\Msol$ and $Z=Z_\odot$ produces $\sim101.4$ SNII, ejects $m_{\rm ej}\sim 899.4\Msol$ of material and expels  $m_{Z,{\rm SNII}}\sim143.8\Msol$ of metals into the ISM (of which newly produced iron and oxygen accounts for $\sim128.4\Msol$).

In addition to the SNII feedback budget discussed above, which can be regarded as \emph{initial} injections of energy and momentum into the ISM, late time evolution of supernova remnants can in principle inject significantly more momentum. During the first $\sim 10-100$ years after a SNII explosion, when SN ejecta move ballistically, the adiabatic Sedov-Taylor (S-T) stage sets in \citep[e.g.][]{Blastwaves1988}, as the swept up inter-stellar material greatly exceeds the ejecta. The shock velocity is high, leading to an approximately adiabatic, energy conserving evolution. After $\sim 10^4\,{\rm years}$, the shock wave slows down sufficiently for the cooling time of post-shock gas to be of the order of or less than the age of the remnant, and an adiabatic assumption is no longer valid.   \cite{Blondin1998} calculated the transition time at which the cooling time equals the age of the remnant ($t_{\rm cool}=t_{\rm SN}$) to be $\approx 2.9\times 10^4\,E_{51}^{4/17}n_0^{-9/17}\,{\rm yrs}$, where $n_0$ is the ambient density and $E_{51}$ the thermal energy in units of $10^{51}~{\rm ergs}$. At this time, the momentum of the expanding shell is approximately
\begin{equation}
\label{eq:ST}
p_{\rm ST}=M_{\rm ST}v_{\rm ST}\approx 2.6\times 10^5\,E_{51}^{16/17}n_0^{-2/17} \Msol\kmsec.
\end{equation}
Note that $p_{\rm ST}$ depends very weakly on the surrounding gas density and linearly on $E_{51}$ and may hence be $\sim 5-15$ times greater than the initial ejecta momentum $p_{\rm SNII}$ in the density range $n=100-0.01\cc$. We regard $p_{\rm ST}$ as an upper limit to what a single SN explosion can generate, as a substantial portion of the energy is lost in shocks (see \S~\ref{sect:shocks}), and the classical S-T solution assumption of a perfectly intact thin shell expanding into a homogeneous medium is almost certainly a simplification. If stellar winds and radiation pressure are sufficiently effective in expelling gas from young star clusters during the first $3-4$~Myrs, hot gas may simply escape the natal cloud via the cleared channels. A spherical model for blast-wave evolution is clearly incorrect in such cases. Keeping this in mind, a scenario of maximally efficient S-T momentum generation can be modelled by replacing our fiducial choice $p_{\rm SNII}$ by $p_{\rm ST}$ \citep[e.g., as is done by][]{ShettyOstriker2012}.

\subsubsection{Supernovae type Ia}
Following \cite{Raiteri1996}, we assume that progenitors of SNIa are carbon plus oxygen white dwarfs that accrete mass from their binary companions. Stellar evolution theory predicts that the binary masses that can given rise to white dwarfs exceeding the Chandrasekhar limit to be in the range of $\sim3-16\Msol$. The number of SNIa events within a star particle, at a given simulation time with an associated time step $\Delta t$, is then 
\begin{equation}
N_{\rm SNIa}=\int_{M_{t_*}}^{M_{t_*+\Delta t}}\hat{\Phi}(M_2){\rm d}M_2,
\end{equation}
where $\hat{\Phi}(M_2)$ is the IMF of the secondary star \citep[][]{GreggioRenzini1983,Raiteri1996}, 
\begin{equation}
\hat{\Phi}(M_2)=A'\int_{M_{\rm inf}}^{M_{\rm sup}}\left(\frac{M_2}{M_{\rm B}}\right)^2 M_{\rm B}^{-2.3} {\rm d}M_{\rm B},
\end{equation}
where $M_{\rm B}$ is the mass of the binary, $M_{\rm inf}={\rm max}(2M_2,3\Msol)$ and $M_{\rm sup}=M_2+8\Msol$. The normalization parameter is set to $A^\prime=0.24\,A$ (see Equation~\ref{eq:IMFK01}). Each explosion is assumed to release $\bar{E}_{\rm SNIa}=10^{51}\,{\rm ergs}$ as thermal energy, hence injecting a total of $E_{\rm SNIa}=N_{\rm SNIa}\bar{E}_{\rm SNIa}$ at each time step. We assume each SNIa to be at the Chandrasekhar limit ($M_{\rm ch}=1.38\Msol$), and that this is  the ejected mass upon explosion leading to ${m}_{\rm SNIa}=N_{\rm SNIa}M_{\rm ch}$. 

We allow each SNIa event to produce $0.76\Msol$ of metal enriched material ($0.13\Msol$ of ${}^{16}$O and $0.63\Msol$ of ${}^{56}$Fe) \citep{Thielemann86}. Note that we explicitly account for late time mass lass of low mass stars ($M<8\Msol$) until the point they exit the main sequence, see \S\ref{sect:massloss}

For the assumed value of $A^\prime$, approximately $15\%$ of all SNe are of type Ia over the lifespan of a $1\Msol$ star (10 Gyr). This is compatible with the notion that $10-20\%$ of the SNe rate in galaxies with ongoing star formation, such as late type spirals (Sbc-Sd), are due to type Ia events \citep{VandenberghMcclure94}.

\begin{figure*}[t]
\begin{center}
\begin{tabular}{cc}
\includegraphics[scale=0.42]{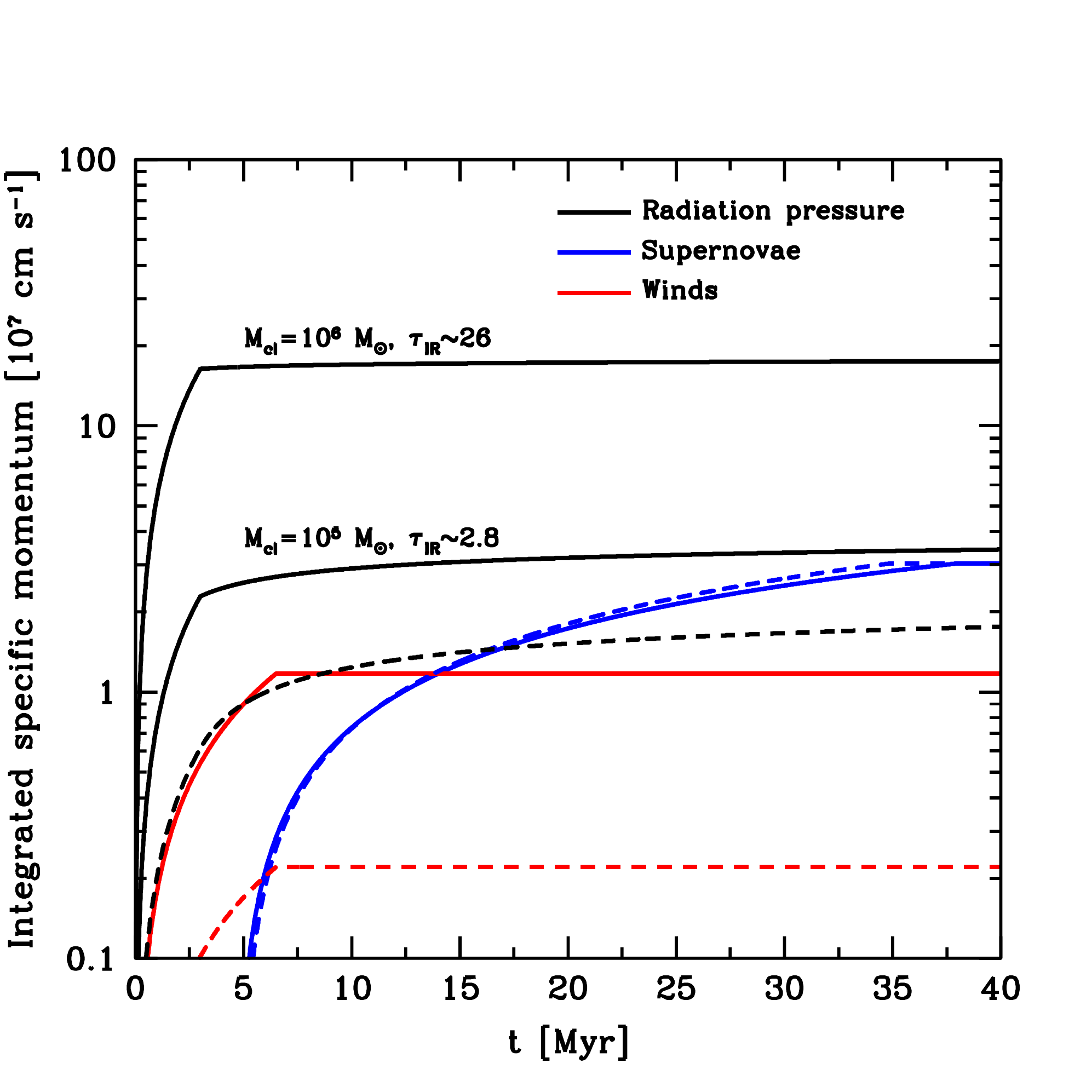}
\includegraphics[scale=0.42]{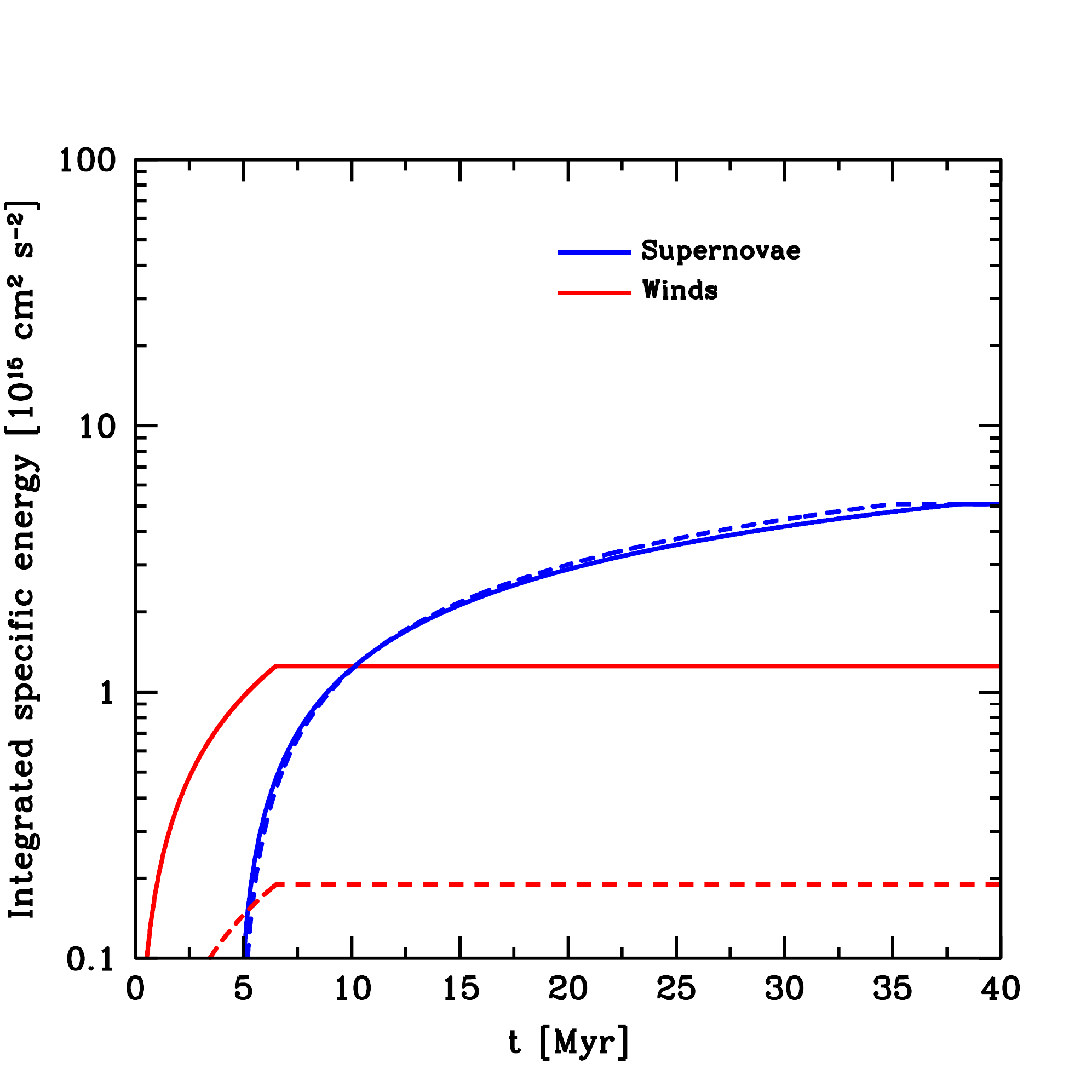}
\end{tabular}
\caption{Left panel: Integrated specific momentum due to radiation pressure (black line), supernovae type II (blue lines) and stellar winds (red lines) for stellar population of $Z=1\,Z_\odot$ (solid lines) and $Z=0.01\,Z_\odot$ (dashed lines) as a function of time. As discussed in \S~\ref{sect:radpressure}, the momentum transferred to gas via radiation pressure depends non-linearly on star population mass. Therefore, we show $p_{\rm rad}$ (Equation~\ref{eq:pradfull}) for $m_*=10^5\Msol$ and $10^6\Msol$. For $m_*=10^5\Msol$, the total injected $p_{\rm rad}$ equals the direct SNe momentum injection for $Z=1\,Z_\odot$, and momentum injected by stellar winds is $\sim 30\%$ of this value. However, radiation pressure and stellar winds have deposited essentially all their momentum before the first SNe explode (here $t\sim4.6\Myr$). This is \emph{always} the case, regardless of metallicity. At $Z=0.01\,Z_\odot$, supernova momentum dominates, as radiation pressure is weakened due to inefficient photon trapping ($p_{\rm rad}\approx 0.6 p_{\rm SNII}$ after 40 Myr). The contribution from stellar winds also becomes negligible. For $m_*=10^{6}\Msol$, the radiation pressure model predicts $\tau_{\rm IR}\sim 26$ (for $Z=1\,Z_\odot$), making the final momentum contribution from radiation almost an order of magnitude greater than all other feedback sources. Right panel: the integrated specific energy injected by the shocked SNe and wind ejecta. The lines correspond to the same metallicities as in the left panel.}
\label{fig:momentum}
\end{center}
\end{figure*}
\subsubsection{Stellar mass loss by low mass stars}
\label{sect:massloss}
Although low mass stars ($M\lesssim 8\Msol$) contribute a negligible amount to the total momentum and energy budget, they shed a considerable amount of mass during the asymptotic giant branch (AGB) phase of their evolution \citep[e.g.][]{Hurley2000}. \cite{Kalirai2008} provides relation between the initial stellar mass and final mass of the remnant in the relevant mass range:
\begin{equation}
M_{\rm final}=(0.109\pm0.007)M_{\rm initial}+0.394\pm0.025\Msol.
\end{equation}
Using the average values, the fraction of mass lost from a star during its lifetime is
\begin{equation}
f_{\rm loss}(M_{\rm initial})=0.891-0.394/M_{\rm initial}.
\end{equation}
Given a star particle of birth mass $m_{*,{\rm ini}}$ and age $t_*$ we calculate, at each time step $\Delta t$, the expelled stellar mass\footnote{\cite{Agertz2011} contained a typo that omitted a factor $M$ from their equation 8. The actual numerical implementation was however correct.} as
\begin{equation}
m_{\rm loss}=\int_{M_{t_*}}^{M_{t_*+\Delta t}}M f_{\rm loss}(M)\Phi(M){\rm d}M.
\end{equation}
The lost stellar mass is added to the gas mass in the corresponding cell. The gas metallicity is also updated to take into account metals added as part of the stellar material, $m_{\rm Z,loss}=Z_*\,m_{\rm loss}$. The mass loss is assumed to be quiescent, i.e., no momentum, other than that of the natal star particle with respect to the ISM, or energy is released. For the IMF in Equation~\ref{eq:IMFK01}, a stellar population loses $\sim25\%$ of its mass  from stars in the mass range $0.5-8\Msol$ during 10 Gyr of evolution.
 
\subsection{Feedback budget comparison}
\label{sect:budget}
In \S\ref{sect:intro} we stated that radiation pressure, stellar winds and SNe have roughly the same momentum injection rate $\dot{p}$. This is shown explicitly in the left-hand panel of Figure~\ref{fig:momentum}, where we plot the time evolution of the integrated specific momentum injected into gas, i.e. $\int p(t)/m_* {\rm d}t$, due to radiation, supernovae and stellar winds for $Z=0.01\,Z_\odot$ and $Z=1\,Z_\odot$ calculated using formulae described above. Note that stellar winds and radiation pressure inject momentum into the ISM immediately after star cluster birth, while SNIIe inject momentum during $t\sim4.6-38\Myr$. The cumulative contribution of stellar winds alone dominates over SNIIe in the first $\sim 14\Myr$ ($6\Myr$) for $Z=1\,Z_\odot$ ($0.01\,Z_\odot$). In the low metallicity case, $\sim 5$ times less momentum is injected via winds into the ISM. As we parametrize the energy release in a similar fashion, the same trends are found for the shocked wind and SNe energy shown in the right-hand panel of Figure~\ref{fig:momentum}.

At solar metallicity, the dominant source of momentum is radiation pressure, reaching the equivalent total specific SN momentum after only 3 Myr (see Equation~\ref{eq:pradfull}), assuming a stellar population of mass $m_*=10^5\Msol$. The result weakens by a factor of $\sim2$ for $Z=0.01\,Z_\odot$ as infrared trapping becomes negligible (note that we only assume photon trapping for $t\leq t_{\rm cl}=3\Myr$ during which cluster stars are assumed to be fully embedded in their natal gas clump). The non-linear behaviour of the strength of radiation pressure with the mass of the stellar population is evident, as shown by comparing results for $m_*=10^5\Msol$ and $10^6\Msol$, where our model (via Equation~\ref{eq:pradfull}) predicts $\tau_{\rm IR}\approx26$ for the latter. This illustrates how radiation pressure can be an important, and even dominant, source of feedback in dense gas associated with young massive star clusters, as it operates at early times before the first SNIIe explode.

Recall that the wind and SNe momenta in Figure~\ref{fig:momentum} refer to the initial \emph{ejecta} momentum and not any late stage momentum generated by an expanding bubble. The momentum expected from the ideal adiabatic Sedov-Taylor phase (Equation~\ref{eq:ST}) is greater than radiation pressure momentum even in the case of a supermassive ($M_{\rm cl}=10^6\Msol$) star cluster. However, as we argued above, it is not clear whether the S-T solution is applicable in the highly inhomogeneous density field of GMCs, especially if gas around young star clusters is partially cleared by early feedback. 

\begin{figure}[t]
\begin{center}
\includegraphics[scale=0.4]{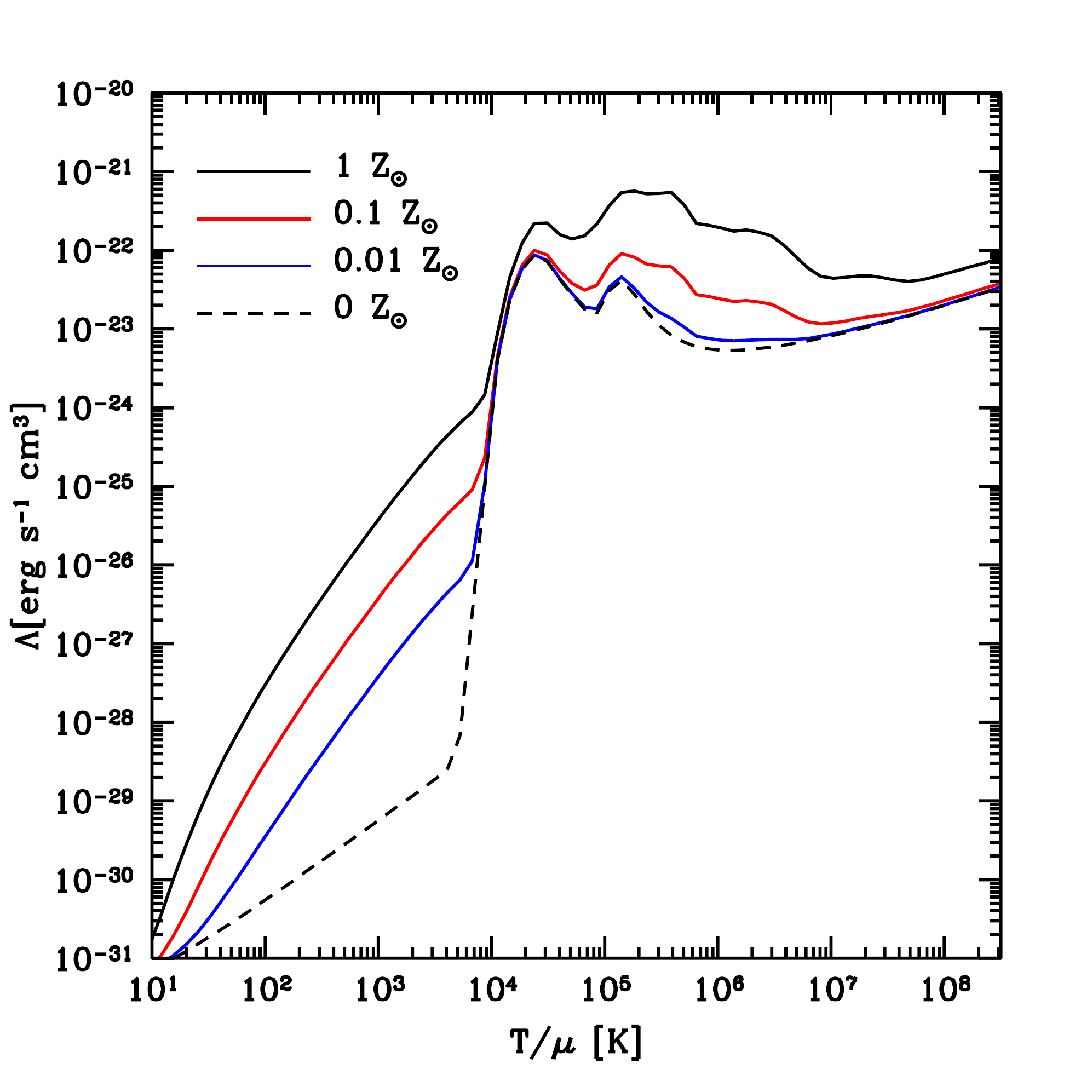}
\caption{The cooling function implemented in the {\small RAMSES} code for $Z=0,0.01,0.1$ and $1\,Z_\odot$ in the absence of a UV background. 
}
\label{fig:coolingfunction}
\end{center}
\end{figure}

\subsubsection{Thermal energy of shocked wind and SNII ejecta}
\label{sect:shocks}
The fate of thermal energy of gas due to thermalized wind and SNII ejecta can be illustrated as follows. Following \cite{sutherlanddopita93}, we define the cooling time scale as $t_{\rm cool}\equiv U/\Lambda_{\rm net}$, where the thermal energy density $U=3nkT/2$ and the net cooling function $\Lambda_{\rm net}=n_e n_i\Lambda_{\rm N}$. Here $n=n_e+n_i$, where $n_e$ and $n_i$ are the number density of electrons and ions respectively, and $\Lambda_{\rm N}$ is the normalized cooling rate in units of ${\rm ergs\,s}^{-1}\,{\rm cm}^3$.

In a fully ionized primordial plasma at $T=10^6\,$K, the normalized cooling rate is $\log(\Lambda_{\rm N})\approx -23.2$ and $\approx-22$ for gas at $Z=1\,Z_\odot$, see Figure~\ref{fig:coolingfunction} where we plot the cooling function used by {\small RAMSES} in the absence of a UV background. In the latter case, the cooling time is

\begin{equation}
\label{eq:tcool}
t_{\rm cool}\approx10^3\left(\frac{100\,{\rm cm}^{-3}}{n_{\rm H}}\right )\,{\rm years},
\end{equation}
and roughly 20 times greater for a pristine plasma. Clearly, the cooling time is very short at average densities relevant for GMCs, and hot gas is quickly radiated away, unless a strong local heating source can maintain it.

The criterion for heating to dominate over cooling can be written as
\begin{equation}
n_{\rm H}^2\Lambda_{\rm N}\leq \rho_*\bar{\Gamma},
\label{eq:heatcrit}
\end{equation}
where $\rho_*$ is the mass density of stars and $\bar{\Gamma}$ is the specific heating rate of the gas in units of ${\rm ergs}\,{\rm s}^{-1}\,{\rm \Msol}^{-1}$. \cite{ceverinoklypin09} argued that a very large fraction of a cell's mass must be converted into stars for feedback heating to overcome the radiative cooling for gas at $T=10^4\K$, where the cooling function peaks. Even at low densities, $n_{\rm H}\sim 0.1\cc$, the stellar-to-gas mass fraction must be above unity. Indeed, by inserting typical values for cooling and heating (see Figure~\ref{fig:sb99}), and scaling to a typical numerical resolution of $\Delta x=40\pc$, Equation~\ref{eq:heatcrit} relation can be written as 
\begin{equation}
\label{eq:heatcool}
n_{\rm H}^2\leq 0.535 \left(\frac{m_*}{10^4\Msol}\right)\left(\frac{40\pc}{\Delta x}\right)^3\left(\frac{10^{-22}}{\Lambda}\right)\left(\frac{\bar{\Gamma}}{10^{34}}\right),
\end{equation}
where the cooling function $\Lambda$ is in units of ${\rm ergs}\,{\rm s}^{-1}\,{\rm cm}^3$ and the specific heating rate $\bar{\Gamma}$ in units of ${\rm ergs}\,{\rm s}^{-1}\,{\rm \Msol}^{-1}$. In a cell of size $\Delta x=40\pc$, a stellar-to-gas fraction of \emph{at least} $m_*/m_{\rm gas}\sim10$ is required for heating to overcome cooling (at $T\sim$ few $10^4\,$K), which is unachievable via star formation alone unless at least $90\%$ of the original cell mass was converted into stars. This is an order of magnitude greater than what is observed in massive GMCs \citep{Evans2009,Murray2011}. As argued by Ceverino \& Klypin, gas cooling rates drop by orders of magnitude at lower gas temperatures, making it possible for thermal feedback to maintain greater pressure gradients between dense star forming regions and the ambient ISM. This leads to expansion of the star forming region that lowers the average density, eventually bringing the medium into a regime where heating can overcome cooling. 

The estimates made above are subject to many caveats. While relevant to understand the fate of thermal energy injected into gas in galaxy formation simulations, the real ISM is multiphase and highly inhomogeneous on the scale of the resolution elements of such simulations. This means that pockets of tenous hot gas may exist within dense gas in a simulation cell, but it also means that estimates of the cooling time are optimistic as they need to include a clumping factor that is expected to be significant in star forming regions. However, it is unclear how efficiently thermal energy should couple to the ISM in realistic settings; \cite{ChoKang2008} demonstrated, using high resolution simulations of SNe explosion in pre-existing wind-blown bubbles, that less than $\sim10\%$ of the shocked thermal energy could be converted into kinetic energy, as the rest is lost in radiative shocks within the bubble.

Keeping these issues in mind, the effect of gas clearing due to pre-SNe momentum feedback may in many situations enhance the effect of feedback, which is one of the main motivations of this work.

\subsection{The star formation recipe}
\label{sect:SFlaw}
In this work we employ a fairly standard prescription star formation based on the star formation rate given by
\begin{equation}
\label{eq:schmidt}
\dot{\rho}_{*}=\frac{\rho_{\rm g}}{t_{\rm SF}}\,\,{\rm for} \,\,\rho>\rho_{*},
\end{equation}
where $\rho_{\rm g}$ is the gas density, $\rho_*$ the threshold of star formation, and $t_{\rm SF}$ the star formation, or equivalently gas depletion, time. Observations indicate that in the local universe $t_{\rm SF}\sim2\Gyr$ \citep{Bigiel2011}, which is a manifestation of the fact that observed galaxies convert their gas into stars quite inefficiently. 

In this work we assume that $t_{\rm SF}=t_{\rm ff}/\epsilon_{\rm ff}$, where $t_{\rm ff}=\sqrt{3\pi/32G\rho}$ is the local gas free-fall time and $\epsilon_{\rm ff}$ is the star formation efficiency per free-fall time. With this assumption Equation~\ref{eq:schmidt} enforces $\dot{\rho}_*\propto \rho^{1.5}$, which is close to the observed \emph{projected} density relation $\Sigma_{\rm gas}\sim\Sigma_{\rm SFR}^n$, where $n\sim1.4$ \citep{kennicutt98}. As noted above, the efficiency of star formation is globally observed to be low, $\epsilon_{\rm ff}\sim 1\%$ \citep{krumholztan07}, and we discuss our adopted values of $\epsilon_{\rm ff}$ in \S\ref{sect:simulations}. 

For now, we would like to note that the efficiency of star formation per free fall is usually kept fixed in  galaxy formation simulations. However, it is likely that this is not the case in observations. In fact, there is ample observational and theoretical evidence for $\epsilon_{\rm ff}$ to depend on scale and environment \citep[e.g.][]{Murray2011,Padoan2012}, which will manifest as stochasticity of star formation efficiency. Such stochasticity can potentially have a strong impact on feedback, because it implies that $\epsilon_{\rm ff}$ can be high in some regions and low in others. The overall star formation would thus be concentrated in fewer star forming sites that have high star formation efficiency, even as the global star formation efficiency averaged over a large patch of ISM is low. We leave an investigation of the effects of such stochastic efficiency on the effects of feedback for future work (Agertz et al. in prep.), and note that this caveat should be kept in mind when interpreting the numerical result presented below.

Recent work by \cite{Gnedin09} and \cite{GnedinKravtsov11}  relate star formation to molecular gas, hence $\rho_{\rm g}\rightarrow \rho_{\rm H_2}$ in Equation~\ref{eq:schmidt}, which can explain why metal/dust poor galaxies at $z\sim 3$, which physically should be more prone to H$_2$ destruction via UV dissociation, show deviations from the $z=0$ K-S relation \cite{GnedinKravtsov2010}. \cite{GnedinKravtsov11} demonstrated that the density at which molecular fraction reaches $50\%$ can be approximated as 
\begin{equation}
\label{eq:gnedin09}
n_*\approx 25\,(Z/Z_\odot)^{-1}\cc,
\end{equation}
which we adopt in all of our simulations as the threshold for star formation.  In addition to the density threshold we also use the temperature threshold by only allowing  star formation to occur in cells of  $T<10^4\K$. No other conditions or thresholds are used. 

To ensure that the number of star particles formed during the course of a simulation is tractable, we sample the Equation~\ref{eq:schmidt} stochastically at every fine simulation time step $\Delta t$. For a cell eligible for star formation, the number of star particles to be formed, $N$, is determined using a Poisson random process \citep{Rasera06,dubois08}
\begin{equation}
P(N)={\lambda_P \over N !} \exp({-\lambda_P}) \, ,
\label{poisson_law}
\end{equation}
where the mean is
\begin{equation}
\lambda_P=  \left ( {\dot{\rho}_* \Delta x^3 \over m_*}\right ) {\Delta t }.
\label{poisson_param}
\end{equation}
Here $\dot{\rho}_*$ is the adopted star formation rate (Equation~\ref{eq:schmidt}), and $m_{*,\rm min}$ is the chosen unit mass of star particles. In this work we adopt $m_{*,\rm min}=\eta\rho_*\Delta x_{\rm max}^3$, where $\eta=0.1$, and $\rho_*$ is taken from Equation~\ref{eq:gnedin09} at solar metallicity. This yields $m_{*,\rm min}\sim10^4\Msol$ for a typical resolution of $\Delta x_{\rm max}=50\pc$. When the Poisson process produces $N>1$ star particles in a cell at a single star formation events, we bin these into one stellar particle of mass $Nm_*$.

\section{Numerical implementation of feedback}
\label{sect:implementation}
The efficiency of stellar feedback depends not only on its magnitude, but also on specifics of implementation in a given numerical code \citep[see, e.g.,][and references therein]{Aquila}. In this work we are mainly interested in gauging the impact of stellar feedback at the state-of-the-art resolution of modern galaxy formation simulations without resorting to ad hoc suppression of cooling \citep{Gerritsen1997PhDT, ThackerCouchman2000, Stinson06, Governato07, Agertz2011} or hydrodynamical decoupling of gas elements \citep{Scannapieco2006, OppenheimerDave2006}.

We choose to inject energy and momentum directly into computational cells as follows. Over a simulation time step $\Delta t$, we calculate the thermal energy release ($E_{\rm tot}$), as well as the associated mass of ejecta  ($m_{\rm tot}$) and metals ($m_{\rm Z,tot}$). These quantities are deposited in the 27 cells surrounding the star particle, although we have also carried out most of our experiments using nearest grid point approach without significant differences to the final results\footnote{One may add further sophistication to this approach by considering supernovae explosions as discrete events, hence only applying $E_{\rm SNII}+E_{\rm SNIa}$ when an integer number of explosions occur during over the time-step \citep[see e.g.][]{Hopkins2012}}. We explore two different methods to deposit momentum:
 
\paragraph{(1) Momentum "kicks"} Over a simulation time step $\Delta t$, the momentum $p_{\rm tot}=\dot{p}_{\rm tot}\Delta t$ is directly deposited isotropically in the 26 cells surrounding the grid cell nearest to star particle.

\paragraph{(2) Non-thermal pressure} The momentum injection rate $\dot{p}_{\rm tot}$ can be thought as a non-thermal pressure corresponding to momentum flux through cell surface $P_{\rm nt}=\dot{p}_{\rm tot}/A$, where the area $A$ is the surface are of a cell ($A=6\Delta x^2$), or an arbitrary computational region, containing a young star particle. This pressure is calculated at every time step and is added to the thermal pressure, $P_{\rm thermal}$  to give total pressure $P_{\rm eff}=P_{\rm therm}+P_{\rm nt}$ that enters in the Euler equation. We describe this technique in detail in Appendix~\ref{sect:nonthermalP}.
\\ \\
The first method is qualitatively similar to what was considered by \citet{NavarroWhite93}, although these authors compute the momentum corresponding to a fraction of injected SNII energy, while we specifically compute the momentum injection due to various specific processes that generate momentum.   
The advantage of the first implementation method is its simplicity, as the second method requires minor modifications to the Riemann solver in the case of the MUSCL-Hancock scheme \citep{Toro1999} adopted by the {\small RAMSES} code. On the other hand, the first method does not explicitly affect the cell containing the feedback producing star particle, which will be evacuated in the case of a pressure-approach. We adopt the first method as our fiducial choice, but we present results of both implementations in \S\ref{sect:simulations}.

Strong heating and/or momentum deposition in diffuse regions can lead to extremely large temperatures and velocities. To avoid this, we disallow feedback if cell temperature is $T\gtrsim 5\times 10^8\K$ and limit momentum feedback to deliver maximum kicks of $\vel=1000\kmsec$.

The effect of momentum feedback is weakened when star particles occur in neighboring regions, or even computational cells, as momentum cancellations will occur \citep[see e.g.][]{Socrates2008}. \cite{Hopkins2011} discussed this effect in their SPH simulations (see their figure A1 and associated text). They maximized the effect of feedback by depositing momentum isotropically from the cloud's center of mass found by an FOF technique. In addition, momentum was deposited in a probabilistic way that ensured that each affected SPH particle would receive a velocity kick at the local cloud escape velocity. If momentum was added gradually around each stellar particle, akin to our current method, Hopkins et al. found that feedback limited star formation less efficiently (by factor of $\sim 5$ in the measured star formation histories). This effect should be kept in mind as an implementation uncertainty.

\subsection{Increasing the impact of hot gas by delayed cooling}
\label{sect:hotgas}
As we demonstrate in \S\ref{sect:resscale}, momentum feedback aids in clearing gas out from star forming regions, and runaway heating can occur in some regions. However, it is still not guaranteed that the evolution of hot gas is accurately captured due to resolution effects (see discussion in \S~\ref{sect:shocks}). In addition to relying solely on early momentum feedback to clear out dense gas, we also consider the two following methods to capture the maximum effect that thermal energy from SNII may have on their surroundings.

The concept of allowing for an adiabatic feedback phase in galaxy scale simulations has been proposed by several authors, \cite[see e.g.][]{Gerritsen1997PhDT,Stinson06}, and is widely utilized in the community \citep{Governato07,Agertz2011,Brook2012}. However, the specific implementations assume the duration of this phase to be much longer than the $\sim10^4\,$ years expected from analytical arguments. \cite{Stinson06} proposed a scheme in which SNe energy is deposited in a region of size $R_{\rm SP}=10^{1.74}E_{51}^{0.32}n_0^{-0.16}\tilde{P}_{04}^{-0.20}\pc$, where $\tilde{P}_{04}=10^{-4}P_0k_{\rm B}^{-1}$ and $P_0$ and $n_0$ are the ambient pressure and density, and cooling is disabled for $t_{\rm max}=10^{6.85}E_{51}^{0.32}n_0^{0.34}\tilde{P}_{04}^{-0.70}\,{\rm years}$. However, the time scale $t_{\rm max}$ corresponds to the survival time of the low-density cavity \citep{McKeeOstriker1977}, \emph{not} the adiabatic phase of SNe. Furthermore, as supernovae energy in the Stinson et al. implementation is delivered at every time step, as in the method described in \S\ref{sect:SNII}, most of the gas in the star forming region will behave adiabatically for $\sim 40\Myr$, assuming a minimum SNII mass of $8\Msol$.

Having noted that cooling suppression models typically exaggerate the effect of SNII energy they are meant to mimic, we consider the effects of one such model below in a subset of our simulations, and compare it with results of simulations with no delay of cooling. When a star particle forms, we assign the time variable $t_{\rm cool}$ to a scalar in the cell containing the particle. This scalar field is passively advected with the hydro flow. At every time step, the variable is updated as $t_{\rm cool}^{t+\Delta t}=t_{\rm cool}^t-\Delta t$. For every cell where $t_{\rm cool}>0$, cooling is disabled. This method approximates the delay of cooling implemented in SPH codes \citep[e.g.,][]{Stinson06} within the Eulerian hydrodynamics context.\footnote{Note however that we assign the cooling delay time $t_{\rm cool}$ to the gas present in the local cell at star particle birth, while \citet{Stinson06} assign $t_{\rm cool}$ to SPH particles available within the blast wave radius $R_{\rm SP}$ at every $\Delta t$ for the duration of SNII explosions. The gas particles affected by delayed cooling at the end of the SNII phase in the Stinson et al. approach are not necessarily the same particles that were present at birth. Furthermore, our $t_{\rm cool}$ variable is allowed to mix, leading to delayed cooling in cells previously not associated with the young star particle's birth cell. } We explore the effects of delayed cooling using two values of $t_{\rm cool}$: 10 and 40 Myr. The latter is, as argued above, the duration of SNe feedback for stars $\gtrsim8\Msol$.

\begin{figure*}[t]
\begin{center}
\begin{tabular}{cc}
\includegraphics[scale=0.4]{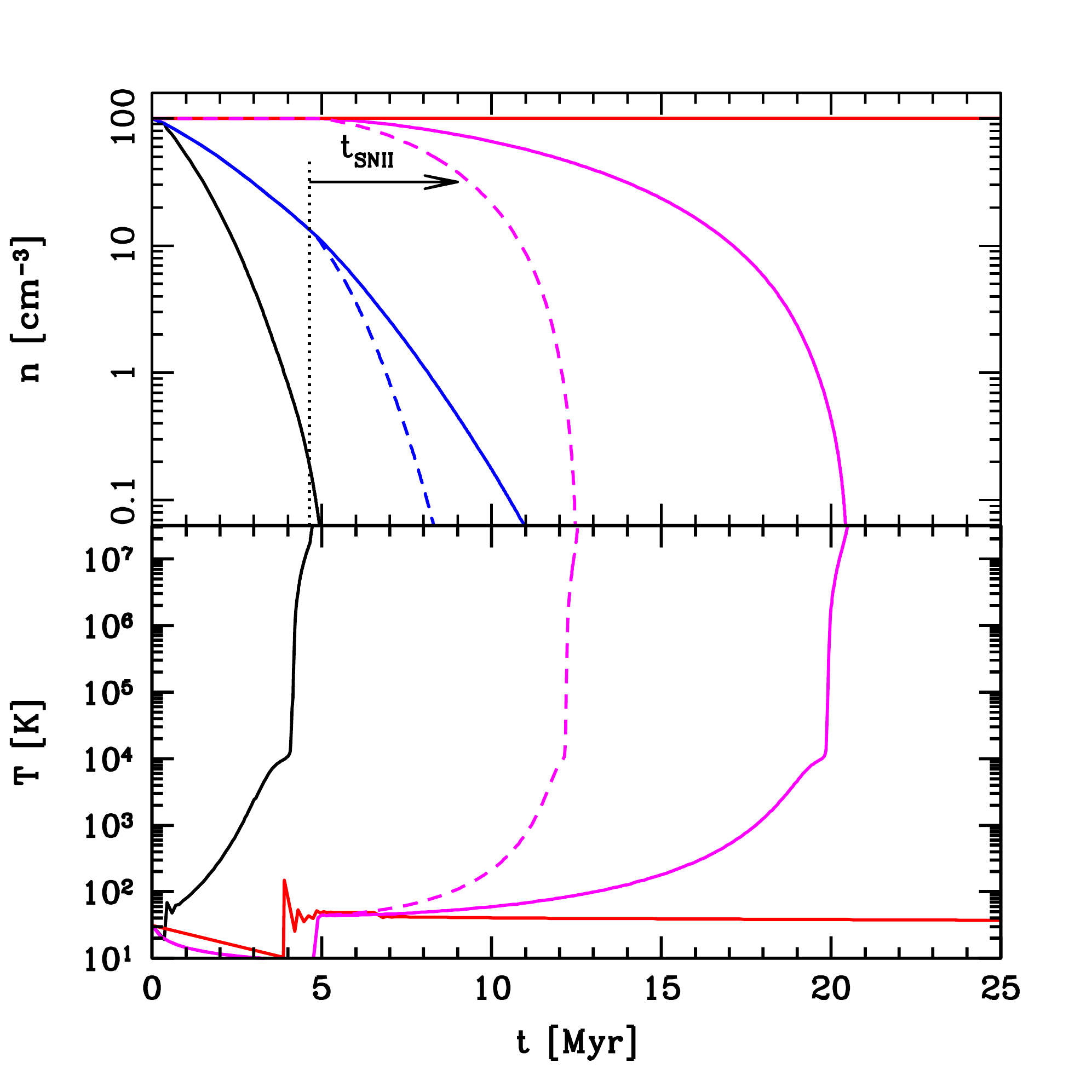}
\includegraphics[scale=0.4]{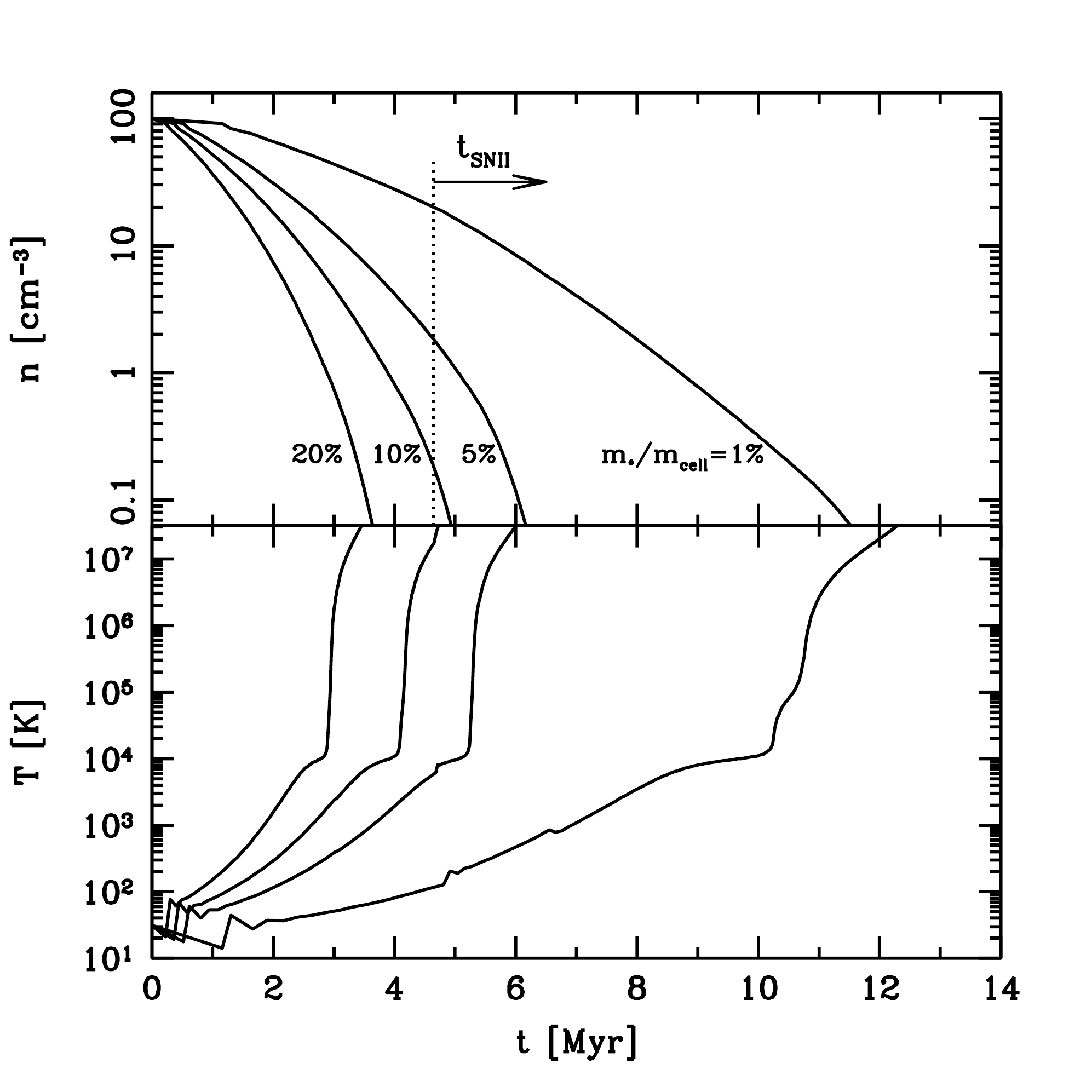}\\
\end{tabular}
\caption{Left panel: evolution of gas density and temperature in computation cell containing star particle for the {\small ALL} (black solid line), {\small MOMENTUM}  (blue solid line), {\small SNMOM} (magenta solid line) {\small ENERGY} simulations (red solid line), see table~\ref{table:cellsim}.  Dashed lines indicate if Sedov-Taylor momentum was used instead of simply the ejecta momentum for SNIIe. Note that simulations including momentum injection from stellar wind and radiation pressure can clear out enough gas for runaway heating before the first SNe explosion occurs ($\sim 4.6\Myr$). The predicted transition to a runaway heating regime (Equation~\ref{eq:heatcool}) agrees well with the results of simulations. Purely thermal energy feedback ({\small ENERGY} run) has almost no effect on gas density and temperature. SNe feedback with momentum can clear out the gas, enabling run-away heating, but considerably slower than in the {\small ALL} run. Right panel: Gas density and temperature evolution in cell containing star particle in the {\small ALL} simulation for different fraction of gas mass converted into stars. The black solid lines show, from right to left, stellar mass fraction $f_*=1,5,10$ and $20\%$. The predicted transition to regime of runaway heating occurs before the first SN explosion for $f_*>10\%$. Note that runaway heating and efficient clearing of gas occur even when only 1\% of gas is turned into stars.
}
\label{fig:evac}
\end{center}
\end{figure*}

\subsection{Feedback energy variable}
\label{sect:fbvar}
We also investigate a scenario in which some fraction of the feedback energy is evolved as a separate energy variable $E_{\rm fb}$, which is passively advected with the hydro flow and only couples directly to the hydrodynamic flow as an effective pressure in the Euler equations. We assume that this energy dissipates over a timescale $t_{\rm dis}$, which is assumed to be longer than the cooling time $t_{\rm cool}$ predicted by the cooling rates in dense star forming gas, see Equation~\ref{eq:tcool}. This approach can be viewed as accounting for the effective pressure from a multiphase medium, where local pockets of hot gas exert work on the surround cold phase. Alternatively, it may be viewed as feedback driven turbulence \citep{Springel2000}, although proper treatment of subgrid turbulence requires not only addition of turbulent pressure, but also significant modifications to the equations solved by the code in order to accurately model turbulent stresses and dissipation \citep{Iapichino2011,Schmidt2011}, which is beyond the scope of this paper

In practice, at every time step we assume that a fraction $f_{\rm fb}$ of the total thermal feedback energy $E_{\rm tot}$ is added to the feedback energy $E_{\rm fb}$, and the remaining ($1-f_{\rm fb}$) enters the thermal energy of the gas. We experiment with $f_{\rm fb}=0.1$ and 0.5, where the lower value is motivated by the radiative SN-driven bubble simulations of \cite{ChoKang2008}. The pressure associated with $E_{\rm fb}$ enters into the sound-speed calculation, as well as in the Riemann solver. During the cooling step, dissipation is modelled as $E_{\rm fb}^{t+\Delta t}=E_{\rm fb}^{t}\exp(-\Delta t/t_{\rm dis})$ in every gas cell. The retention of feedback energy is here rather different than in delayed cooling method described above; in the latter, the cooling delay operates over fixed time only in the gas present in a star particle's birth region.

$E_{\rm fb}$ will only be important in local dense star forming gas, where most of the thermal energy is radiated away due to high \emph{average} density. In diffuse regions the energy budget will be dominated by the surviving thermal energy which dissipates consistently on its proper cooling time scale. We assume the dissipation time scale to be comparable to the decay time of supersonic turbulence, i.e. of order of the flow crossing time \cite[e.g.][]{Ostriker2001}. Massive GMCs typically have sizes of $\sim 10\pc$ and velocity dispersions of $\sim 10\kmsec$, leading to a crossing time of $t_{\rm cr}\sim 1\Myr$.  At the scale of the disk, where the cold gas layer thickness is an order of magnitude thicker, one may argue for $t_{\rm cr}\sim 10\Myr$. We hence consider feedback energy dissipation time in the range $t_{\rm dis}=1-10\Myr$.

\cite{Teyssier2012} recently demonstrated that a feedback scheme employing a separate energy variable, similar to what is described above, is quite efficient and has a significant effect on the star formation history, and dark matter density profile, of an isolated dwarf galaxy. 

\begin{figure*}[t]
\begin{center}
\begin{tabular}{cc}
\includegraphics[scale=0.43]{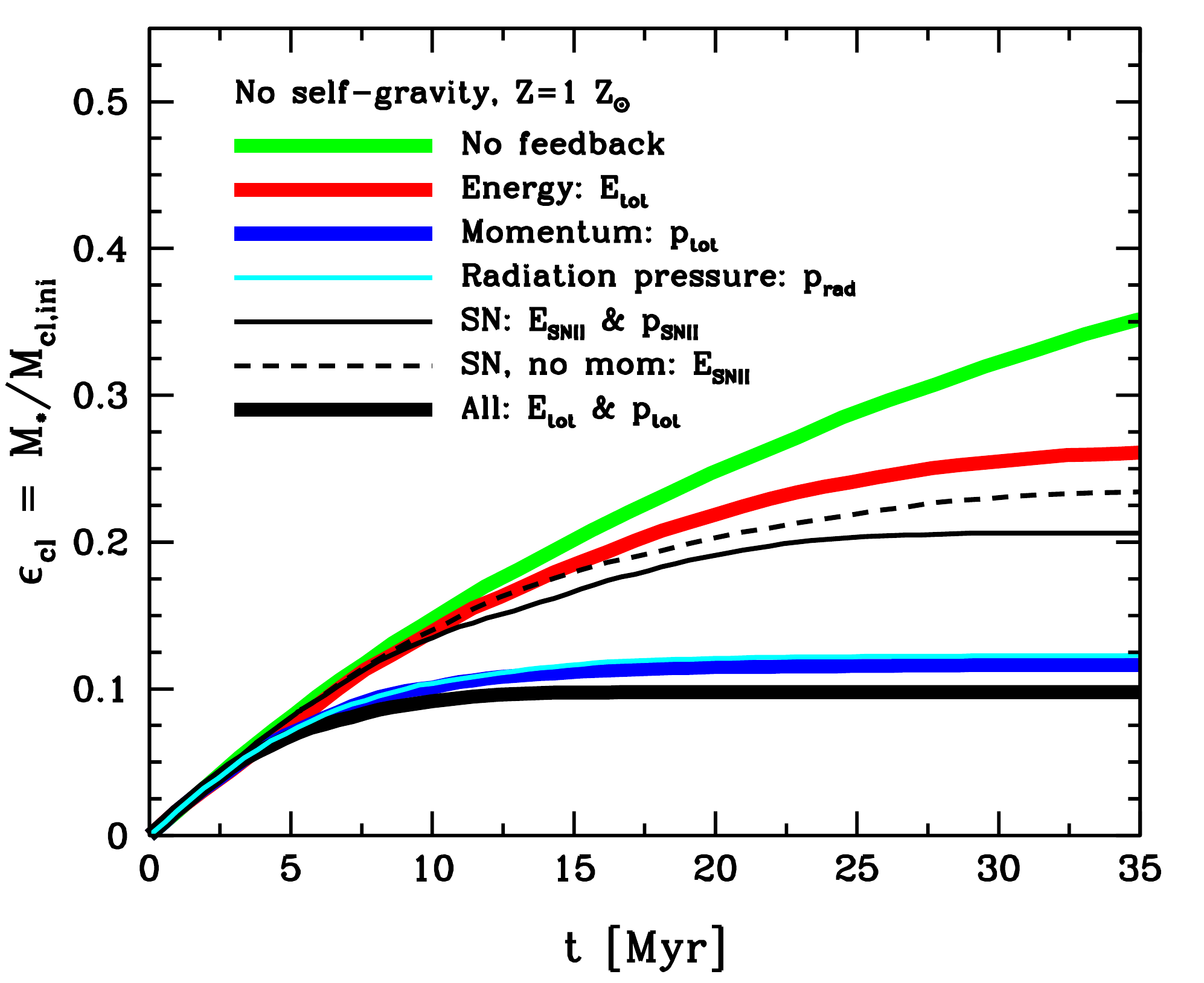}
\includegraphics[scale=0.43]{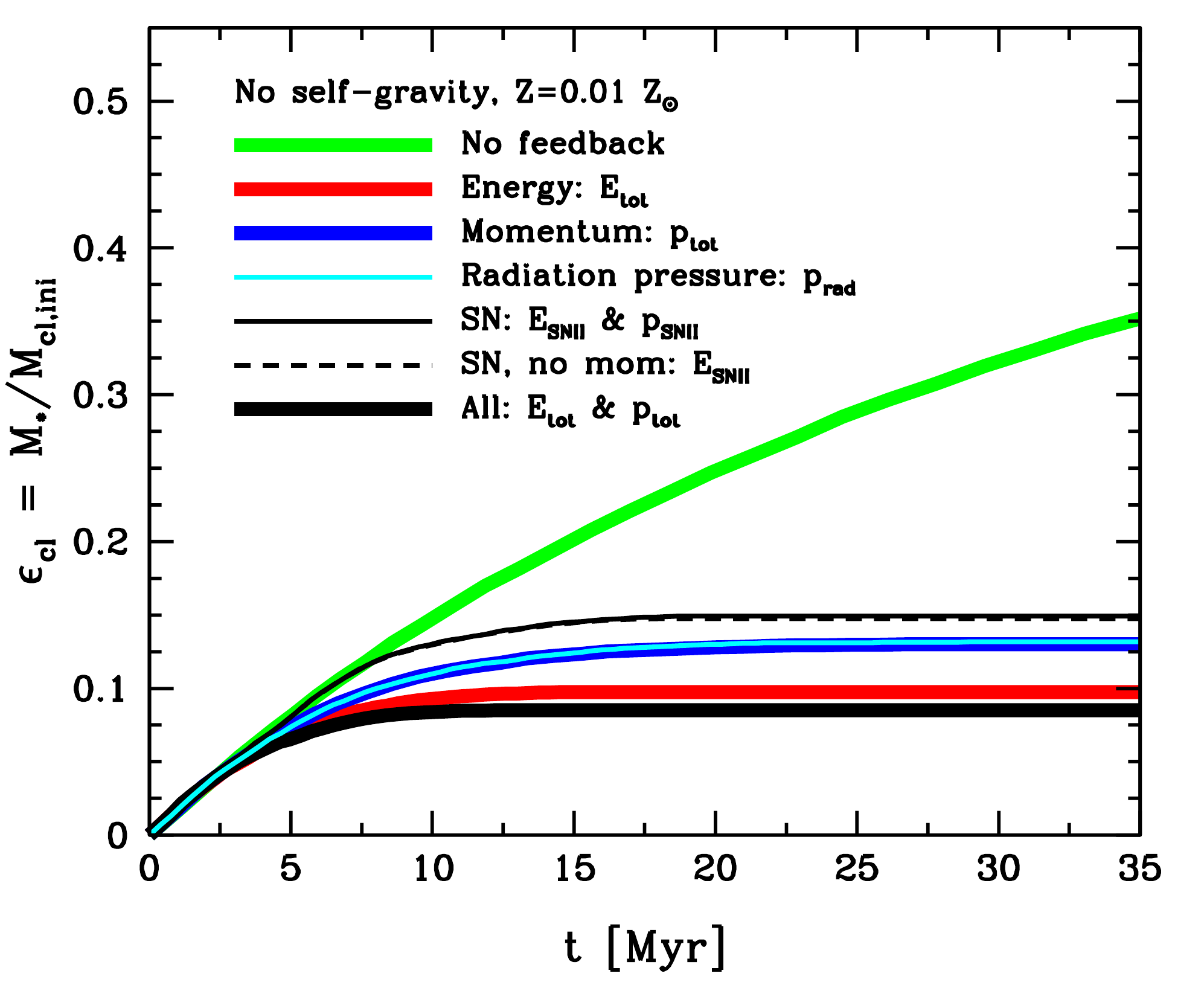}\\
\end{tabular}
\caption{Cloud star formation efficiency for gas metallicities of $Z=1\,Z_\odot$ (left) and $Z=0.01\,Z_\odot$ (right) in the non self-gravitating cloud simulation. Efficient cooling in the more enriched cloud reduces the effect of thermal energy on star formation. The opposite is true for momentum feedback, dominated by radiation pressure, which by itself limits $\epsilon_{\rm cl}$ to $\sim0.1$. In the low metallicity cloud, gas cooling rates are lowered by more than an order of magnitude, making thermal energy feedback the dominant factor limiting star formation.
}
\label{fig:cloudeff}
\end{center}
\end{figure*}

\section{Simulations}
\label{sect:simulations}
In this section we use idealized simulations of gas clouds and isolated galactic disks to gauge the effect of different prescriptions for stellar feedback described in the previous sections on the local and global efficiency of star formation (the $\Sigma_{\rm SFR}-\Sigma_{\rm gas}$ relation), as well as on the structural properties of galactic disks. A more extensive analysis of processes such as outflows, and study of feedback implementations in cosmological galaxy formation simulations will be presented in future work (Agertz et al. in prep). The simulations considered here have resolution similar to the resolution of state-of-the-art cosmological simulations, and the results should hence be directly applicable to interpretation of results in cosmological runs. Specifically, we restrict the spatial resolution to reach minimum cell sizes of $\Delta x\sim 10-100\pc$.

\subsection{Effect of feedback at the resolution scale}
\label{sect:resscale}
\begin{table}[t]
\caption{Feedback in a cell simulations}
\label{table:cellsim}
\begin{tabular}{ll}
\hline
Simulation & Feedback \\ 
\hline
ALL & $E_{\rm tot}$ \& $p_{\rm tot}$, see Equation~\ref{eq:FBarray}\\
MOMENTUM & Only momentum: $p_{\rm tot}$ \\
MOMENTUM\_ST & Only momentum: $p_{\rm tot}$, where $p_{\rm SNII}=p_{\rm ST}$ \\
ENERGY & Only energy: $E_{\rm tot}$ \\
SN & Only SNII energy: $E_{\rm SNII}$ \\
SNMOM & SNII energy and momentum: $E_{\rm SNII}$, $p_{\rm SNII}$\\
SNMOM\_ST & SNII energy and momentum: $E_{\rm SNII}$, $p_{\rm ST}$\\
\hline
\end{tabular}
\end{table}

It is instructive to first study the impact of feedback in a typical star forming computational cell. To this end, we place a single star particle of mass $m_*$ in a cell of size $\Delta x=40\pc$ within a periodic box of fixed resolution and homogeneous gas density $\rho_{\rm gas}=100\,m_{\rm H}\cc$ of solar metallicity. The initial gas temperature matters little, as it settles to a few $10\K$ after one time step. We adopt a series of (cell) stellar mass fractions $f_*=m_*/m_{\rm tot}\in\{1,5,10,20\}\%$, consistent with observations of massive GMCs \citep{Evans2009,Murray2011}. In the case of $f_{*}=10\%$, $m_*=1.6\times 10^4\Msol$. We are interested in studying the effect of feedback on the scale of individual simulation cells, where it will be applied in actual galaxy simulations. At such scales, the gas self-gravity is weak due to the softening of forces on the scale of a couple of cells, and is not hence not calculated properly in the actual simulations. For simplicity, we choose not to include self-gravity in these tests. 

This setup is evolved for 30 Myr using the different feedback implementations shown in table \ref{table:cellsim}. In all tests we employ momentum deposition via a non-thermal pressure term in the Riemann solver (method 2 in \S~\ref{sect:implementation}) rather than via ``kicks,'' as we want to measure the impact on the central cell containing the star. The infrared optical depths relevant for the above stellar mass fraction, as approximated via Equation~\ref{eq:prad2a}, become $\tau_{\rm IR}\approx 0.39,0.49,0.55,0.96$, i.e. at most a factor of two boost compared to the single scattering ``$L/c$-regime''.

In the left panel of Figure~\ref{fig:evac} we show the gas density and temperature evolution for the runs with $f_{*}=10\%$. The evolution strongly depends on the form of feedback employed; while all momentum based feedback sources can evacuate the cell, energy-only feedback ({\small ENERGY} and {\small SN} run) has no effect
\footnote{This result is somewhat at odds with results of \cite{ceverinoklypin09}, who found that purely thermal feedback from winds and SNe could effectively over-pressurize gas of similar characteristics, leading to gas evacuation. Part of the difference stems from how thermal energy is injected; in our simulations, energy is deposited to the gas at every time step, heating it to $\sim 10^4\K$. When the new gas state is passed to the cooling routine, all energy is lost over one time step, bringing the dense gas back to a few $\sim 10\K$. Ceverino \& Klypin considered thermal feedback via a heating term in the cooling routine, which when balanced against cooling led to a larger equilibrium temperature.}, illustrating the common overcooling problem. If the initial SNe momentum, $p_{\rm SNII}$, is included ({\small SNMOM}), gas is pushed out of the cell and the heating criterion of Equation~\ref{eq:heatcool} is satisfied after $\sim 20\Myr$, leading to temperatures $>10^8\K$. When all momentum sources of feedback ({\small MOMENTUM}) are included, the gas is effeciently evacuated from star forming cell, reaching $n\sim0.1\cc$ after only $\sim10\Myr$. Simulations adopting the more evolved Sedov-Taylor momentum ($p_{\rm ST}$) for each SNII result in an even faster evacuation. Not surprisingly, the strongest effect on density and temperature is found in the {\small ALL} run, in which runaway heating set in only after $\sim 3.7\Myr$ due to stellar winds and radiation pressure alone. 

In the right hand panel of Figure~\ref{fig:evac} we show the evolution of the {\small ALL} run for different values $f_*$. Runaway heating is achieved for all values of $f_*$, even for $f_*=1\%$ after $\sim10\Myr$. For $f_*>10\%$, this occurs before the first SNIIe explode.

We conclude that the implemented subgrid feedback prescriptions, especially early momentum injection, greatly enhance the ability of feedback to disperse and heat gas in star forming cells, even when only $\sim 1\%$ of gas is turned into stars. In more realistic setups, a patch of gas will continue to form stars until the star cluster has destroyed its surrounding or depleted all of the gas above the star formation threshold. We explore these scenarios in the next section.  

\begin{figure*}[t]
\begin{center}
\begin{tabular}{ccccc}
\includegraphics[scale=0.3]{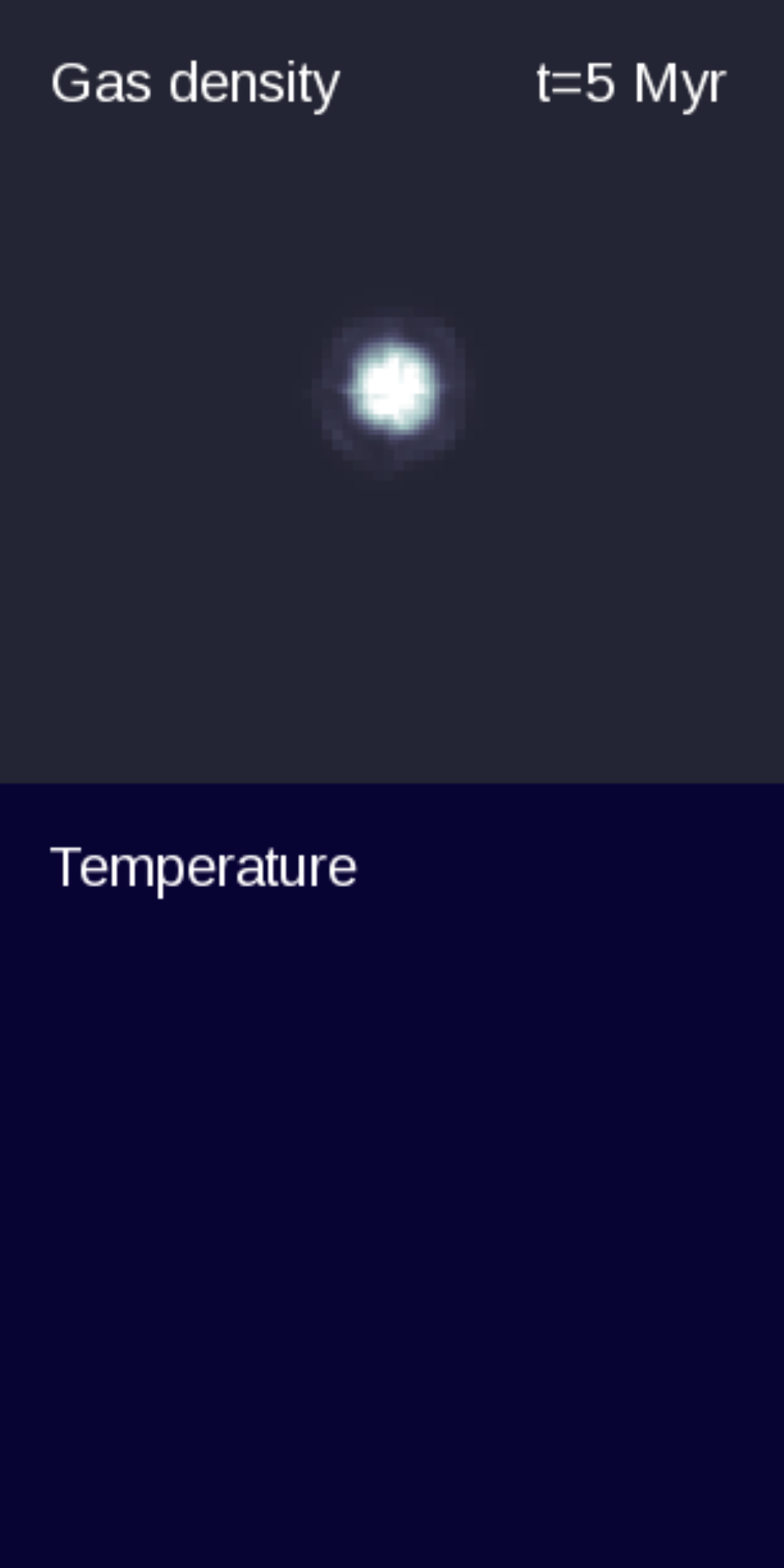}
\includegraphics[scale=0.3]{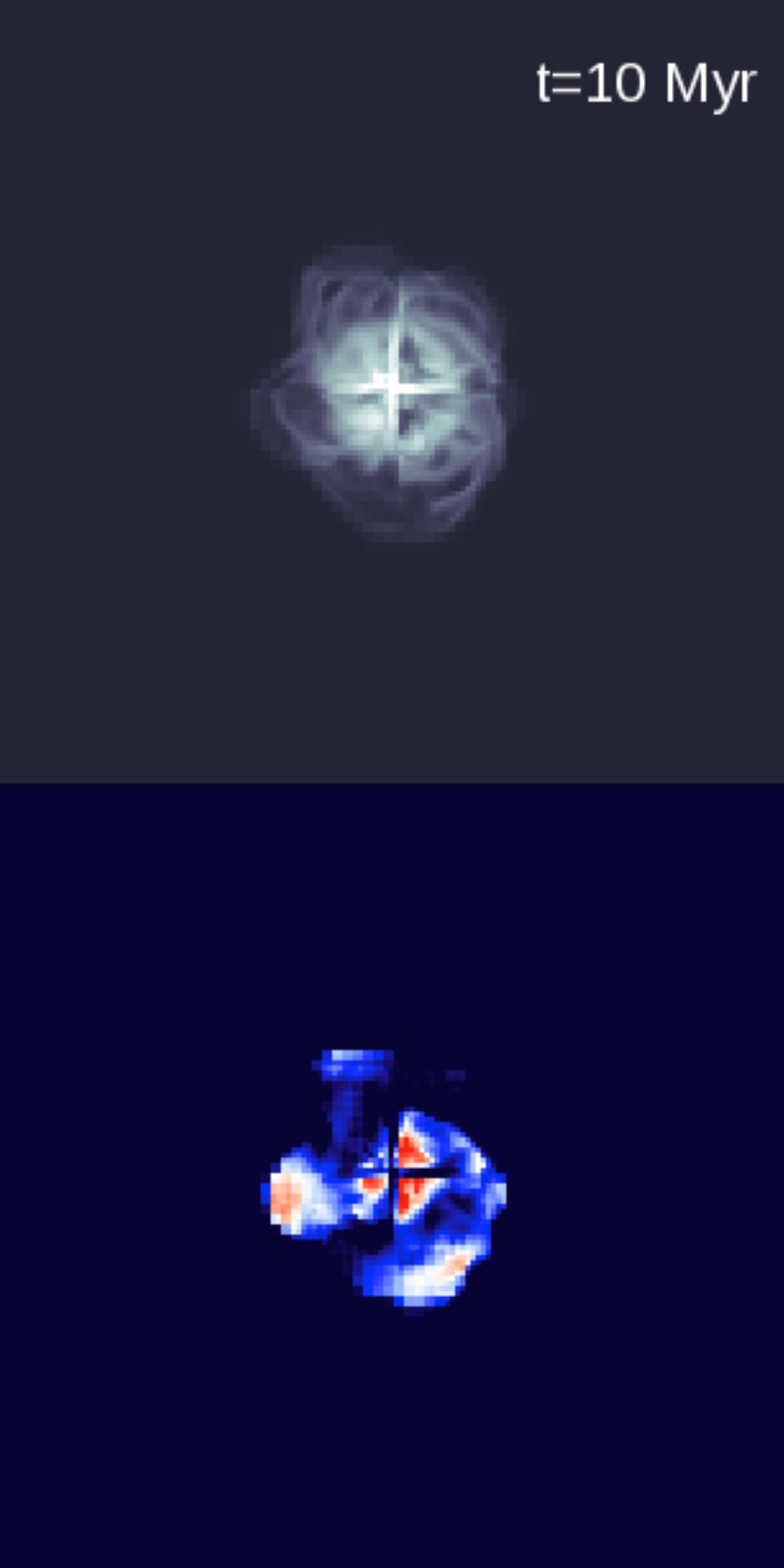}
\includegraphics[scale=0.3]{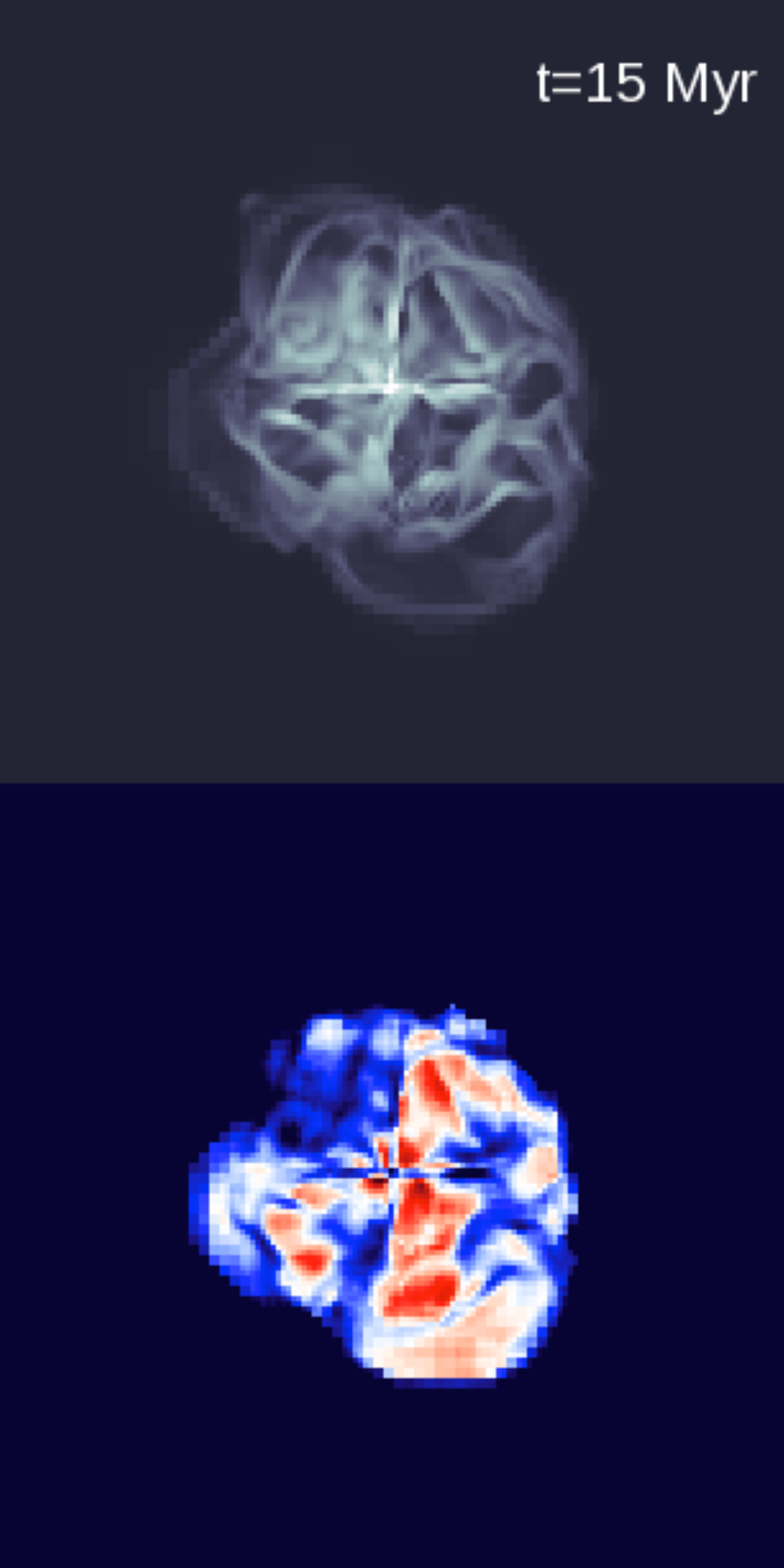}
\includegraphics[scale=0.3]{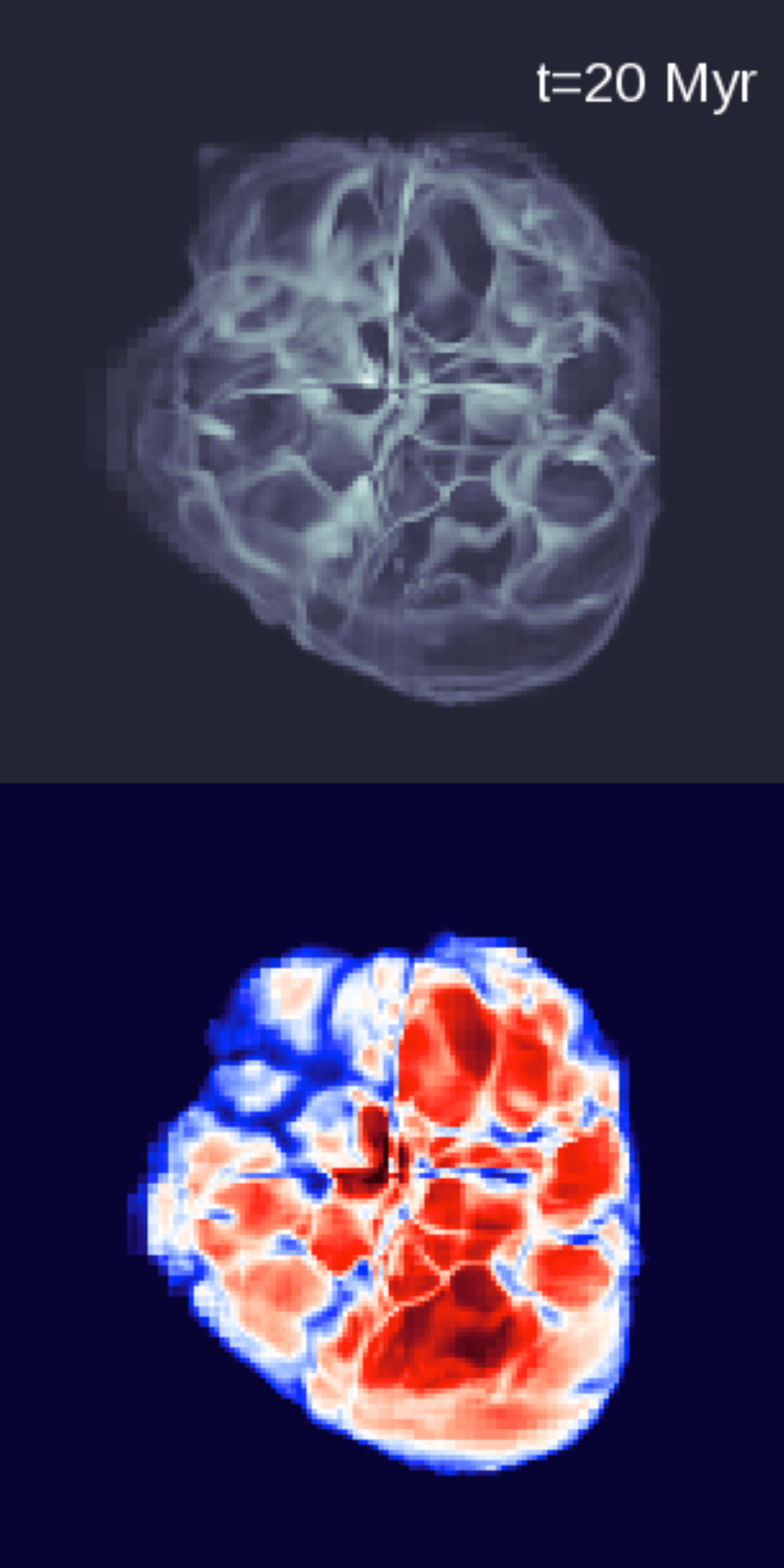}
\includegraphics[scale=0.3]{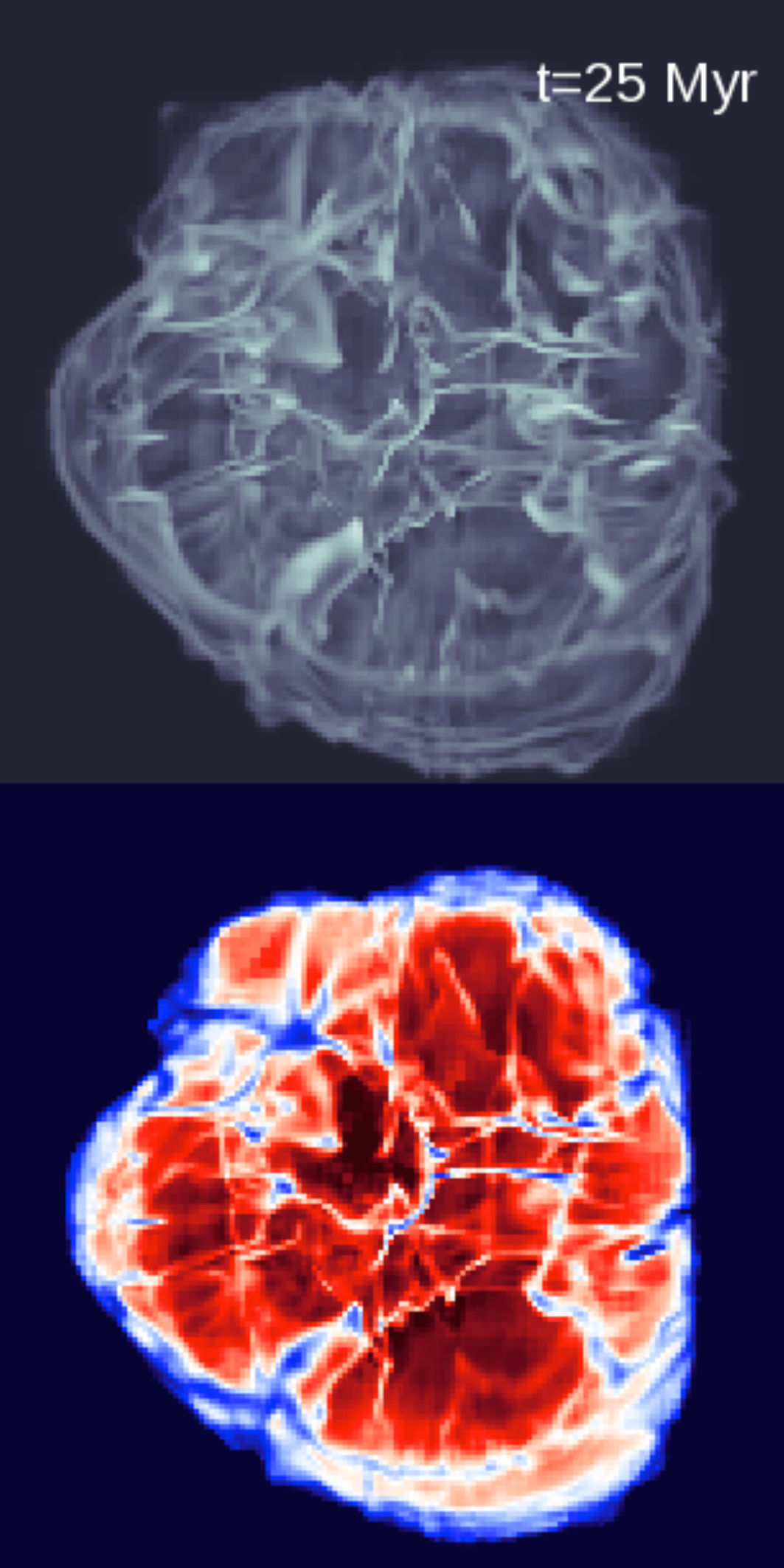}\\
\end{tabular}
\caption{Mass-weighted projected density (top) and temperature (bottom) of the gas in the "All" simulation of a non-gravitating isolated cloud. Each panel is $1.5\kpc$ across. Pre-SN feedback via stellar winds and radiation pressure pushes the cloud apart, thereby increasing cloud porosity. SNe later explode in a more tenuous medium capable of maintaining greater gas temperatures. Hot gas at temperatures of $T>10^7\K$ (shown in dark red) pushes on the ambient cold, dense phase.
}
\label{fig:allin}
\end{center}
\end{figure*}

\subsection{Isolated cloud}

\begin{figure}[ht]
\begin{center}
\includegraphics[scale=0.43]{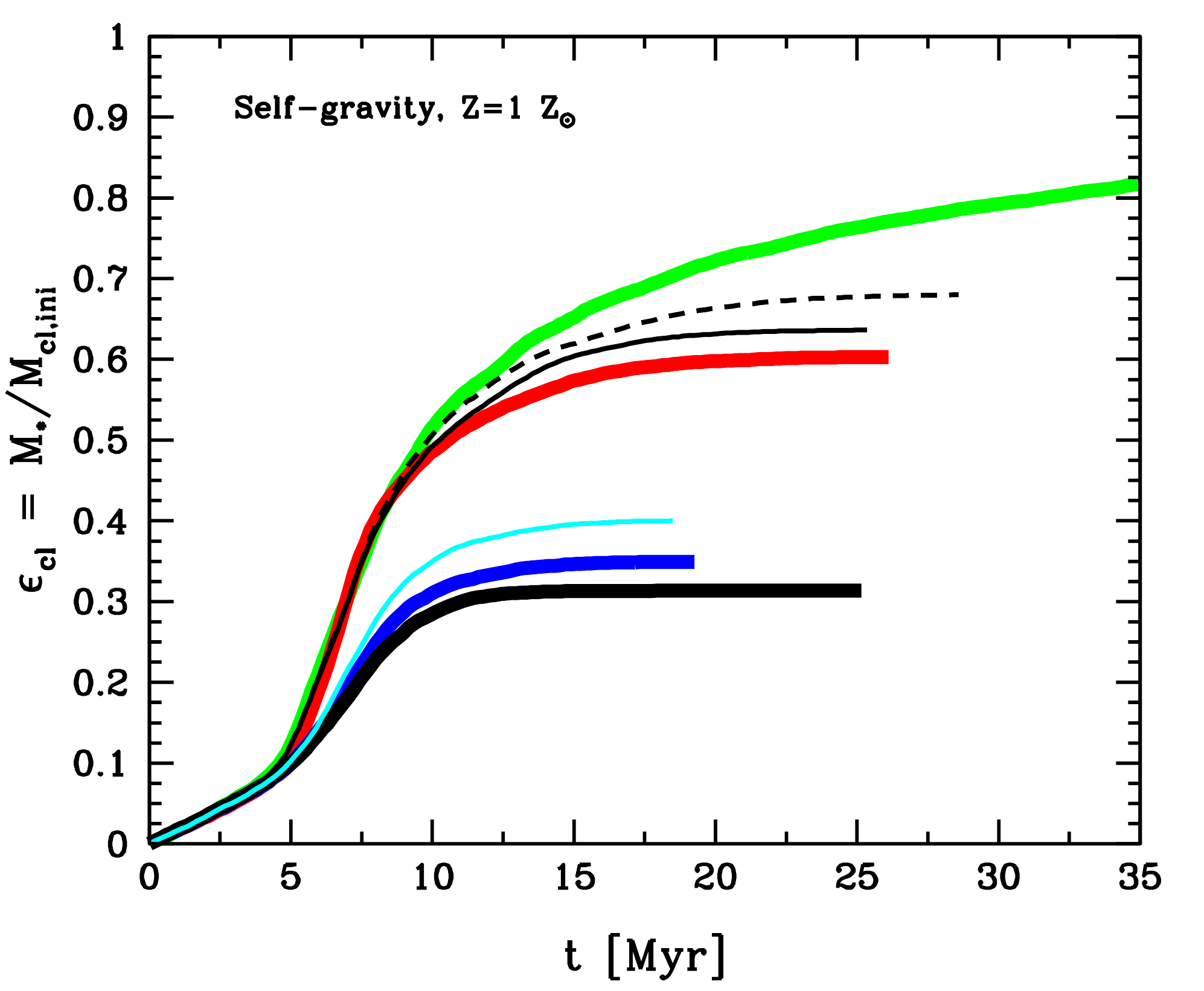}
\caption{Cloud star formation efficiency when self-gravity is included. The line types are the same as in Figure~\ref{fig:cloudeff}. Cloud contraction boosts star formation efficiency at early times compared to the test without self-gravity, but the relative impact of feedback is the same: the star formation efficiency is limited to much lower values when early momentum injection is included (lower black blue and cyan lines).}
\label{fig:cloudeffgrav}
\end{center}
\end{figure}

\begin{figure*}[t]
\begin{center}
\begin{tabular}{cc}
\includegraphics[scale=0.43]{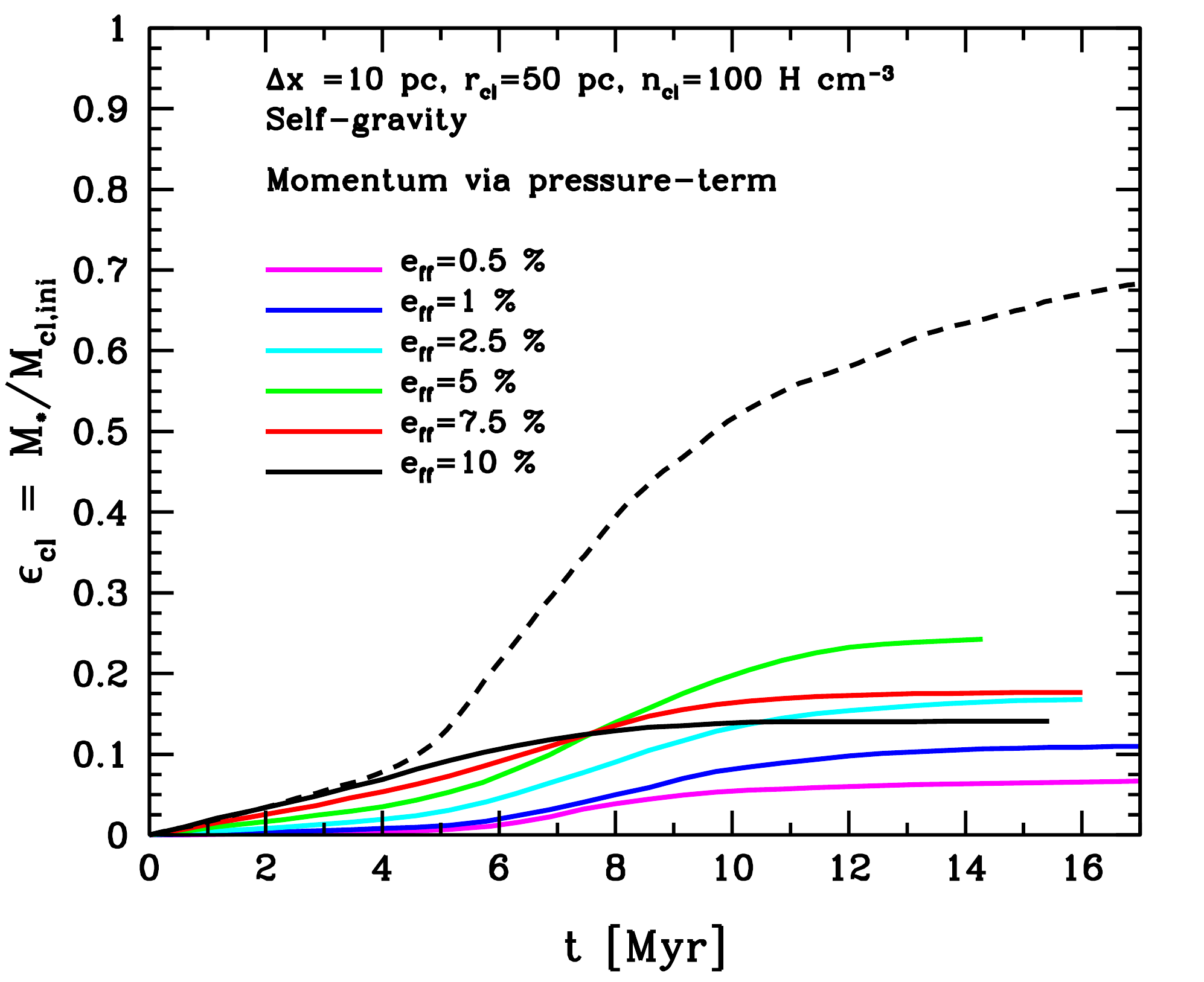}
\includegraphics[scale=0.43]{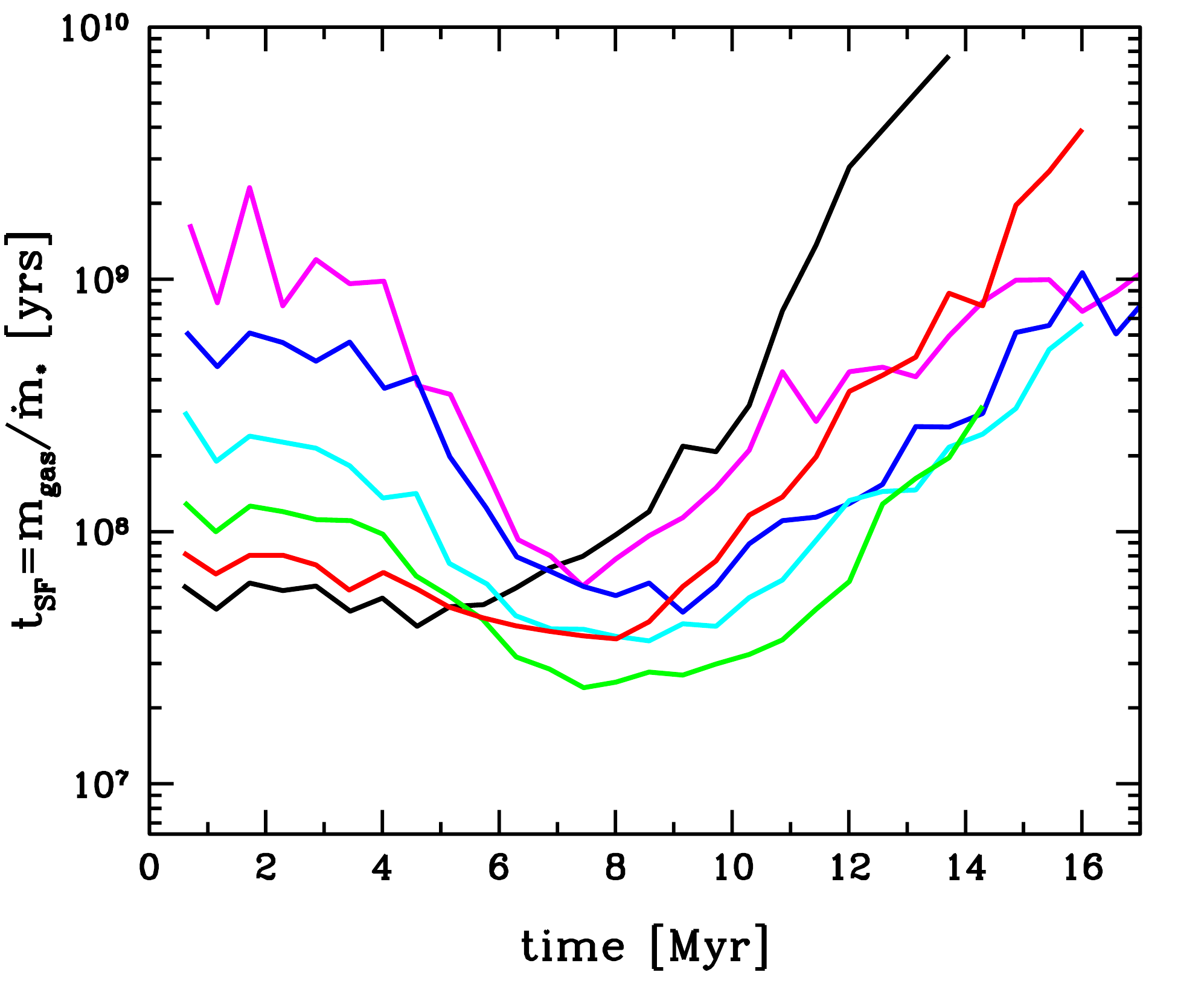}
\end{tabular}
\caption{Left panel: global cloud conversion efficiency $\epsilon_{\rm cl}$ for cloud runs with different assumed values of $\epsilon_{\rm ff}$. Radiation pressure in these tests is modelled as a non-thermal pressure term (see text for discussion) and its inclusion lowers the over all efficiency by roughly a factor of two for $\epsilon_{\rm ff}=10\%$ compared to when momentum is deposited directly to surrounding cells. Note that although $\epsilon_{\rm ff}$ is varied by a factor of 20, the final stellar mass fraction only varies by a factor of two. The dashed line shows $\epsilon_{\rm cl}$ for a simulations with $\epsilon_{\rm ff}=10\%$ without feedback. Right panel: instantaneous gas consumption time scale, $t_{\rm SF}=m_*/\dot{m}_*$. Regardless of the initial $t_{\rm SF}$, which simply reflects the initial conditions and choice of $\epsilon_{\rm ff}$, all clouds reach $t_{\rm SF}=50-100\Myr$ after roughly one free-fall time at which point stellar feedback starts to disrupt the cloud.
}
\label{fig:cloudP}
\end{center}
\end{figure*}

In this idealized test, a spherical cloud of dense cold gas ($n_{\rm cl}=100\,m_{\rm H}\cc$, $T_{\rm cl}=10\K$) of radius $r_{\rm cl}=50\pc$ is placed in pressure equilibrium with a diffuse ambient medium ($n_{\rm ISM}=0.1\,m_{\rm H}\cc$, $T_{\rm ISM}=10^4\K$). Star formation is then allowed to proceed, as described in \S\ref{sect:SFlaw} with $\epsilon_{\rm ff}=10\%$. As we are interested in the behaviour of a marginally resolved ISM, we adopt maximum resolution of $\Delta x=10\pc$. At this resolution, the cloud consists of 552 cells at exactly $n_{\rm cl}=100\,H\cc$, having a total initial gas mass of  $M_{\rm cl}=1.25\times 10^6\Msol$. 

In the following tests, we evolve the cloud with and without self-gravity, which in a very crude way can be seen as limiting cases of cloud virial parameter $\alpha_{\rm vir}$; no self-gravity simply means that unresolved turbulence supports the cloud ($\alpha_{\rm vir}\gtrsim1$) and vice versa. Note that we do not attempt to model details of star formation in giant molecular clouds, which requires more advanced simulation setups. Our main goal is simply to gauge \emph{systematic} differences between different feedback implementations at the resolution level that should be affordable in cosmological simulations in the near future. 

\subsubsection{No self-gravity}
In Figure~\ref{fig:cloudeff} we show evolution of star formation efficiency within the cloud, defined as $\epsilon_{\rm cl}(t)=M_*(t)/M_{\rm cl,ini}$, where $M_*(t)$ is the total stellar mass formed at time $t$ and $M_{\rm cl,ini}$ is the initial cloud gas mass, in the simulations without self-gravity.

For $Z=1\,Z_\odot$ without any feedback, the cloud forms stars unhindered until $\epsilon_{\rm cl}\sim0.75$ when the cell densities fall below the star formation threshold.  Supernovae feedback alone can reduce the overall efficiency to $\epsilon_{\rm cl}\sim0.2$. When no momentum from SNe is accounted for, the efficiency is somewhat larger: $\epsilon_{\rm cl}\sim 0.25$, and the same conclusion holds when all thermal energy (and no momentum) sources of stellar feedback are present. 

The stellar fractions differ significantly when pre-SN momentum feedback is included. Radiation pressure alone sets $\epsilon_{\rm cl}\sim 0.125$, and the conversion efficiency decreases somewhat when momentum from wind and SNe feedback is added. When momentum and energy deposition from all feedback mechanisms is included, the final efficiency approaches $\epsilon_{\rm cl}\sim 0.1$, although with significantly more hot gas present in comparison to pure momentum feedback. The hot gas causes vigorous late time expansion of the star forming region, which is illustrated in the time evolution of the projected density and temperature in Figure~\ref{fig:allin}.

The results are different in the case of low-metallicity gas\footnote{We here adopt a metallicity independent star formation threshold of $\rho_{*}=25\,{\rm cm}^{-3}$ to facilitate a comparison with the $Z=1 Z_\odot$ case.} shown in the right hand-side of Figure~\ref{fig:cloudeff}. As the gas cooling rates are lowered, a purely energy based feedback scheme can lower the efficiency of star formation to $\epsilon_{\rm cl}\sim 0.1$. The effect of radiation pressure is however not much different from the simulation adopting $Z=1\,Z_\odot$, despite $\tau_{\rm IR}$ being 100 times smaller ($\kappa_{\rm IR}\propto Z$). This is because $\tau_{\rm IR}$ plays a minor role in both cases, as stellar masses in the local cells are small ($m_*\lesssim 10^4\Msol$).

\subsubsection{With self-gravity}
In Figure~\ref{fig:cloudeffgrav} we show the cloud star formation efficiency for the self-gravitating cloud. We do not enforce hydrostatic equilibrium as the (unresolved) temperature profiles would immediately be erased by cooling. As the cloud now contracts, the global cloud star formation efficiency becomes greater by more than a factor of three in all simulations. However, the systematic trends measured in the non self-gravitating tests are recovered; pre-SN feedback, and specifically momentum, limits star formation by roughly a factor of two more efficiently than SNe feedback. 

In this setup, a stronger impact of feedback is found when momentum feedback is generated via a non-thermal pressure in the Riemann solver (method 2 in \S~\ref{sect:implementation}). In the left panel of Figure~\ref{fig:cloudP} we show the ``ALL'' simulation adopting free-fall star formation efficiencies in the range $\epsilon_{\rm ff}=0.5-10\%$. The $10\%$ case is here lower by a factor of two compared to momentum feedback via "kicks". Even though we vary the star formation efficiency by a factor of 20, the final global conversion stays within $\epsilon_{\rm cl}\sim7-25\%$, in agreement with observed GMCs \citep{Evans2009,Murray2011}, compared to $>70\%$ when feedback is ignored.

This efficient self-regulation can be understood by studying the star formation time scale, defined as 
\begin{equation}
t_{\rm SF}=\frac{\rho_{\rm g}}{\dot{\rho}_{*}}=\frac{m_{\rm cl,ini}-m_*}{\dot{m}_*},
\end{equation}
plotted in the right hand side of Figure~\ref{fig:cloudP}. The ability for the cloud to contract to higher densities makes it possible to achieve efficient star formation regardless of initial $\epsilon_{\rm ff}$; the star formation time-scale regulates to $t_{\rm SF}\sim 50\Myr$ after which the cloud is destroyed by feedback. The stellar age spread in patch of gas is on the order of $\sim20$ Myr, where the most concentrated cluster of stars formed over a narrow range of a few Myr. This naive model is hence qualitatively in agreement with observed star cluster forming regions in local galaxies e.g. 30 Doradus, where the stars in the massive compact star cluster are younger than $\sim 4\Myr$, while the peripheral stars may be as old as $\sim 30\Myr$ \citep{Demarchi2011}, see also conclusions by \cite{Murray2011} regarding Milky Way GMCs.

It is plausible that effective self-regulation only occurs when simulated star forming gas cloud are resolved sufficiently for  self-gravity to allow for some degree of collapse/contraction. At a cosmological resolution $\Delta x\sim 100\pc$, such collapse may not occur to the same degree as observed in the experiments here, especially as the gas is pressurized artificially at the scale of resolution to prevent spurious fragmentation \citep{truelove97}. 
\begin{table*}[t]
\begin{center}
\caption{Galactic disk simulations of $z=0$ spiral galaxy analogue. See Equation~\ref{eq:FBarray} for notation.}
\label{table:simsummary1}
\begin{minipage}{140mm}
\center
\begin{tabular}{ll}
\hline
\hline
Run & Description \\ 
\hline
\hline
Direct injection runs & \\
\hline
nofb(e001) & No feedback, $\epsilon_{\rm ff}=10\%\,(1\%)$\\
momentum & Only momentum: $p_{\rm tot}$, $\epsilon_{\rm ff}=10\%$\\
energy & Only energy: $E_{\rm tot}$, $\epsilon_{\rm ff}=10\%$\\
prad & Only radiation pressure: $p_{\rm rad}$, $\epsilon_{\rm ff}=10\%$\\
SNnomom & Only SNe energy: $E_{\rm SNII}$, $\epsilon_{\rm ff}=10\%$\\
SN & Only SNe energy and momentum: $E_{\rm SNII}$ \& $p_{\rm SNII}$, $\epsilon_{\rm ff}=10\%$\\
all(e001) & All feedback processes: $E_{\rm tot}$ \& $p_{\rm tot}$ $\epsilon_{\rm ff}=10\%\,(1\%)$\\
all\_tau10 & All feedback processes: $E_{\rm tot}$ \& $p_{\rm tot}$, fixed $\tau_{\rm IR}=10$, $\epsilon_{\rm ff}=10\%\,$\\
all\_tau30 & All feedback processes: $E_{\rm tot}$ \& $p_{\rm tot}$, fixed $\tau_{\rm IR}=30$, $\epsilon_{\rm ff}=10\%\,$\\
\hline
\hline
Simulations adopting delayed cooling & \\
\hline
SN\_dc10 & Only SNe energy: $E_{\rm SNII}$, delayed cooling $t_{\rm cool}=40\Myr$, $\epsilon_{\rm ff}=10\%$\\
SN\_dc40 & Only SNe energy: $E_{\rm SNII}$, delayed cooling $t_{\rm cool}=40\Myr$, $\epsilon_{\rm ff}=10\%$\\
energy\_dc10 & Only energy: $E_{\rm tot}$, delayed cooling $t_{\rm cool}=10\Myr$, $\epsilon_{\rm ff}=10\%$\\
energy\_dc40 & Only energy: $E_{\rm tot}$, delayed cooling $t_{\rm cool}=40\Myr$, $\epsilon_{\rm ff}=10\%$\\
all\_dc10 & All feedback processes: $E_{\rm tot}$ \& $p_{\rm tot}$, delayed cooling $t_{\rm cool}=10\Myr$, $\epsilon_{\rm ff}=10\%$\\
all\_dc40(e001) & All feedback processes: $E_{\rm tot}$ \& $p_{\rm tot}$, delayed cooling $t_{\rm cool}=40\Myr$, $\epsilon_{\rm ff}=10\%(1\%)$\\
\hline
\hline
Runs adopting a feedback energy variable  & \\
\hline
energy\_f05\_t1 & Only energy: $E_{\rm tot}$, feedback energy fraction $f_{\rm fb}=0.5$,  dissipation time $t_{\rm dis}=1\Myr$, $\epsilon_{\rm ff}=10\%$\\
energy\_f05\_t10 &  Only energy: $E_{\rm tot}$, $f_{\rm fb}=0.5$, $t_{\rm dis}=10\Myr$, $\epsilon_{\rm ff}=10\%$\\
all\_f05\_t1 & All feedback: $E_{\rm tot}$ \& $p_{\rm tot}$, $f_{\rm fb}=0.5$, $t_{\rm dis}=1\Myr$, $\epsilon_{\rm ff}=10\%$\\
all\_f05\_t10(e001) & All feedback: $E_{\rm tot}$ \& $p_{\rm tot}$, $f_{\rm fb}=0.5$, $t_{\rm dis}=10\Myr$, $\epsilon_{\rm ff}=10\%(1\%)$\\
all\_f01\_t1 & All feedback: $E_{\rm tot}$ \& $p_{\rm tot}$, $f_{\rm fb}=0.1$, $t_{\rm dis}=1\Myr$, $\epsilon_{\rm ff}=10\%$\\
all\_f01\_t10 & All feedback: $E_{\rm tot}$ \& $p_{\rm tot}$, $f_{\rm fb}=0.1$, $t_{\rm dis}=10\Myr$, $\epsilon_{\rm ff}=10\%$\\
all\_f01\_t40 & All feedback: $E_{\rm tot}$ \& $p_{\rm tot}$, $f_{\rm fb}=0.1$, $t_{\rm dis}=40\Myr$, $\epsilon_{\rm ff}=10\%$\\
\hline\\
\end{tabular}
\end{minipage}
\end{center}
\end{table*}

\subsection{Disk galaxy}
\label{sect:disks}

Following \cite{Hernquist1993} and \cite{Springel2000} \citep[see also][] {SpringelMatteoHernquist2005} we create a particle distribution representing a late type, star forming spiral galaxy embedded in an NFW dark matter halo \citep{nfw1996,nfw97}. The halo has a concentration parameter $c=10$ and virial circular velocity, measured at overdensity $200\rho_{\rm crit}$, $\vel_{\rm 200}=150\kmsec$, which translates to a halo virial mass $M_{\rm 200}=1.1\times 10^{12}\Msol$. The total baryonic disk mass is $M_{\rm disk}=4.5\times10^{10}\Msol$ with $20\%$ in gas. The bulge-to-disk mass ratio is $B/D=0.1$. We assume exponential profiles for the stellar and gaseous components and adopt a disk scale length $r_{\rm d}=3.6\kpc$ and scale height $h=0.1r_{\rm d}$ for both. The bulge mass profile is that of \cite{Hernquist1990} with scale-length $a=0.1r_{\rm d}$. 

We initialize the gaseous disk analytically on the AMR grid assuming an exponential profile. The galaxy is embedded in a hot ($T=10^6\K$), tenuous ($n=10^{-5}\cc$) gas halo enriched to $Z=10^{-2}Z_\odot$, while the disk has solar abundance. We conduct all simulations at a maximum AMR cell resolution of $\Delta x=70\pc$, typical of current state-of-the-art galaxy formation simulations carried out to $z=0$. 

We systematically vary the different sources of stellar feedback operating in the simulations, and conduct additional tests which include thermal feedback via phenomenological approaches described in \S~\ref{sect:hotgas}. Table~\ref{table:simsummary1} presents details of all simulations considered in the following analysis. All runs adopt the standard star formation prescription outlined in \S\ref{sect:SFlaw}, and we generally adopt a star formation efficiency per free fall time of $\epsilon_{\rm ff}=10\%$. We note that this value of efficiency is an order of magnitude larger than the average values derived globally for kiloparsec patches of gas or in individual clouds \citep[e.g.][]{krumholztan07,bigiel2008}. However, as we show below, runs with feedback and large free-fall efficiencies produce normalizations of the Kennicutt-Schmidt relation quite close to observations \citep[see also][]{Hopkins2011}. 

\begin{figure*}[t]
\begin{center}
\begin{tabular}{cc}
\includegraphics[scale=0.4]{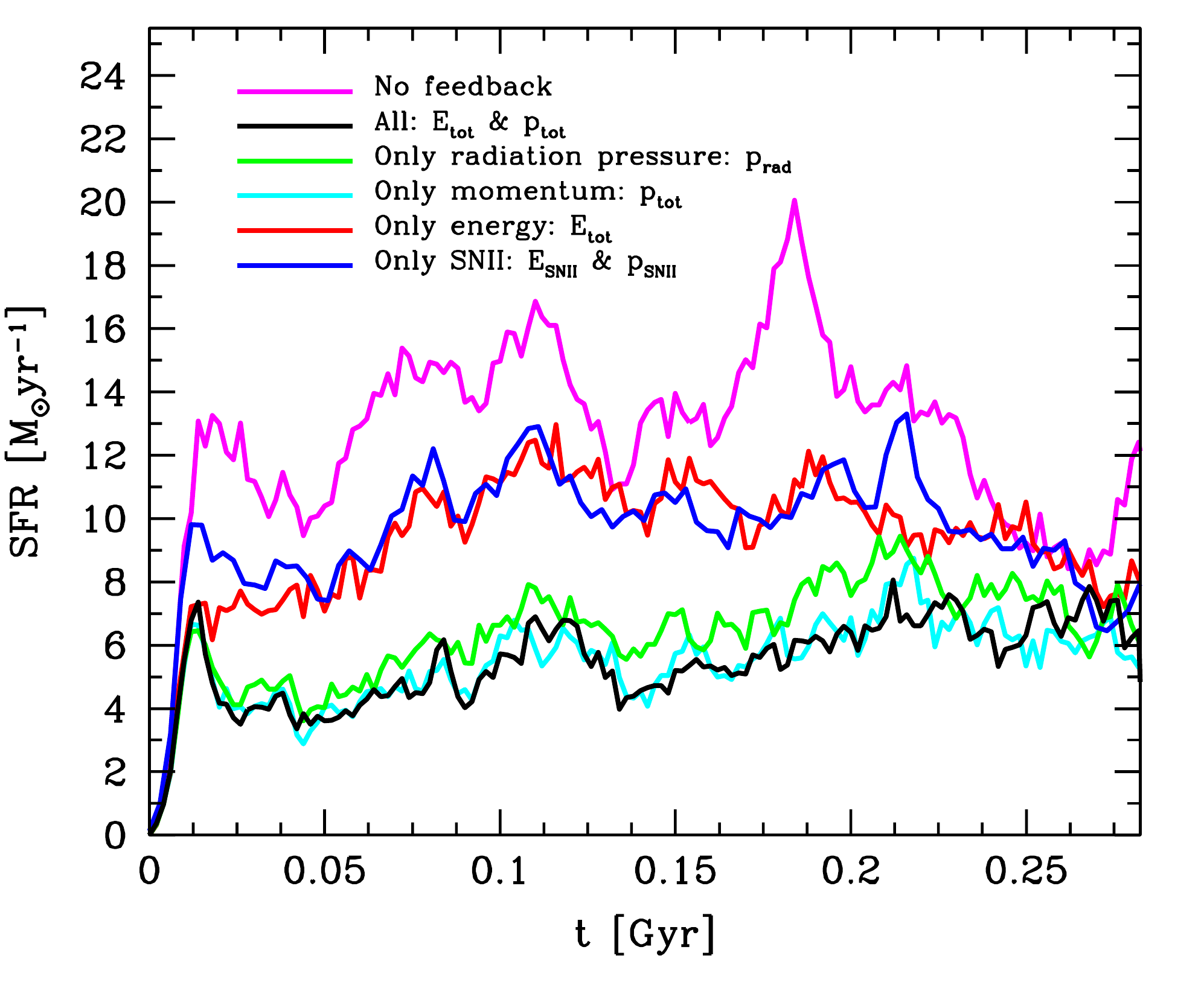}
\includegraphics[scale=0.4]{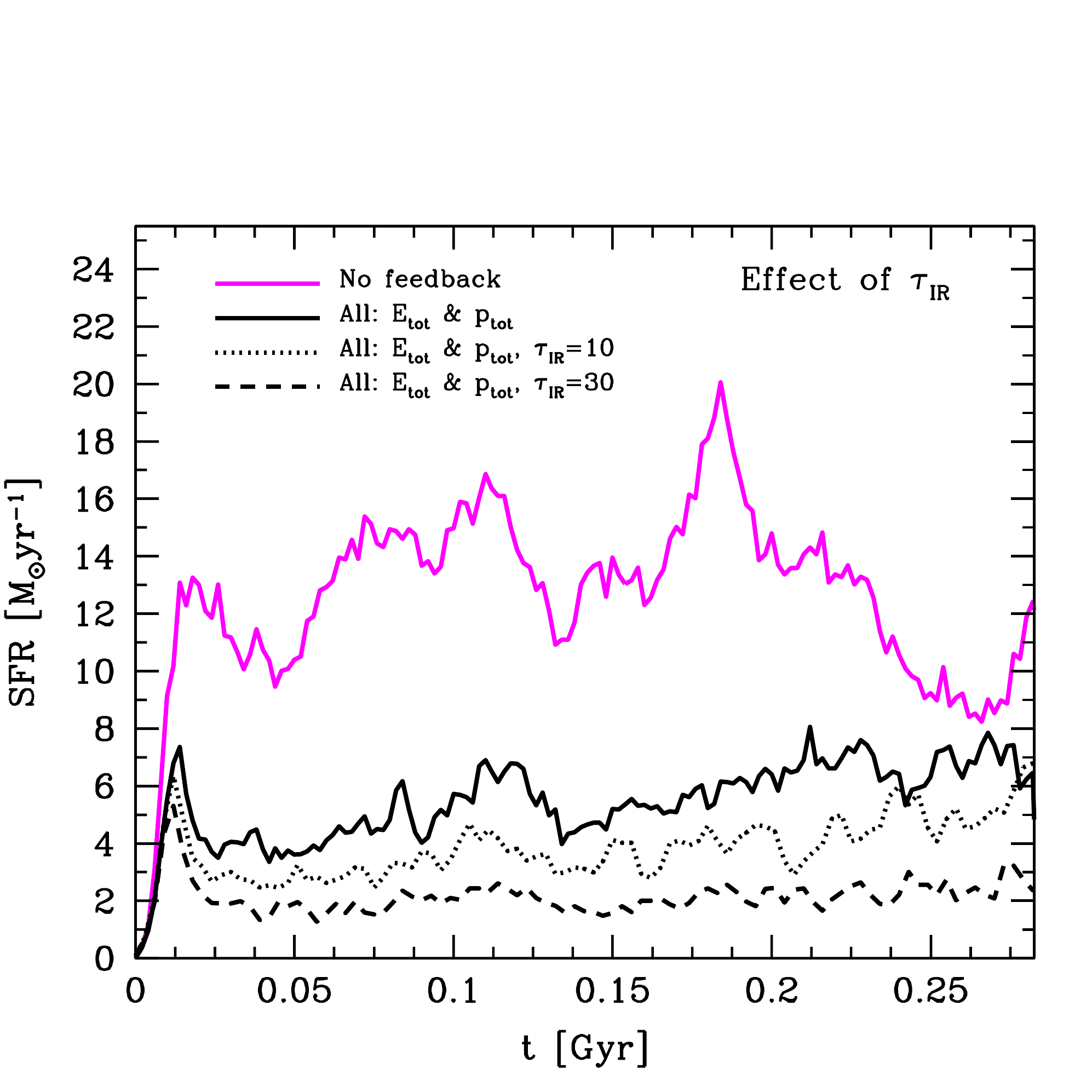}\\
\includegraphics[scale=0.4]{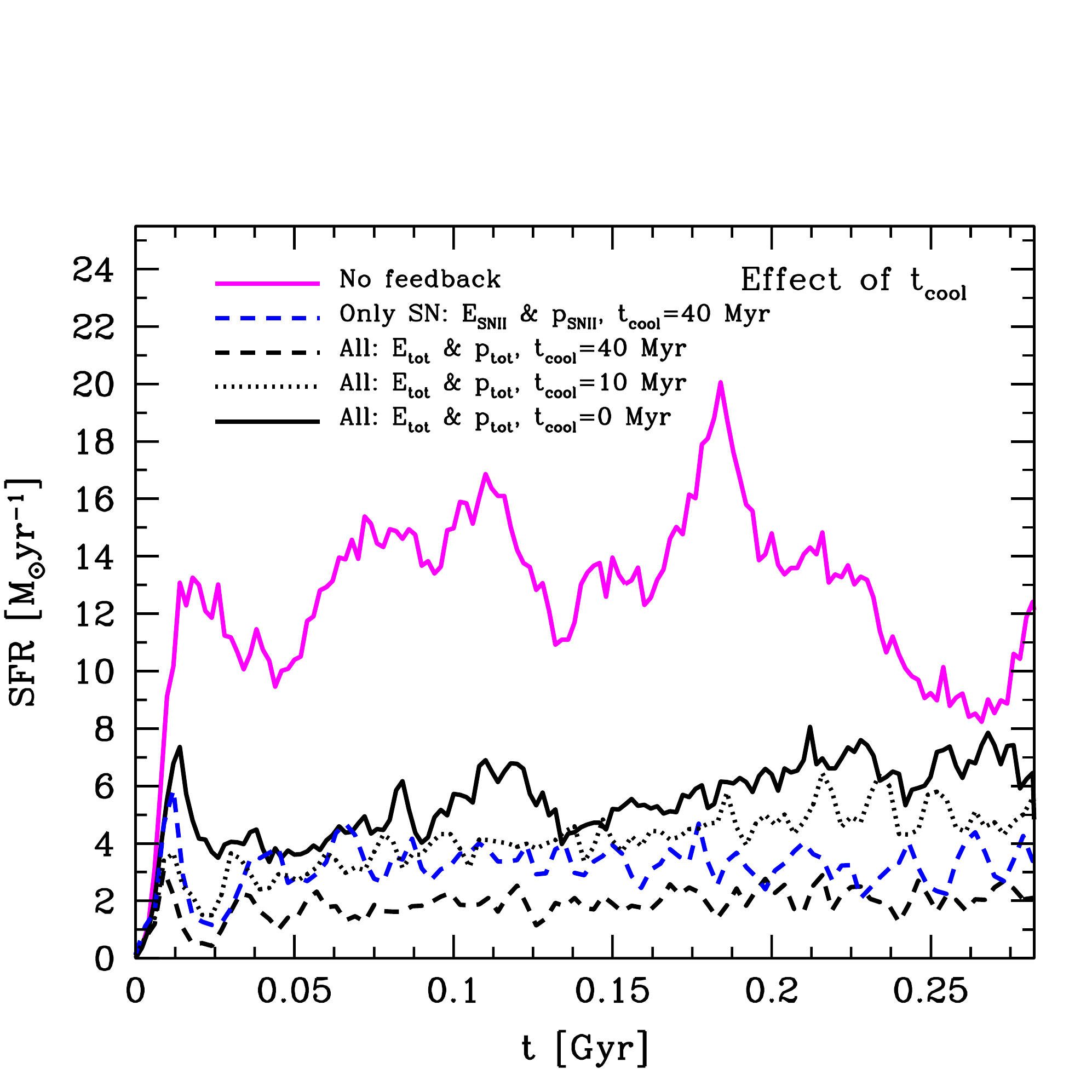}
\includegraphics[scale=0.4]{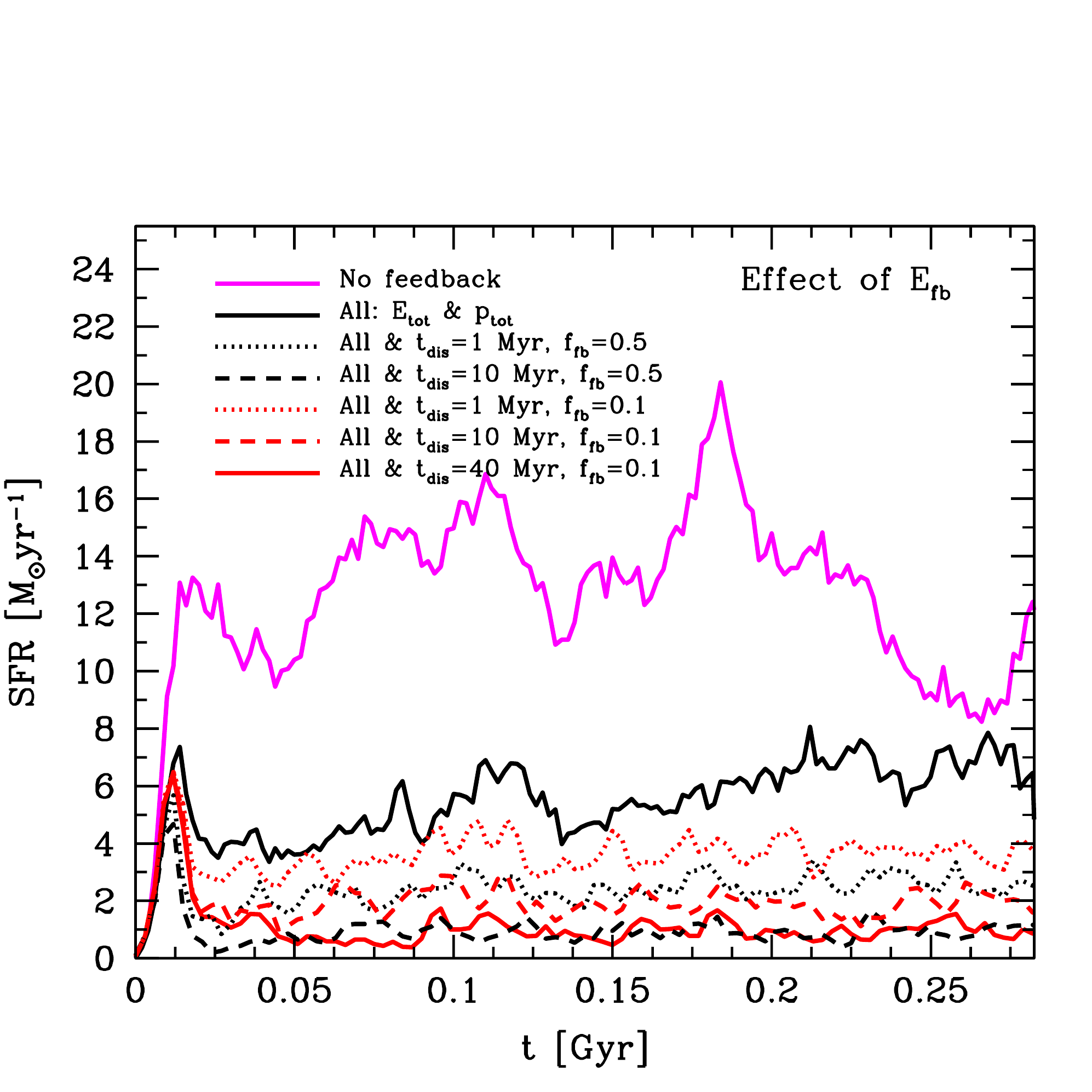}\\
\end{tabular}
\caption{Star formation histories for the isolated galactic disk simulation. Top left: the impact of various feedback sources, in the ``straight injection'' implementation of feedback. The strongest suppression of star formation is in simulations that include early momentum injection, especially momentum due to radiation pressure. Thermal energy feedback has a sub-dominant effect due to short cooling times in dense gas. Top right: the impact of increasing the infrared optical depth to a fixed value of $\tau_{\rm IR}=10$ or 30. $\tau_{\rm IR}=10$ reduces the SFR by $\sim 30\%$, while boosting radiation pressure using $\tau_{\rm IR}=30$ suppresses the SFR by another factor of 2-3 compared to the fiducial ``All'' run, and a factor of $\sim 5-10$ compared to the case of no feedback. Bottom left: the impact of delaying cooling in the local gas around newborn star particles for $t_{\rm cool}=10$ and 40 Myr. Note that delaying cooling for such values of $t_{\rm cool}$ results in a similar suppression of SFR compared to the radiation pressure momentum injection with large values of $\tau_{\rm IR}$. For example, ${\rm SFR}\approx 2\Msolyr$ for $t_{\rm cool}=40\Myr$, which is similar to the SFR for run with $\tau_{\rm IR}=30$ shown in the top right panel. Bottom right: the impact of assigning some fraction $f_{\rm fb}$ of the feedback energy to an energy variable $E_{\rm fb}$ that dissipates on longer timescales $t_{\rm dis}$ than expected from cooling in the dense gas. Even if only 10\% of the energy is assumed to dissipate over $t_{\rm dis}=1\Myr$, SFR is suppressed by $\sim 30\%$. If the  energy fraction is increased to $f_{\rm fb}=0.5$, and/or dissipation occurs over longer timescale $t_{\rm dis}\gtrsim 10\Myr$, we find a significant impact on star formation, as SFR approaches a steady rate of $\sim 1\Msolyr$.
}
\label{fig:SFR}
\end{center}
\end{figure*}

\subsubsection{Star formation histories}
Figure~\ref{fig:SFR} shows the star formation histories in disk simulations presented in table~\ref{table:simsummary1}. The top left panel presents the impact of direct feedback injection, i.e.  without any phenomenological approach to thermal energy. We find the same trend in star formation rate as in the isolated cloud test. Simulations that include only thermal energy or SNe have a minor effect on the star formation history compared to no feedback, while the inclusion of momentum lowers the SFRs by up to a factor of three. This process is mainly due to early, pre-SN feedback, especially radiation pressure. After a few orbital times all simulations regulate to roughly the same SFRs, although at different gas fractions.

It is instructive to compare our results with the recent work by \cite{Hopkins2011Prad}. These authors reported average infrared optical depths of $\langle\tau_{\rm IR}\rangle\sim10-30$ in their simulated Milky Way-like galaxy\footnote{These models marginally resolve the collapse of individual GMCs, although not their internal structure. The optical depth $\tau_{\rm IR}$ in their simulations steadily increases in the star forming clouds until feedback halts the gravitational collapse. The reported optical depths refer to the average values, used in the feedback scheme, at the moment when particles are stochastically chosen to receive a feedback velocity ``kick.''}. Our model, on the other hand, predicts more modest average values in the range $\langle\tau_{\rm IR}\rangle\sim 2-6$. The actual values of $\tau_{\rm IR}$ in dense gas surrounding young, embedded star clusters are highly uncertain both because we do not know covering fraction of absorbing dusty gas \citep[see, e.g.,][]{Krumholztau2012} and because dust temperatures used in calculations of $\tau_{\rm IR}$ are assumed to be high, $T_{\rm d}>100$~K, while the optical depth can be much lower if dust temperatures are much lower because $\tau_{\rm IR}\propto T^2_{\rm d}$ \citep{semenov_etal03}.

To understand how significantly larger values of $\tau_{\rm IR}$ affect our results, we perform two "All" simulation using fixed optical depths $\tau_{\rm IR}=10$ and 30. As shown in the right panel of Figure~\ref{fig:SFR}, increasing $\tau_{\rm IR}$ further suppresses SFR by $\sim 30\%$ for  $\tau_{\rm IR}=10$, and by a factor of 2-3 for $\tau_{\rm IR}=30$. The latter case renders SFRs $\sim 5-10$ times lower than in the case of no feedback. 

In the bottom left panel we present the impact of disabling cooling in the gas surrounding newly born star particles. The SFR in runs with $t_{\rm cool}=10$ and $40\Myr$ is suppressed by amount similar to the runs with high $\tau_{\rm IR}$ values discussed above. A significant suppression (by a factor of two) can be achieved via SNe alone, provided gas cooling is disabled for extended periods of time, $t_{\rm cool}=40\Myr$. 

The effect of treating a fraction $f_{\rm fb}$ of the feedback energy as an auxiliary energy variable $E_{\rm fb}$ that dissipates on a timescale $t_{\rm dis}$, longer than expected from cooling in the dense gas, is shown in the bottom right panel of Figure~\ref{fig:SFR}. Even for a modest $f_{\rm fb}=10\%$ dissipating over $t_{\rm dis}=1\Myr$, SFRs can be affected by $\sim 30\%$. As the energy fraction is increased to $f_{\rm fb}=0.5$, and/or dissipation occurs over longer time scale $t_{\rm dis}\gtrsim 10\Myr$, we find a significant impact on the SFHs, and SFRs approach a steady $\sim 1\Msolyr$. As discussed in \S~\ref{sect:shocks}, up to $90\%$ of SNe energy may be lost in radiative shocks within wind-blown bubbles \citep{ChoKang2008} in a few Myr. However, as the above simulations indicate, even this amount of preserved energy has a non-negligeble effect on star formation rate. We view this as an indication that some form of sub-grid treatment of feedback energy may be required, even in the presence of pre-SN feedback sources, due to the unresolved ISM phases and gas motions.

We note that these results should only be viewed as indicative, as the effect of feedback can in general  depend on  metallicity, ISM pressure, depth of potential well, accretion rates etc., which we plan to explore in future work.

\begin{figure*}[t]
\begin{center}
\begin{tabular}{cc}
\includegraphics[scale=0.36]{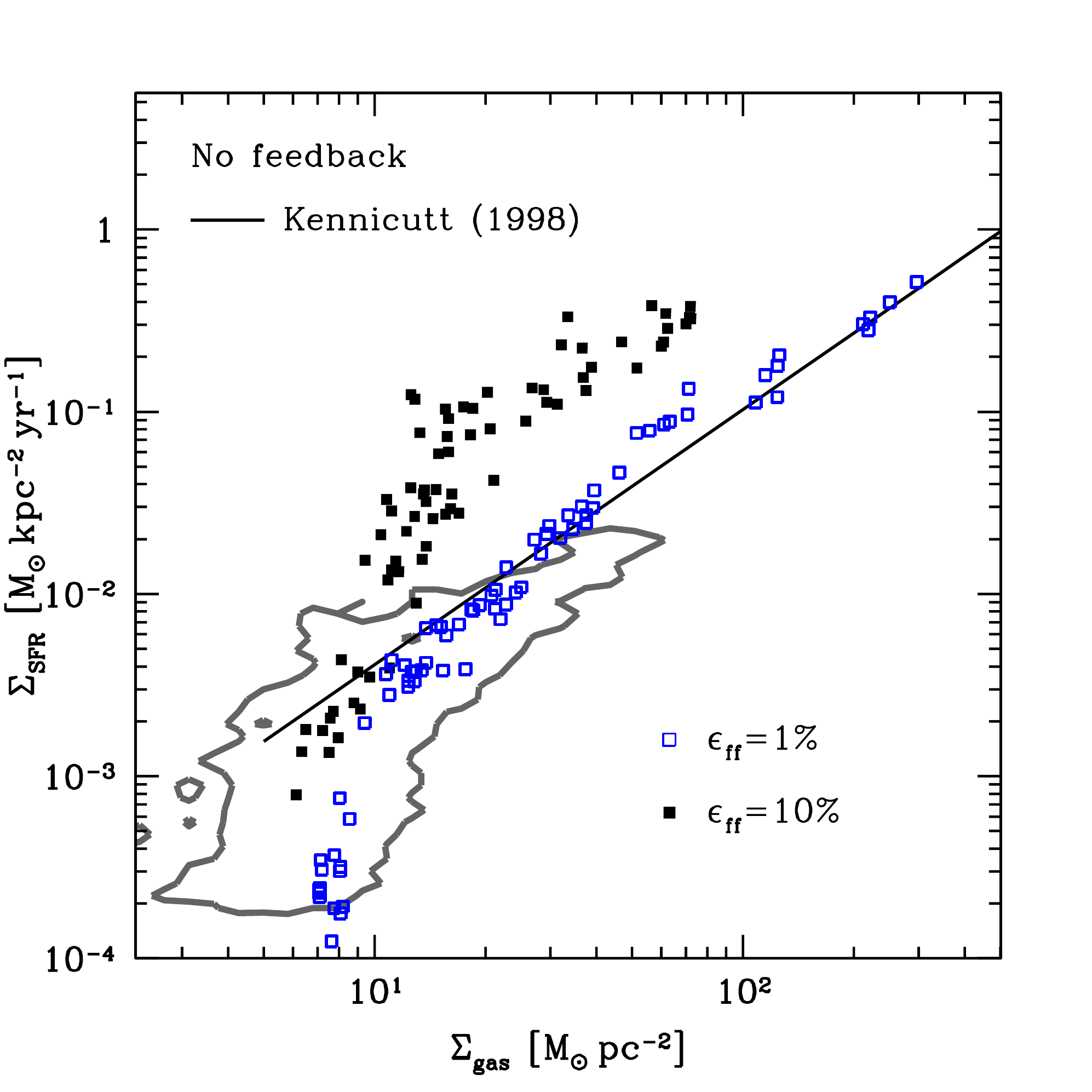}
\includegraphics[scale=0.36]{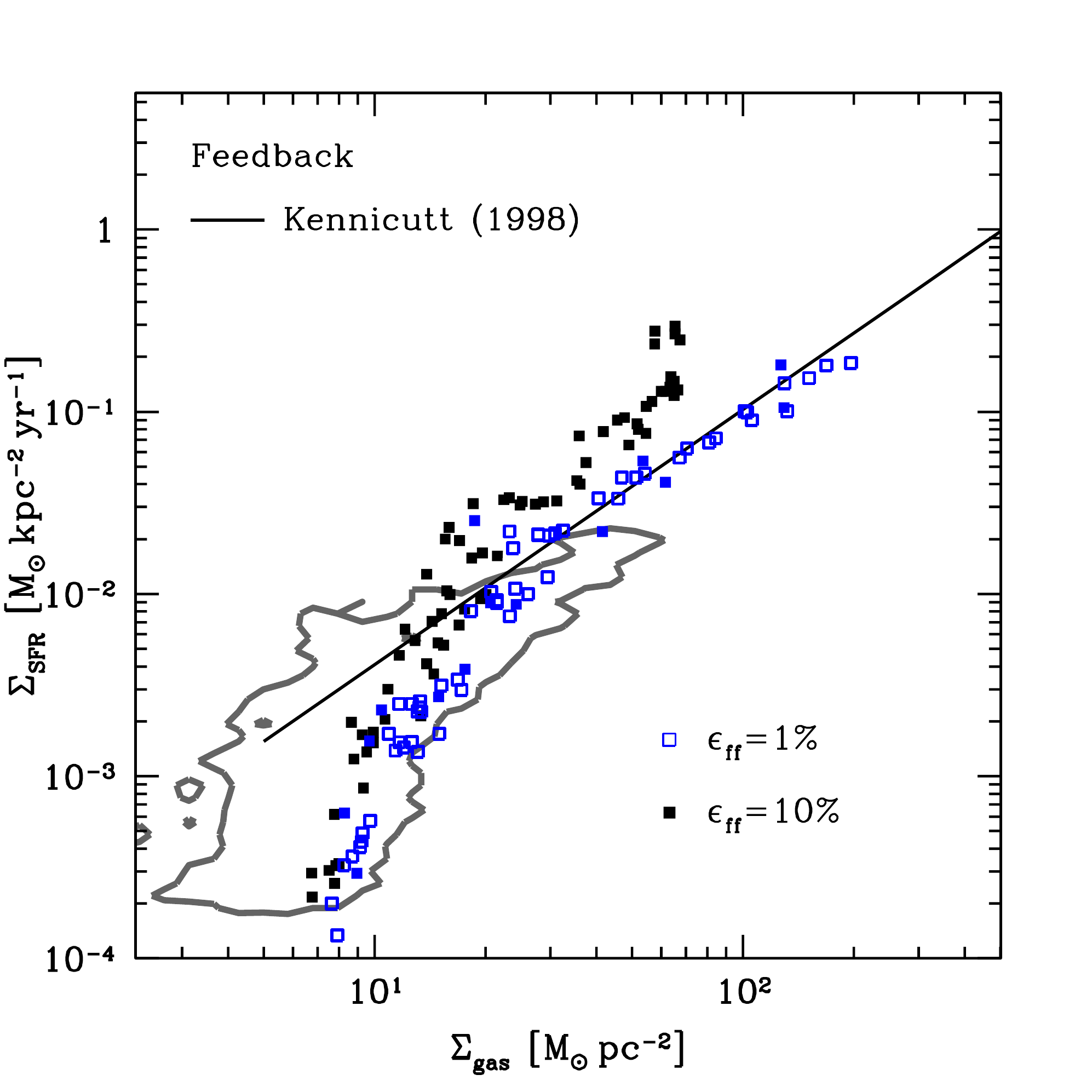}\\
\end{tabular}
\caption{The impact of feedback implementation and assumed star formation efficiency on the $\Sigma_{\rm SFR}-\Sigma_{\rm gas}$ relation. The left panel shows runs with no feedback, while the right panel shows runs with ``All'' feedback implementation (see table \ref{table:simsummary1}). In both panels the two sets of points show runs with two different assumed star formation efficiencies $\epsilon_{\rm ff}=1\%$ to $10\%$. The points correspond to the average disk values in azimuthal bins of width $\Delta r=720\pc$, and are calculated from simulation snapshots in the time range $240-300\Myr$. The black solid line shows the galactic scale averaged data from \cite{kennicutt98} and the contour lines the distribution of sub-kpc sized patches in the sample of nearby galaxies by \citep{bigiel2008}. In runs with no feedback, the normalization of the relation scales linearly with the assumed value of $\epsilon_{\rm ff}$, while in runs with feedback the amplitude of the relation changes by a factor of at most two for values of $\epsilon_{\rm ff}$ that differ by a factor of ten. Star formation in the central parts of the galaxy, here points with the largest values of $\Sigma_{\rm gas}$, is not affected by feedback to the same extent as the rest of the disk and the difference in amplitude for runs with different $\epsilon_{\rm ff}$ persists in these regions. 
}
\label{fig:KSfb}
\end{center}
\end{figure*}

\begin{figure*}[t]
\begin{center}
\begin{tabular}{ccc}
\includegraphics[scale=0.3]{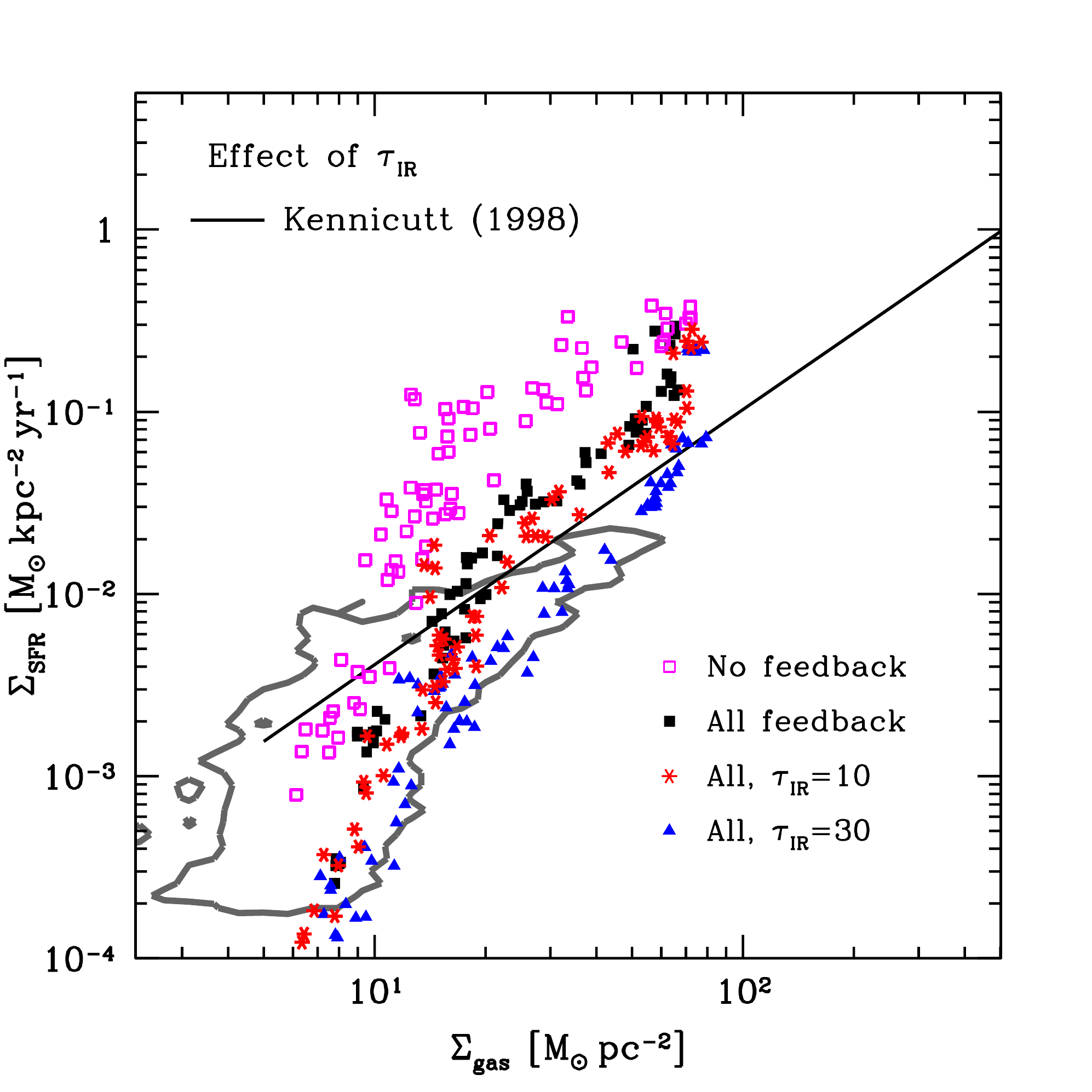}
\includegraphics[scale=0.3]{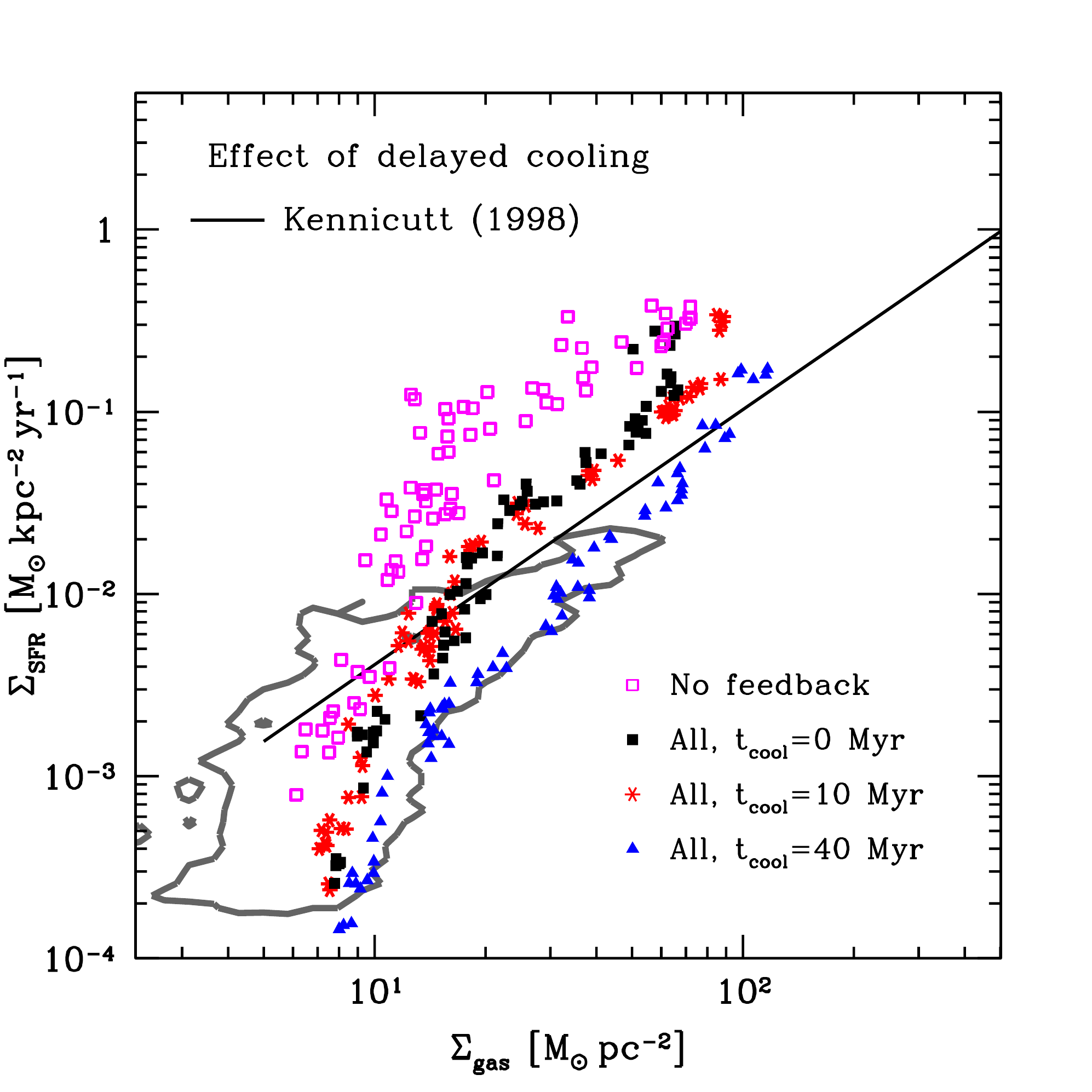}
\includegraphics[scale=0.3]{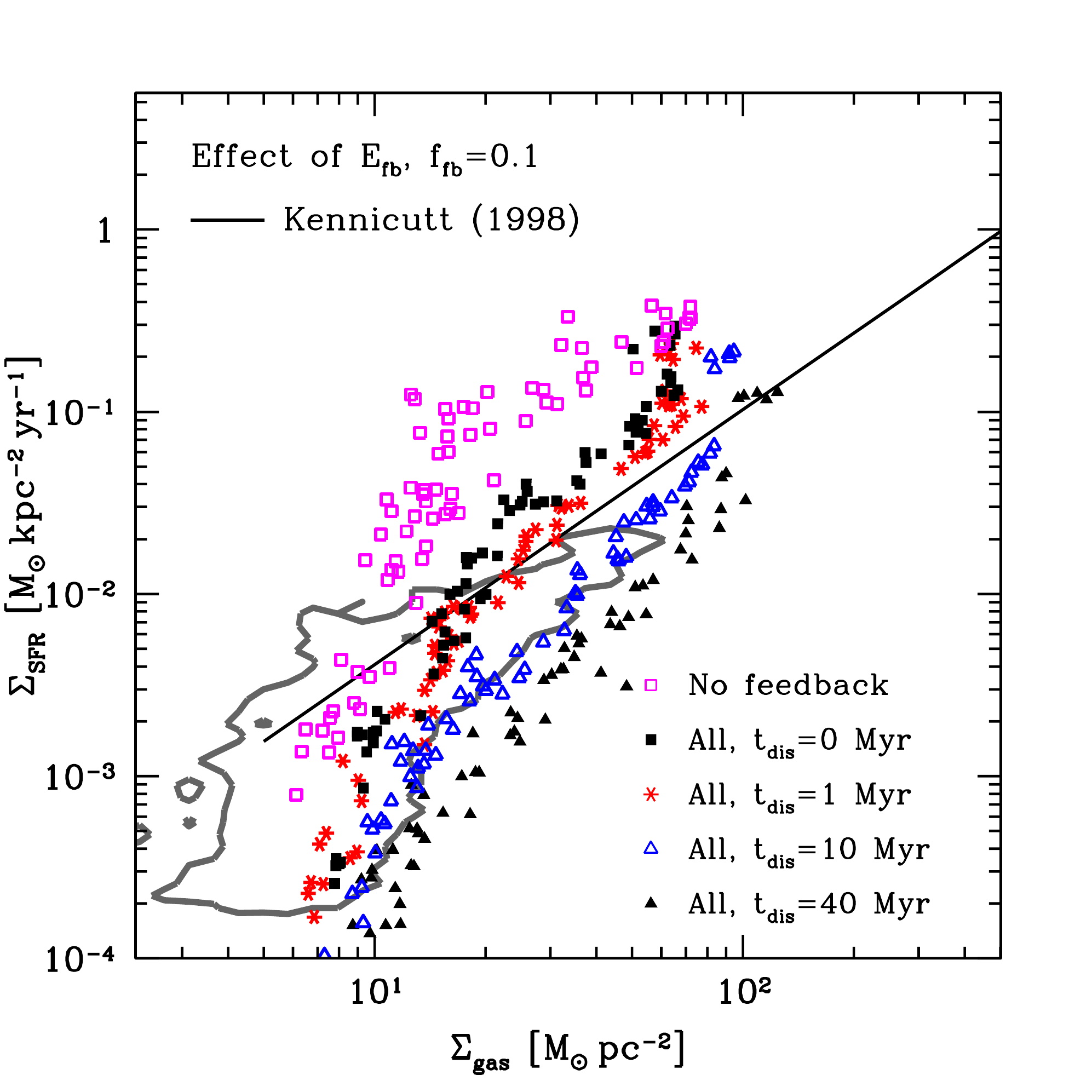}\\
\end{tabular}
\caption{The impact of different implementations of feedback on the $\Sigma_{\rm SFR}-\Sigma_{\rm gas}$ relation. The left panel shows the effect of varying the strength of radiation pressure momentum injection, the middle panel shows effect of delaying cooling around newly born star particles, and right panel shows effect of  treating a fraction of thermal feedback energy as a separate energy variable. Data points show azimuthally averaged values adopting bin sizes of $\Delta r=720\pc$, and are calculated from simulation snapshots in the time range $240-300\Myr$. The observational data points are described in the caption of Figure~\ref{fig:KSfb}. Larger values of $\tau_{\rm IR}$, cooling delay time $t_{\rm cool}$, or energy dissipation time $t_{\rm dis}$, can lead to a similar suppression of normalization of the $\Sigma_{\rm SFR}-\Sigma_{\rm gas}$ relation. The investigated feedback methods show a factor of $\sim 20$ spread in the normalizations of the $\Sigma_{\rm SFR}-\Sigma_{\rm gas}$ relation, which shows that this relation can be a useful tool in constraining parameters of feedback models.
}
\label{fig:KScomp}
\end{center}
\end{figure*}

\subsubsection{The $\Sigma_{\rm SFR}-\Sigma_{\rm gas}$ relation}
Figure~\ref{fig:KSfb} shows how the Kennicutt-Schmidt (KS) relation is affected by the change of star formation efficiency per free-fall time in the presence, and absence, of feedback. All data points refer to quantities averaged over azimuthal bins of width $\Delta r=720\pc$, and are calculated from simulation snapshots in the time range $240-300\Myr$. Shown is also the THINGS data from \cite{bigiel2008}\footnote{Surface densities are corrected by a factor of 1.36 to account for helium.} and the galaxy-scale average relation from \cite{kennicutt98}. The \cite{bigiel2008} relation is derived for kilo parsec sized patches, and is hence a more comparison to our simulated data.

Without feedback, simulations adopting $\epsilon_{\rm ff}=1\%$ are consistent with the \cite{kennicutt98} relation. However, at high $\Sigma_{\rm gas}$ the adopted non-linear star formation relation ($\dot{\rho}_*\propto\rho^{1.5}$) over-shoots the observed, less steep relation of \cite{bigiel2008}. In runs with no feedback, the normalization of the $\Sigma_{\rm SFR}-\Sigma_{\rm gas}$ relation scales linearly with the assumed value of $\epsilon_{\rm ff}$, while in runs with feedback (the ``All'' model) the amplitude of the relation changes by a factor of at most two for values of $\epsilon_{\rm ff}$ that differ by a factor of ten.  However, data points at the largest values of $\Sigma_{\rm gas}$, corresponding to the galactic center in the analyzed simulation snapshots, are less affected by feedback and the difference in amplitude for runs with different $\epsilon_{\rm ff}$ persists in these regions. We note that in runs with $\epsilon_{\rm ff}=1\%$, the KS relation with and without feedback is similar. 

The dependency of feedback model parameters on the KS relation is shown in Figure~\ref{fig:KScomp}, in which different panels show the effect of increasing the strength of radiation pressure, delaying cooling for longer times, and increasing the contribution/duration of feedback energy using a second energy variable. Overall, the sensitivity to the parameters is fairly weak: the KS relation is similar for models in which dissipation of SNII energy is slowed down by delay of cooling or via using second energy variable for  $t_{\rm cool}\leq 10\Myr$ or $t_{\rm dis}\leq 1\Myr$, and for models with early momentum injection with optical depth up to $\tau_{\rm IR}=10$. As parameters are dialed up to even larger values ($\tau_{\rm IR}=30$, $t_{\rm cool}=40\Myr$, or $t_{\rm dis}\gtrsim 10\Myr$), normalization of the KS relation is significantly suppressed. 

These results show that our fiducial feedback model (``All'')  \emph{at the adopted resolution level}, results in star formation rates comparable to the runs in which cooling is delayed or SNe energy is dissipated on a controlled time scale. The results also show that normalization of the KS relation can be used to constrain the plausible range of values of parameters, or at least exclude the most extreme values. 

\begin{figure*}[t]
\begin{center}
\includegraphics[scale=0.65]{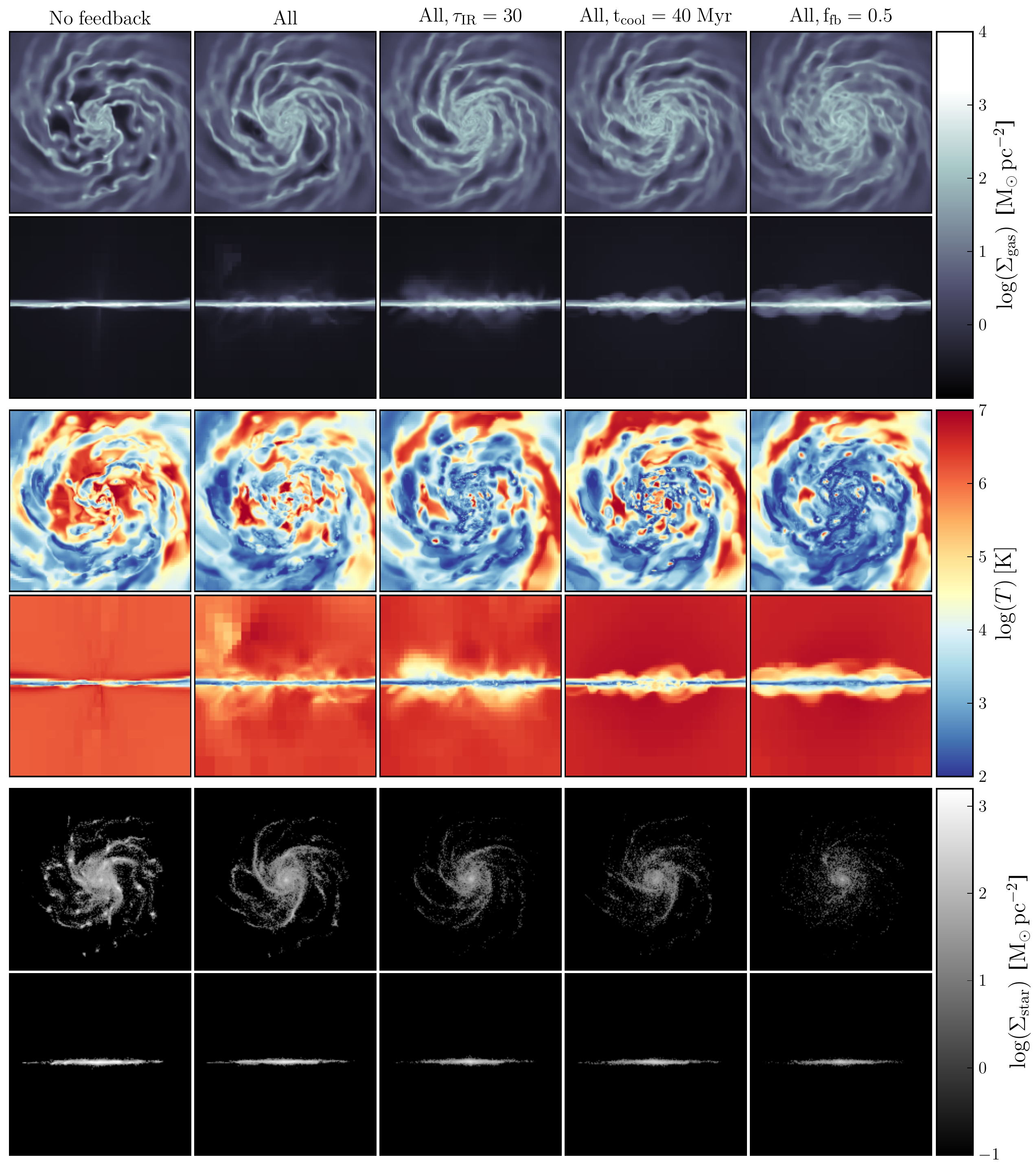}
\caption{Face-on and edge-on maps of the galactic disk at $t=200\Myr$ showing gas surface density (top), mass weighted average gas temperature (middle) and stellar surface density (bottom). The face-on plots are calculated within $z\pm 1.5\kpc$ of the disk to avoid excess halo material. Each panel is 24 kpc across. The temperature is calculated as $\int \rho T/\int\rho$, where the integral is performed along each pixel sightline within $z\pm 150\pc$. The map of stellar distribution only includes star particles formed after the start of simulation, and does not include the star particles present in the initial conditions.
}
\label{fig:maps}
\end{center}
\end{figure*}

\subsubsection{Visual comparison}
In Figure~\ref{fig:maps} we show face-on and edge-on maps at $t=200\Myr$ of the gas surface density, mass-weighted average temperature within $z=\pm 150\pc$ of the disk, and stellar surface density of five of the simulations from table \ref{table:simsummary1}: "nofb", "all", "all\_tau30", "all\_dc40" and "all\_f05\_t10". The two former runs are our fiducial runs with and without feedback, and the latter three represent efficient feedback implementations. 

In runs without feedback, dense star forming clumps of gas form out of spiral arms, and remain intact throughout the simulation until star formation depletes most of their gas, or the clumps sink to the disk center. This run thus produces very massive star clusters clearly visible in the stellar surface density map. In the "all" simulation, gas clumps do not form or are effectively dispersed and gas distribution in this run is considerably less clumpy. Consequently, massive star clusters are not produced, and this effect is even more pronounced in the three example of efficient feedback.

All simulations feature a highly multiphase medium. Large holes filled with hot coronal gas at $T\sim10^6\K$ forms between the cold gas associated with the spiral arms in all simulations. This effect is less prominent in the simulations incorporating feedback, as cold gas is pushed out of star forming regions, resulting in a larger filling factor of cold material. This effect is especially apparent in the face-on temperature map of the "all\_f05\_t10". The edge-on maps of density and temperature in all feedback runs show that fountains and outflows of both cold and warm gas ($T\sim10^4-10^5\K$) and hot gas ($T>10^7\K$) are present close to the disk plane. The efficient feedback runs all feature a more porous ISM, with prominent pockets of hot gas forming within spiral arms, as seen in the face-on density and temperature maps.
 
This illustrates that specific details of feedback implementations do matter in determining qualitative structural properties of the ISM and even stellar distribution. We quantify the differences in density and temperature structure of the ISM in these runs by considering the corresponding probability distributions in the next section.  
\begin{figure*}[t]
\begin{center}
\begin{tabular}{cc}
\includegraphics[scale=0.38]{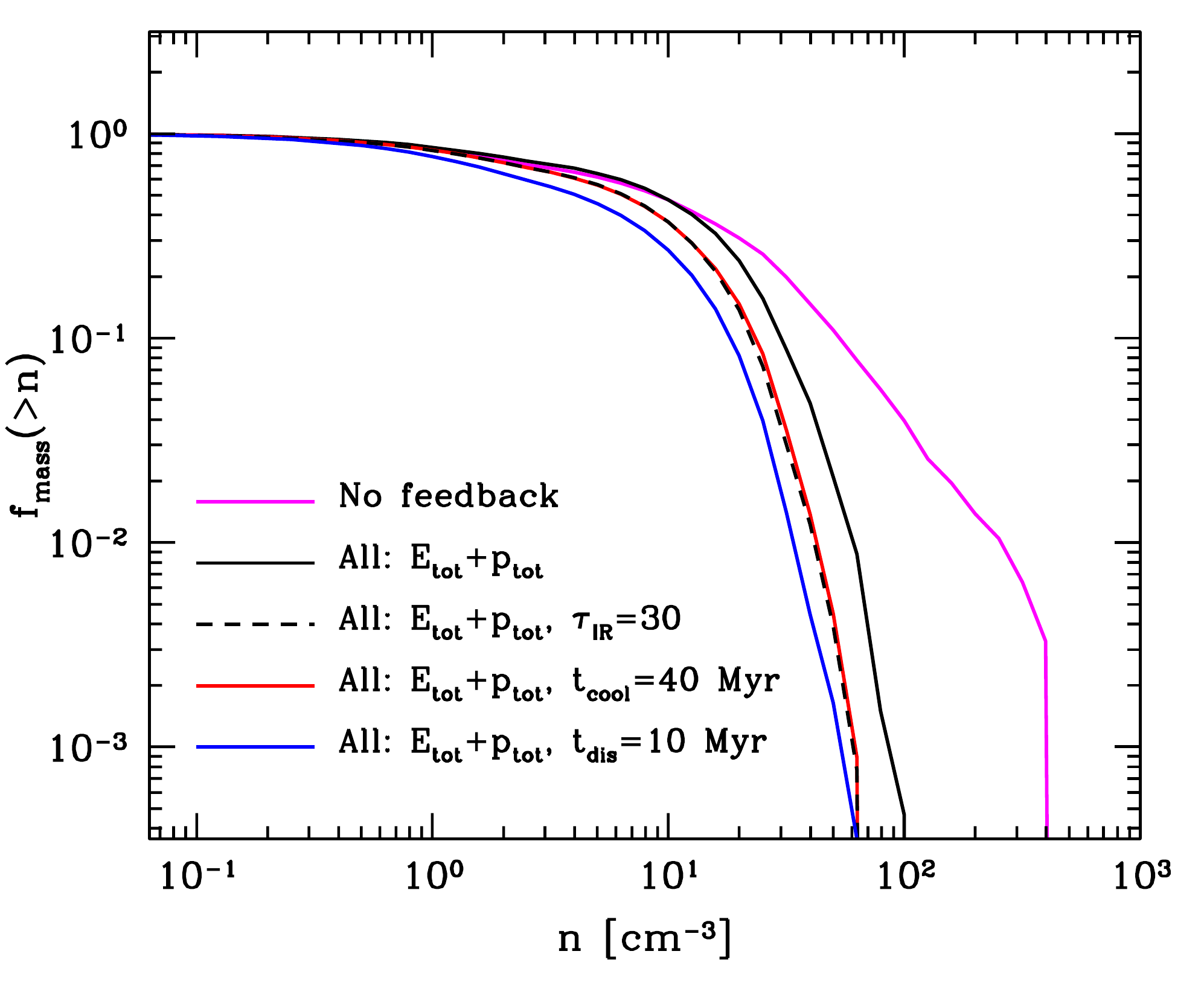}
\includegraphics[scale=0.38]{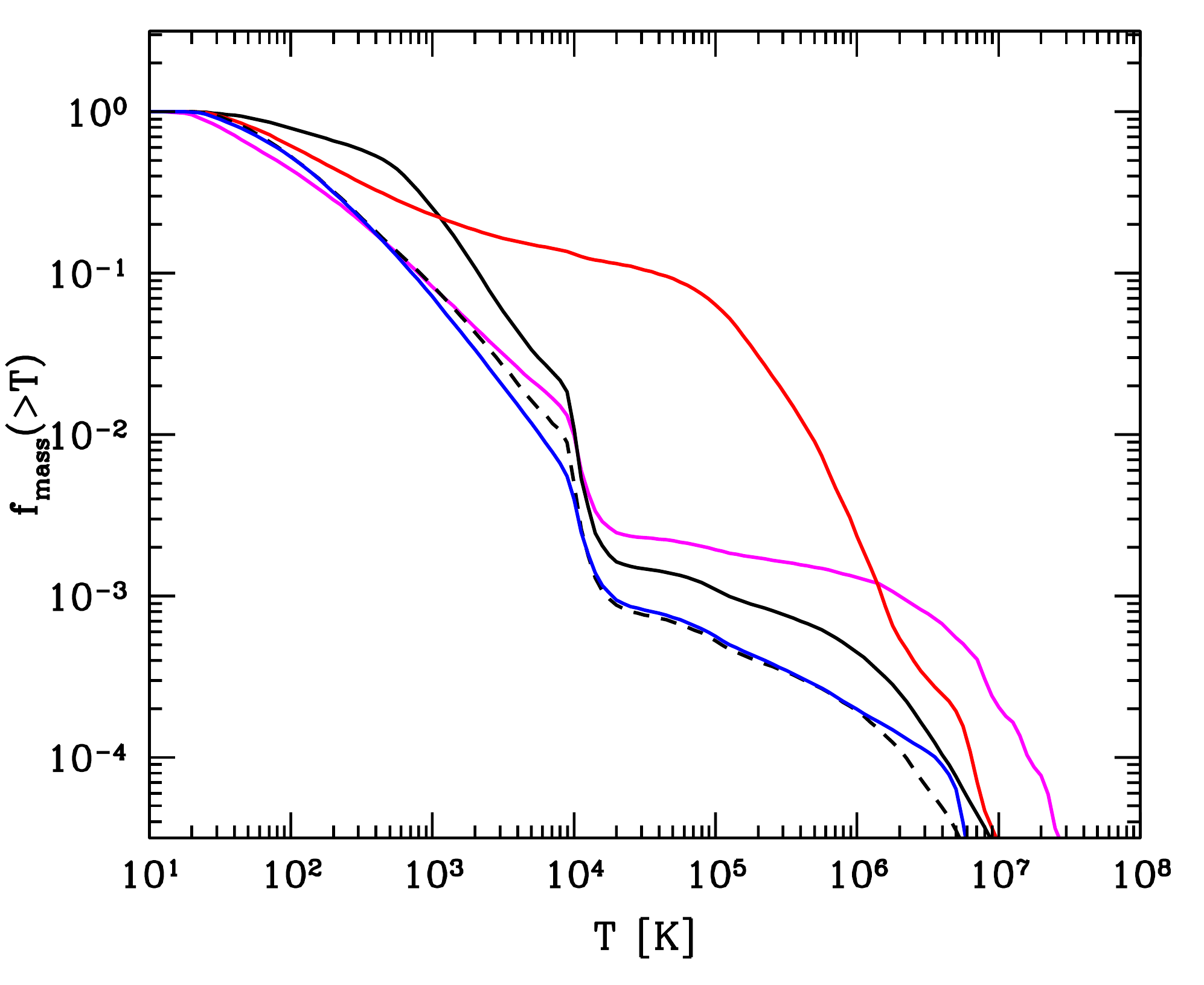}\\
\end{tabular}
\caption{The cumulative mass fraction of the ISM above a given density $n$ (left panel) and temperature $T$ (right panel).  The prominent tail at densities in excess of $n>100\,\cc$ in the simulation without feedback is due to a population of dense, long-lived gas clumps. In the presence of feedback, such gas clumps are effectively dispersed which significantly reduces the fraction of gas at such densities. The temperature structure shows significant differences between different runs. In the run with delayed cooling significant  $\approx 10\%$ of the disk gas is heated to temperature in excess of $T\sim10^5\K$, while this fraction is only $\sim 0.1\%$ in other runs. The "all" run has more mass around $T\sim10^3\K$, which is associated with embedded star particles heating the ISM to warm temperatures. In the case of strong feedback, dense gas is not heated, but dispersed, and diffuse gas is heated to very high temperatures. This mechanism has little effect on the cumulative mass function, as the hot phase is negligible by mass.
}
\label{fig:cumul}
\end{center}
\end{figure*}
\begin{figure*}[t]
\begin{center}
\begin{tabular}{cc}
\includegraphics[scale=0.38]{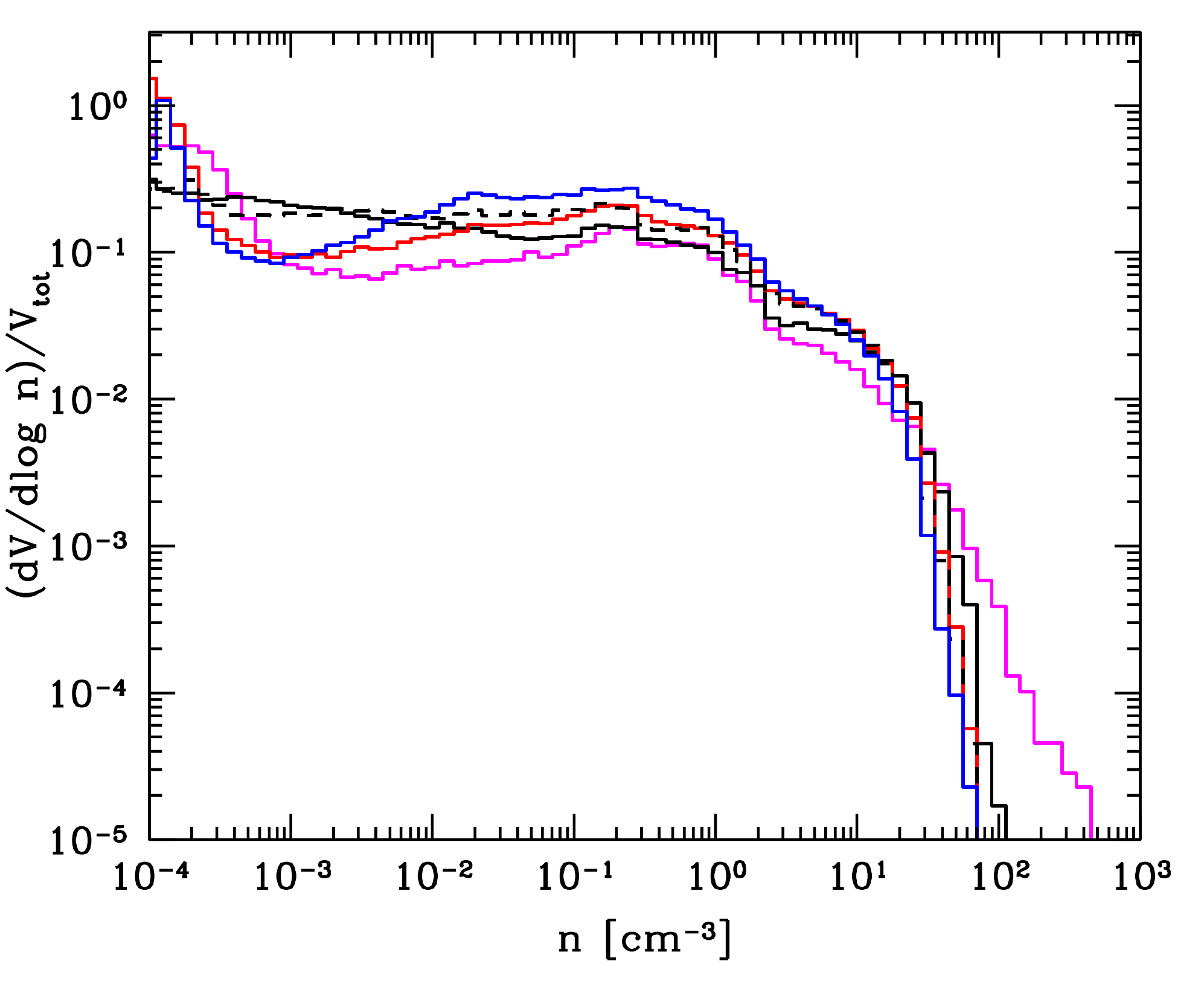}
\includegraphics[scale=0.38]{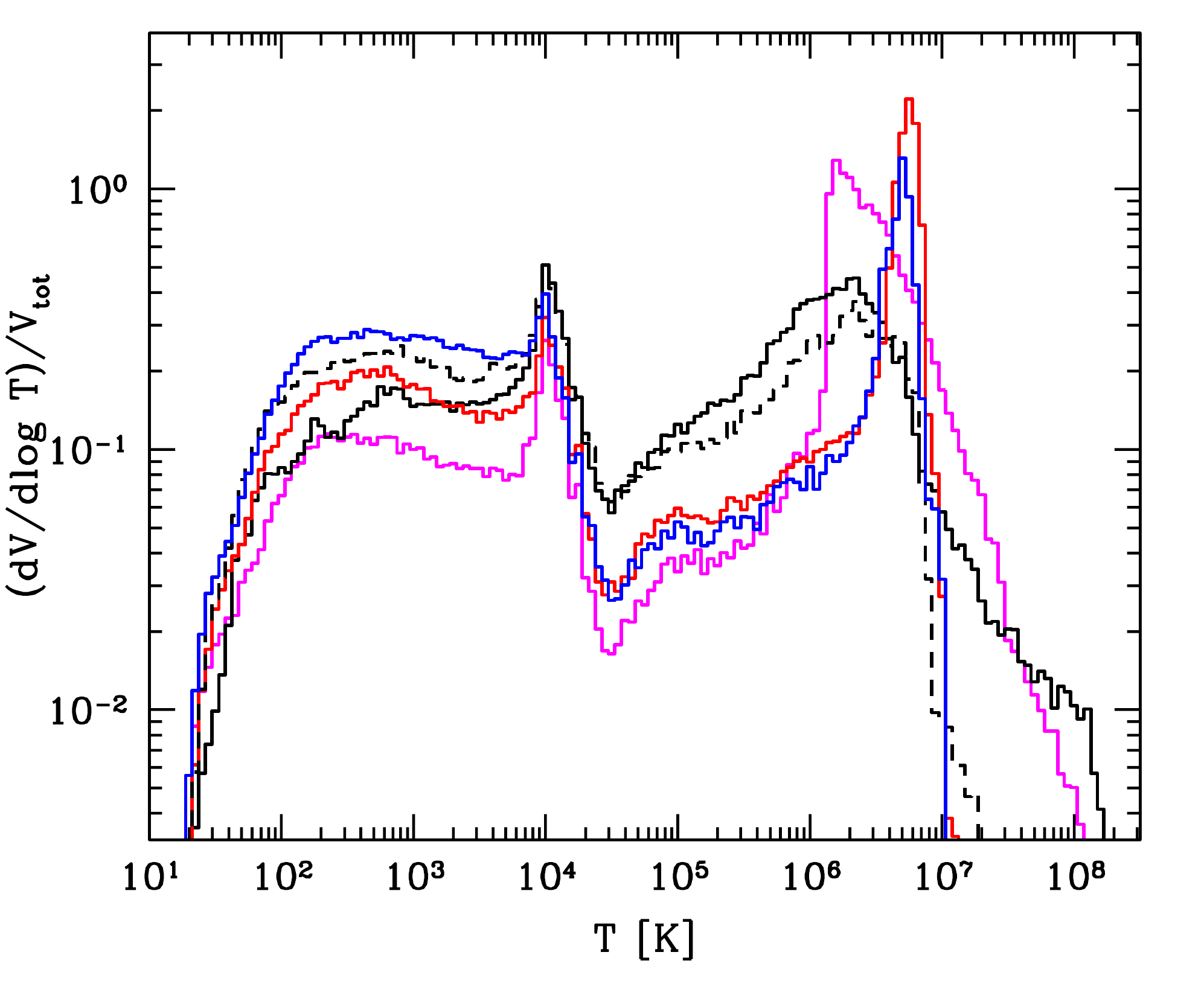}\\
\end{tabular}
\caption{The probability distribution functions (PDFs) of density (left panel) and temperature (right panel) at $t=200\Myr$. The PDFs are measuring a fraction of volume at a given density or temperature. The line types are the same as in Figure~\ref{fig:cumul}.  The multiphase structure of the ISM is evident in all simulations. 
}
\label{fig:PDF}
\end{center}
\end{figure*}

\subsubsection{Structure of the interstellar medium}
The visual differences discussed above are quantified in Figure~\ref{fig:cumul}, where we show the cumulative mass fraction above a given density and temperature at $t=200\Myr$. All simulations are analyzed in the regions shown in Figure~\ref{fig:maps} within a distance of $\pm 0.5\kpc$ of the disk plane. In the case of no feedback, the existence of dense gas clumps is manifested in the tail of the density distribution at $n>100\,\cc$. 
The density and temperature distributions in runs with feedback are  qualitatively similar; the high-density tail at $n\approx100\,\cc$ is suppressed as gas in star forming regions is efficiently dispersed. Simulation with a second feedback energy variable has the least amount of dense gas, as could be deduced from its  SFR in the bottom-right panel of Figure~\ref{fig:SFR}. We note that the details of the high-density tail, as well as the the dispersal process, likely depend on the choice of star formation density threshold and numerical resolution. The distributions presented here are useful in interpreting trends of the KS relation normalization discussed above. For example, it is clear that runs with efficient feedback have SFR comparable to the run with no feedback and ten times lower $\epsilon_{\rm ff}$ because they simple have less dense gas. 

The temperature structure in the right panel also reveals significant differences between feedback schemes. In runs with delayed cooling, $\sim 10\%$ of the disk's gas mass is at $T\gtrsim10^5\K$, which is two orders of magnitudes greater than in the other runs.  This can be seen in the temperature map in Figure~\ref{fig:maps}, where the central region features a hole of hot, ionized, but dense, gas formed out of percolating star forming regions of feedback ejecta. However, all runs have a comparable fraction of gas in the hot coronal phase ($T>10^6$~K). For comparison, in the Milky Way disk $\lesssim 1\%$ of the gas mass is thought to be in the hot phase \citep[e.g.][]{ferriere01}. The prominent bump in the "all" run around $T\sim10^3\K$ is associated with embedded star particles heating the ISM to warm temperatures. In the strong feedback models "all\_tau30" and "all\_f05\_t10", feedback disperses dense gas disperses more efficiently, and heating occurs in the diffuse rather then dense phase, which is why there is no significant mass contribution in the warm or hot phase from these runs in this figure.

Figure~\ref{fig:PDF} shows the density, $({\rm d}V/{\rm d}\log n)/V_{\rm tot}$, and temperature PDFs, $({\rm d}V/{\rm d}\log T)/V_{\rm tot}$, defined as a fraction of disk volume in a given density or temperature range. A log-normal PDF is not a good description to the density PDF in our simulations contrary to results of \citet[][]{wada07}, although it may be possible to describe the PDFs as super-positions of several log-normal distributions corresponding to different gas phases \citep{RobertsonKravtsov08}. The figure shows that the run without feedback has the most dense gas, but the smallest amount of tenuous gas at $n<0.1\,\cc$. Interestingly, the run with delayed cooling has less tenuous gas of density $n\sim10^{-2}-10^{-3}\cc$ than our fiducial run. This indicates that feedback models with early feedback injection can efficiently create both a diffuse ionized warm phase and a tenuous coronal phase without resorting to artificially delaying gas cooling. 

The multiphase structure of the ISM is apparent in the temperature PDF, where all simulations show signatures of a three phase ISM \citep{mckeeostriker77}, connected by gas at intermediate temperatures. Without feedback, the gas cools down to a very thin disk (only a few cells in vertical height) with a substantially lower contribution to the volume in the cold phase ($T<10^4\K$) compare to runs with feedback, which all feature thicker cold gas disks due to feedback driven turbulence. In addition, more cold gas is lost in star formation events when feedback is absent. This discrepancy is especially apparent when comparing to the most efficient feedback run, "all\_f05\_t10". The hot ($T\sim10^6\K$) tenuous gas phase is present in all runs, although vigorous heating in "all\_dc40" and "all\_f05\_t10" creates pockets of gas at $\sim10^7\K$, which vent out of the disk to the surrounding corona. As can be seen in Figure~\ref{fig:maps}, the circum-galactic medium is more structured in "all" and "all\_tau30", which is evident from the wider distribution of gas at $T\sim10^4-10^8\K$.

\begin{figure}[t]
\begin{center}
\includegraphics[scale=0.4]{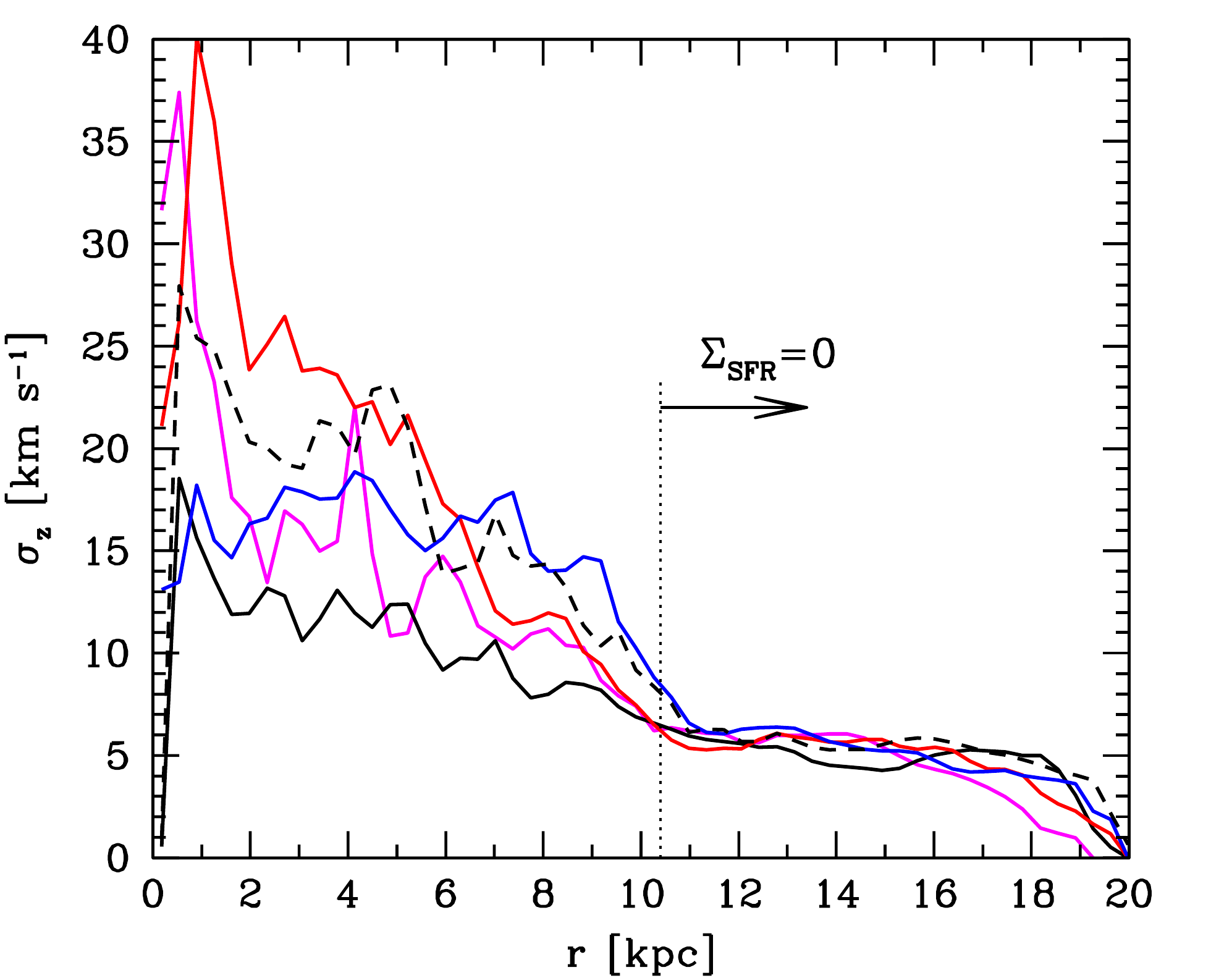}
\caption{The vertical line-of-sight velocity dispersion $\sigma_{\rm z}(r)$. The lines are described in Figure~\ref{fig:cumul}. All simulations show a radially declining dispersion profile, settling on $\sigma_z\sim 5\kmsec$ in the outer parts of the disk ($r\gtrsim 10\kpc$) where $\Sigma_{\rm SFR}=0$. The central increase in turbulent dispersions occurs even in the case of no feedback (magenta line), illustrating the propensity of gravitational instabilities in generating random motions. As feedback is boosted, the velocity dispersions increase significantly towards the central, star forming, part of the galaxy. In the case of delayed cooling (red line), the dispersions are almost twice as high as in the standard "all" run (black line), and a similar trend, although slightly weaker, is found for $\tau_{\rm IR}=30$ (black dashed line), or a separate feedback energy variable is adopted (blue line).
}
\label{fig:sigmaz}
\end{center}
\end{figure}
\subsubsection{Velocity dispersion profiles}
We quantify the level of turbulent gas motions in the disks via the mass weighted, vertical line-of-sight velocity dispersion profile $\sigma_{\rm z}(r)$, shown in Figure~\ref{fig:sigmaz} for the gas cold component ($T<10^4\K$). Such profiles can be observed in real galaxies and comparisons of model results and observations can help to constrain parameters of feedback models. Indeed, we could expect that models with the most efficient feedback generate stronger gas motions, which should be manifested in larger velocity dispersions. The figure shows that significant velocity dispersion declining with increasing radius is produced in all runs. Such declining dispersion profiles are indeed observed in spiral galaxies for the neutral HI gas \citep[e.g.][]{meurer96,petric07,Tamburro2009}. The fact that significant velocity dispersion is observed in the run with no feedback, indicates that most of the motions are due to disk instabilities and not due to feedback per se. In fact, velocity dispersion in the inner regions is even somewhat smaller in our fiducial run with the ``All'' feedback model. This difference is probably due to formation of massive gas clumps which can more efficiently stir the gas as they move around and merge with each other in the weaker feedback runs.  Nevertheless, the largest velocity dispersions, in the inner 10 kpc of the disk, are observed in runs with delayed cooling and large $\tau_{\rm IR}$, i.e. models with the most efficient feedback. 

Using the THINGS galaxy sample, \cite{Tamburro2009} analyzed the radial HI velocity dispersion, $\sigma_{\rm HI}$, and star formation rate surface density profiles and found positive correlation between the kinetic energy of HI and the SFR. The increase in $\sigma_{\rm HI}$ at smaller radii indeed correlates with an increase in star formation activity, both in observations and simulations, but so does the level of shear and strength of disk self-gravity. Gravitational instabilities can generate a significant base line level of turbulence even without any contribution from feedback \citep{Agertz09}, as illustrated in Figure~\ref{fig:sigmaz}. Observations indicate a characteristic plateau of $\sigma_{\rm HI}\sim 10\kmsec$ in galaxies with a globally averaged $\langle \Sigma_{\rm SFR}\rangle \lesssim10^{-3}-10^{-2}\Msolyr\kpc^{-2}$ \citep{dib06}, above which stellar feedback becomes the more dominant driver of the observed HI velocity dispersions \citep[as  shown numerically by][]{Agertz09}. The propensity for different feedback models to generate turbulent velocity dispersions in ISM gas may therefore manifest more strongly in starbursting systems. We leave an investigation of the velocity dispersion dependence on feedback parameters and star formation surface density for a future study.

\section{Discussion and Conclusions}
\label{sec:conclusions}
In this paper we have presented a new model for stellar feedback that explicitly considers the injection of both momentum and energy in a time resolved fashion. In particular, we have calculated the time dependent momentum and energy budget from radiation pressure, stellar winds, supernovae type II and Ia, as well as the associated mass and metal loss for all relevant processes. We present a novel prescription for modeling the early (pre-SNII) injection of momentum due to stellar winds and radiation pressure from massive young stars. These stellar feedback processes were implemented and tested in the AMR code {\small RAMSES}.  We have also examined and compared the effects of feedback in this new implementation and other popular recipes on properties of simulated galactic disks.

Using idealized simulations of star forming patches of gas and star forming spiral galaxies, we study how each stellar feedback source affects the overall rate of galactic star formation, as well as density, temperature, and velocity structure of the ISM. We find that early pre-SN injection of momentum is an important ingredient, which qualitatively changes the effectiveness of stellar feedback. In a given stellar population, supernovae explode only after $\sim 4\Myr$, while essentially \emph{all} momentum and energy associated with radiation pressure and stellar winds are deposited in the first $3-4\Myr$. We show that such momentum injection disperses dense gas in star forming regions, which drastically increases the impact of subsequent SNII energy injection, even when no delay of cooling is assumed. Our simulations of massive ($M\sim 10^6\Msol$) star forming clouds indicate that momentum based feedback alone can limit the global cloud star formation efficiency to $\epsilon_{\rm cl}\sim 10\%$. In absence of the pre-SN momentum feedback, we recover the classical over-cooling problem for stellar feedback \citep{Katz92, NavarroWhite93}, as gas cooling times are short in the dense star forming ISM ($t_{\rm cool}\sim 10^3\,{\rm years}$), and star formation is less affected by feedback.

In a simulated Milky Way-like galaxy, we find that star formation rates, and the normalization of the Kennicutt-Schmidt relation, are significantly affected by inclusion of stellar feedback. Interestingly, we find that the normalization of the Kennicutt-Schmidt relation is less sensitive to the assumed star formation efficiency per free-fall time ($\epsilon_{\rm ff}$) in schemes with efficient feedback due to self-regulating effect of feedback on density and temperature PDFs within interstellar medium of simulated galaxies. An order of magnitude change in $\epsilon_{\rm ff}$ only results in only a factor of two increase in the KS relation normalization.

Our results illustrate the importance of not only accounting for the entire momentum and energy budget of stellar feedback, but also to inject momentum and energy at the appropriate stages of stellar evolution. A similar conclusion was recently reached by \cite{Hopkins2011} \citep[see also][]{Hopkins2012structure} based on high-resolution SPH simulations. 
In this paper we show how this effect can be incorporated at the resolution typical for state-of-the-art galaxy formation simulations.

Although the qualitative trends illustrated by our results are clear, it is not obvious whether the effects of feedback, especially the survival and impact of shocked winds and SNe ejecta, are modelled correctly. This is because any subgrid feedback scheme by necessity is implemented at scales close to the resolution of the simulations, where numerical effects play a role. In our experiments we find that even if only $\sim10\%$ of thermal feedback energy is retained for $1-10\Myr$ \cite[as suggested by e.g.][]{Thornton1998,ChoKang2008}, stored and followed using a separate energy variable, this energy has a significant effect on star formation rates, the ISM density structure and turbulent velocity dispersions. 

Comparing different feedback prescriptions, we find that the recipe presented in this paper results in effects on galactic star formation rate and interstellar medium structure similar to the results of feedback schemes with a delay of feedback energy dissipation if the infrared optical depth in star forming regions is sufficiently high ($\tau_{\rm IR}\gtrsim 10$). This conclusion is consistent with the results of \cite{Hopkins2011}. 

\cite{Hopkins2011} reported average values of $\langle \tau_{\rm IR}\rangle\sim 10-30$ in their isolated ``Milky Way'' SPH simulation. This value refers to the typical $\tau_{\rm IR}$ adopted as an SPH particle is kicked by their feedback scheme. In the empirically-motivated subgrid model for radiation pressure momentum presented in this paper, such high values of $\tau_{\rm IR}$ are achieved only around massive star clusters,  $M_{\rm cl}\gtrsim 10^6\Msol$, which are rare in our simulations of galactic disks. The actual values of IR optical depth around young clusters are quite uncertain. If large values, $\tau_{\rm IR}\gtrsim 30$, indeed are appropriate, this can be incorporated as a normalization constant in our relation in Appendix~\ref{sect:Pradmodel} (i.e. $\eta_2$ in Equation~\ref{eq:prad2a}). We note the that in some situations, even momentum from single scattering of photons (i.e. without IR trapping, $\tau_{\rm IR}\approx 0$) can have a significant effect \citep{Wise2012,Chattopadhyay2012}. The regimes in which such feedback is efficient remain to be explored and clarified. 

The above conclusions are based on the simulations conducted at spatial resolutions typical of what is affordable by current cosmological simulations of galaxy formation, i.e. $\sim 10-100\pc$. We have not demonstrated numerical convergence in this work, and we do not necessarily expected this to be easily achieved; as resolution improves, the density PDF changes as self-gravitating gas can collapse to higher densities and the gas dissipates energy at a higher rate. This leads to a shorter star formation time scale (as $t_{\rm SF}\sim\rho^{-0.5}$, Equation~\ref{sect:SFlaw}) and hence an increased rate of star formation. Numerical convergence can in principle be achieved by imposing a pressure floor via a polytropic equation of state, $P\propto \rho^{\gamma}$, where $\gamma=2$, similar to what is necessary to avoid artificial fragmentation \citep{truelove97}. In this case we impose, by hand, a floor to the allowed minimum Jeans mass, for which convergence in principle is achievable. However, in this case simulations may converge to an incorrect (and arbitrary) result, if the polytropic equation of state does not capture the actual thermodynamic properties of ISM realistically.  
 
It is clear that any implementation of the star formation--feedback loop requires thorough testing against observations, such as the Kennicutt-Schmidt relation, velocity dispersion profiles of gas etc. We plan to carry out such tests using the implementations of the feedback models described in this paper, as well as different implementations of star formation recipes, in self-consistent cosmological galaxy formation simulations in future work.  

\acknowledgements
We thank Romain Teyssier for fruitful discussions. OA acknowledge the support of the Kavli Institute for Cosmological Physics at the University of Chicago through grants NSF PHY-0551142 and PHY-1125897 and an endowment from the Kavli Foundation and its founder Fred Kavli. OA and AK are grateful for the hospitality of Overflow Coffee Bar, where many ideas for this work came into being.

\bibliographystyle{apj}
\bibliography{feedback.bbl}

\appendix
\section{The subgrid model for radiation pressure momentum}
\label{sect:Pradmodel}
One of the main difficulties in modelling momentum transferred to gas by radiation pressure is in proper accounting for contribution of momentum due to multiple scatterings of infrared photons by dust grains. In the implementation of \citet{Hopkins2011Prad}, an iterative clump finding algorithm was used to identify star forming clouds. All stars within the cloud radius transfer momentum to the gaseous components according to Equation~\ref{eq:radpressure}, where the infrared optical depth $\tau_{\rm IR}=\kappa_{\rm IR}\Sigma_{\rm gas}$. Here $\Sigma_{\rm gas}$ is the gas surface density and Hopkins et al. adopt a constant opacity $\kappa_{\rm IR}\approx 5\ \rm cm^2\, g^{-1}$, which is appropriate for dust temperatures of $T_{\rm d}\gtrsim 100$~K.

The surface density of gas in star forming clump was calculated directly from simulations as $\Scl=\Mcl/(\pi\Rcl^2)$, where $R_{\rm cl}$ is the radius given by the clump finding routine. Hopkins et al. report average values (at gas particle launch) of $\langle \tau_{\rm IR}\rangle\sim10-30$ in the Milky Way environment, and $\sim30-100$ in high redshift disk analogues. Values of this magnitude are a direct outcome of clouds collapsing to the point where densities are high enough for $\tau_{\rm IR}$ to halt the process, leading to cloud collapse. However, this process is highly sensitive to simulation resolution and other numerical effects. Moreover, surface density of real star forming clumps depend on internal processes within these regions, such as supersonic turbulence and feedback, which will not be resolved even with the $\sim$ parsec resolution. This uncertainty hence propagates into the calculation of the momentum transfer via reprocessing of IR radiation by dust. 

Below we discuss an alternative way of estimating $\tau_{\rm IR}$ based on observed properties of young star clusters and molecular clumps, which can readily be implemented in simulations adopting spatial a resolution of $\Delta x\sim 10-100\pc$.

\begin{figure}[t]
\begin{center}
\includegraphics[scale=0.5]{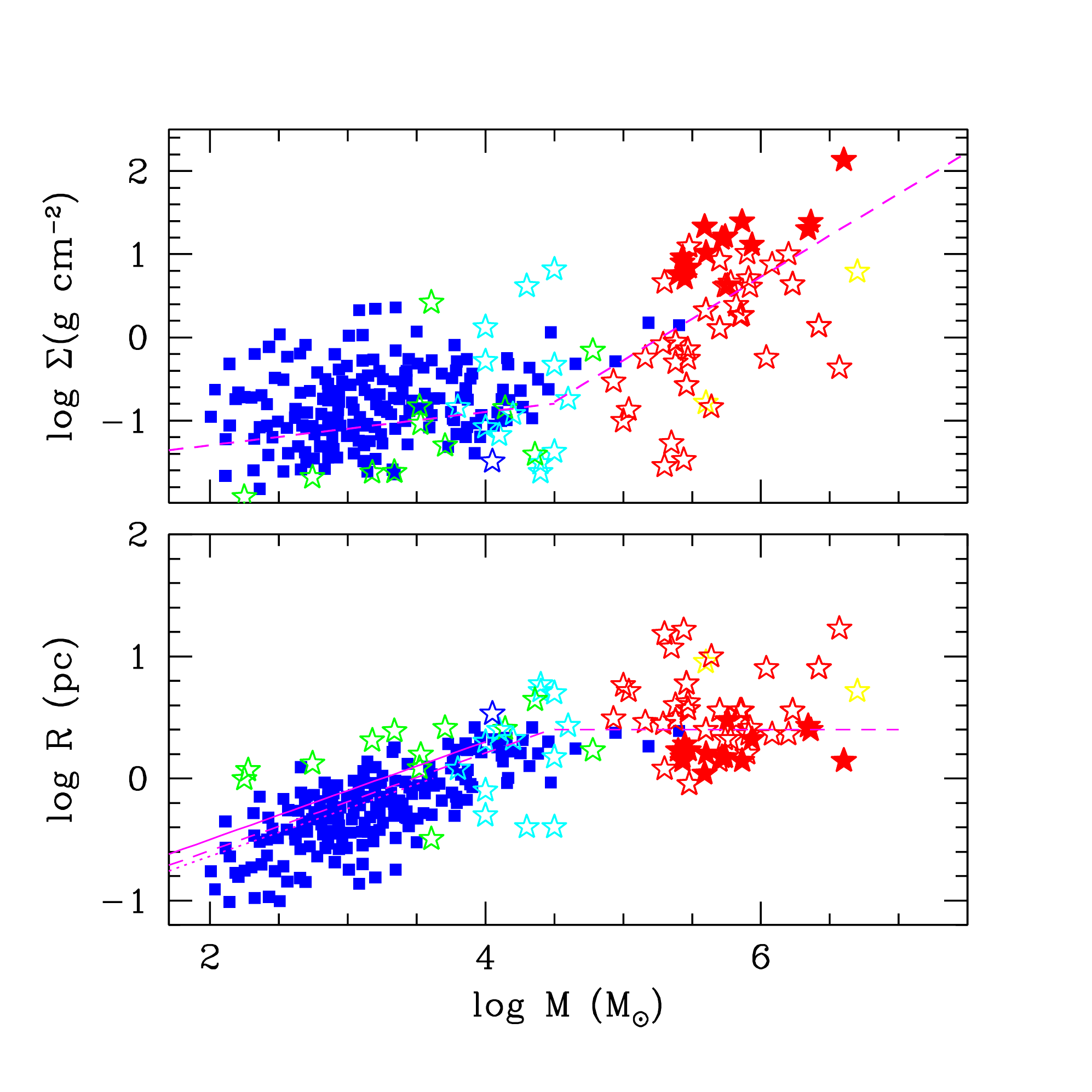}
\caption{Surface density versus mass (top panel) and half-mass radius versus mass (bottom panel) for the molecular clumps in the Milky Way \protect\citep[blue squares, from the compilation by][in their Figure 1]{fall_etal10} and young star clusters \citep[stars, from compilation by][]{portegies_etal10}. The green stars show clusters in the LMC \citep{mackey_gilmore03a} and SMC, while cyan stars show star clusters within Milky Way. The other star symbols show clusters in other galaxies, including starbursts such as M82 \citep[solid red stars,][]{krumholz_matzner09} and the Antennae galaxies. Note that surface densities are estimated within the half mass radius: $\Sigma=M/(2\pi R^2)$. The broken magenta dashed lines show power law approximation to the clusters given by Equation~\ref{eq:mr}. Solid line in the bottom panel is power law fit to the MW low-mass star clusters from \protect\citet{lada_lada03}, while dotted line is fit to the mass-radius of MW clumps from \protect\citet{dib_etal10}.
}
\label{fig:mr}
\end{center}
\end{figure}

\begin{figure}[t]
\begin{center}
\includegraphics[scale=0.5]{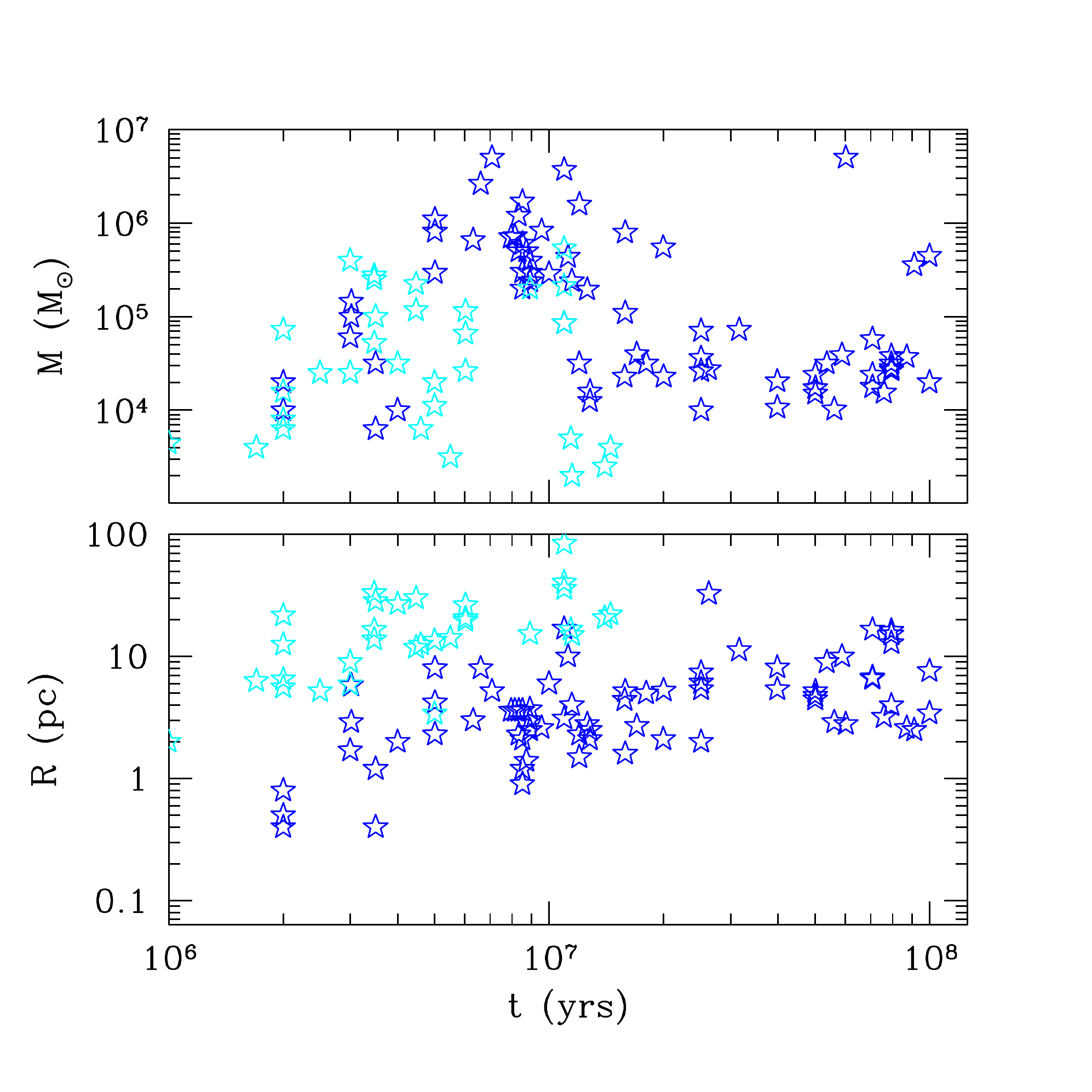}
\caption{Mass (top panel) and half-mass radius (bottom panel) of young star clusters as a function of their age using cluster compilation of \citet{portegies_etal10}. The blue stars show clusters with ages longer than three dynamical times, while cyan stars show clusters with ages shorter than three dynamical times. The latter are classified as stellar associations. 
}
\label{fig:mt}
\end{center}
\end{figure}
\subsection{Surface densities of observed star forming clumps}

Observed molecular clouds have sizes of $\sim 5-100$~pc, average
densities of $\sim 100\ \rm cm^{-3}$, and surface densities of
$\Sigma_{\rm GMC}\sim 50-200\ \rm
M_{\odot}\,pc^{-2}$ \citep[e.g.,][]{bolatto_etal08,fukui_kawamura10}. They
have complex internal structure with gas density ranging over several
orders of magnitude. This structure is thought to arise due to
supersonic turbulent flows and gravitational contraction during cloud
formation \citep[e.g.,][]{Padoan1997,Li2004,Kritsuk2007}.

Star clusters form in {\it clumps} which have densities of
$>10^3-10^4\rm cm^{-3}$, while individual stars form in high-density
cores of even higher density within the
clumps \citep[$>10^5\rm cm^{-3}$, e.g.][]{lada_lada03}.  Clumps in the Milky Way
molecular clouds \citep[e.g.,][]{williams_etal94,williams_etal00,lada_lada03} have
masses of up to few $\times 10^5\Msun$ and their surface densities
range from $\sim 150\Msunpc2$ to $\sim 15000\Msunpc2$ (or $\sim
0.03-3\rm g\,cm^{-2}$). Based on the structural 
parameters of more massive young star clusters, their parent gas clumps had even higher surface densities. Young ($t<10\Myr$) star clusters  Milky Way clusters  with masses of order $\sim10^4\Msol$ are have half light radii of $r_{\rm h}\sim 0.5-2\pc$ \citep[see compilation by][]{portegies_etal10}. Observations of young clusters in external galaxies find $r_{\rm h}\sim2\pc$ independent of luminosity or cluster mass \citep[M51, ][]{Scheepmaker2007} and young super star clusters in star burst galaxies e.g. M82 show similar properties \citep{McCradyGraham2007}.

The bottom panel of Figure~\ref{fig:mr} shows the relation between half-mass radius and mass for molecular clumps in the Milky Way \citep[from the compilation shown in Figure 1 of][]{fall_etal10} and for young (age $<2\times 10^7$~yrs) star clusters in the Milky Way and other nearby galaxies  from the compilation of \citet{portegies_etal10} as a function of their mass. Different lines show relations derived for star clusters and clumps in several recent studies, as described in the figure caption. The corresponding surface densities of the clumps and star clusters are plotted in the upper panel, and shows that although scatter is substantial, clumps and clusters 
in the MW follow a similar relation at masses $\lesssim 10^5\ \rm\Msun$: $R\propto M^{\alpha}$ with $\alpha\approx 0.3-0.5$. This implies that it is reasonable to assume that radii and masses of young clusters are a good reflection of the corresponding properties of their parent molecular clumps. 

For clusters of mass $\gtrsim 10^5\ \Msun$ the relation flattens ($\alpha\approx 0$), although the scatter is large. The broken dashed line shows an approximation to the observed behavior of clumps and clusters: 
\begin{eqnarray}
\Rcl&=&\left(\frac{\Mcl}{3000\Msun}\right)^{0.4}{\rm pc},\ \ \ {\rm for}\ \Mcl<3\times 10^4\Msun,\\
\Rcl&=&2.5\ {\rm\ pc},\ \ \ {\rm for}\ \Mcl\geq 3\times 10^4\Msun.
\label{eq:mr}
\end{eqnarray}
The corresponding data and lines for surface densities defined as $\Scl=\Mcl/(2\pi\Rcl^2)$ are shown in the upper panel of the figure. For $\Mcl<3\times 10^4\ \Msun$ the surface densities are generally $\Sigma\lesssim 1\ \rm g\, cm^{-2}$, while for more massive clusters they can reach $\Sigma\gtrsim 10\ \rm g\, cm^{-2}$. For the latter values of surface densities, the optical depth $\tau_{\rm IR}\gtrsim 50$ if dust temperatures are warm ($T_{\rm d}\gtrsim 200$~K), as illustrated in Figure~\ref{fig:tauobs}.

One complication to considering radii and masses of observed young
clusters is that they can evolve with time from the time of their
birth, when most of the radiative pressure feedback has
occurred. Indeed, observations show some weak correlation of cluster
sizes with age \citep[e.g., see][for a recent review; in particular
their Figure~8]{portegies_etal10}. Figure~\ref{fig:mt} shows the masses and half-mass
radii of young star clusters in the Local Group and beyond, from the compilation of \cite{portegies_etal10}, as a function of their age. While some correlation of radii with age is apparent, the mass-age panel shows that interpretation of this correlation is not straightforward. While youngest clusters do have somewhat smaller ages, they also have smallest masses. 

Interestingly, the figure shows that there are quite a few observed
massive clusters ($\Mcl\gtrsim 10^5\ \Msun$) with ages $\sim
5-20\times 10^6$~yrs. Such clusters are largely missing, however, at
smaller ages. This is likely because most of the distant clusters of
such mass are deeply embedded in gas and dust and therefore have been
missed in observations. Some of the massive young stellar clusters are
indeed observed to be deeply embedded in dust \citep[$A_V\sim
9-10$~mag][]{gilbert_etal00,gilbert_graham07}. The most massive
cluster in the Antennae galaxies, for example, is very faint in the
optical band, but is one of the brighest sources in the IR. This
indicates that conditions for substantial radiation pressure do exist
in the natal gas clumps.

Another interesting feature of the figure is the fact that there are
almost no massive clusters with ages $\gtrsim 20$~Myr. Unless there is
some selection effect, this probably indicates that massive clusters
dissolve as a result of dispersal of their parent clump gas and
subsequent stellar mass loss.

In summary, it is not obvious that structural parameters of massive clusters at age $\sim 10^7$~yrs are significantly different from the parameters of these clusters at birth. It is thus premature to apply an age correction to the data shown in the Figure~\ref{fig:mr}.

\subsection{A subgrid model for radiation pressure momentum} 
As the above discussion illustrates, a direct implementation of radiation
pressure momentum injection is not feasible in galaxy formation simulations, as it requires a resolved density structure of star forming clouds at parsec scales. For more general application, it is useful to develop a subgrid model based on empirical knowledge of structure and physics of star forming molecular clumps, which could be valid for different spatial resolutions. The proposed model is local in it is nature, acting only in the local resolution elements surrounding young stars, and therefore does not account for continuous acceleration of gas far outside of galactic disks, as recently proposed by \cite{Murray2011}, and modelled in the radiative transfer simulations by \cite{Wise2012}.

In this model, any star particle formed by star formation recipe in simulations is regarded as an ensemble of star clusters, with an associated ensemble of natal molecular clumps onto which radiation pressure acts at early times.  Such model can then be used to calculate the total 
momentum input from stars in all clumps within star forming cell by integrating over the clump mass function.

We start off by defining the rate of momentum deposition imparted by radiation pressure from young stars on a molecular clump as 
\begin{equation}
\dot{p}_{\rm cl}=(\eta_1+\eta_2\tau_{\rm IR})\frac{L(t)}{c},
\label{eq:pdotcl}
\end{equation}
where $L(t)$ is the bolometric luminosity of stars in a star cluster of age $t$, and as before $\tau_{\rm IR}=\kappa_{\rm IR}\Sigma_{\rm cl}$ is the IR optical depth. The first term describes the direct radiation absorption/scattering and should in principle be $\propto [1-\exp{(-\tau_{\rm UV})}]$, but given that UV optical depth is always very large in dense star forming regions, $\eta_1\approx 1$. The optical and UV photons heat dust particles in surrounding gas and IR photons radiated by dust can transfer additional momentum if gas is optically thick to the IR radiation. The second term thus specifies momentum transferred via multiple scatterings of IR photons re-radiated by dust particles \citep[see, e.g.][]{gayley_etal95}, where $\eta_2$ is added to parametrize possible modifications to the adopted $\tau_{\rm IR}$. As for $\eta_1$, our fiducial choice is $\eta_2=1$, although factors of a few maybe be motivated due to grid smearing/cancelations for large advection velocities, as discussed below. 

By integrating $\dot{p}_{\rm cl}$ over the clump mass function, we obtain the total imparted momentum rate from all star clusters onto their natal clumps,
\begin{equation}
\label{eq:ptot}
\dot{p}_{\rm tot}=\int^{\Mclmax}_{\Mclmin} \dot{p}_{\rm cl}\psi(\Mcl)d\Mcl,
\end{equation}
where the minimum and maximum clump masses ($\Mclmin$ and $\Mclmax$) set the normalization of the cluster mass function. We approximate the observed mass function of molecular clumps by a power law 
\begin{equation}
\psi(\Mcl)=A_{\rm cl}\Mcl^{-\beta},
\label{eq:massfunc}
\end{equation}
with $\beta\approx 1.7\pm 0.2$ \cite[][]{kramer_etal98} similar
to the power law slope of the molecular clouds themselves \citep[e.g.,][]{fukui_kawamura10}. The latter can be approximated by a Schechter like function with exponential cutoff. The mass function of young star clusters can also be approximated by the Schechter form \citep[][see \citeauthor{portegies_etal10} \citeyear{portegies_etal10} for a review]{zhang_fall99,degrijs_etal03,bik_etal03,cresci_etal05,mccrady_graham07}. Different studies that sample different parts of cluster mass function and often approximate it with a simple power law, can get somewhat different values of the slope. Nevertheless, the mass function of star clusters is generally found to be quite similar in shape to that of the molecular clouds and clumps. The similarity of mass function slopes
implies that star formation efficiency in clumps,
$\ecl\equiv M_{\rm *,cl}/\Mcl$ (where $M_{\rm *,cl}$ is mass of a star cluster a given
clump of mass $M_{\rm cl}$ forms), is approximately independent of clump mass \citep{fall_etal10}.  

We interpret a formed star particle of mass $m_*$ as the total
mass of an ensemble of star clusters formed with the constant
efficiency $\ecl$ from molecular clumps with a mass function given by
Equation~\ref{eq:massfunc}. Although the actual mass function may have Schechter form, for our purposes we can approximate it as a power law (i.e., small-mass end of the Schechter function) with effective maximum clump mass, $\Mclmax$. 
The choice of this maximum mass is related to the normalization of the cluster mass function and we will assume that the maximum mass of star cluster $\Mstclmax=\ecl\Mclmax$ is
\begin{equation}
\Mstclmax=\mu_{\rm max}m_*.
\label{eq:mstclmax}
\end{equation}
Motivation for such relation comes from observation that masses of
young (ages of $<10^7$~yrs) star clusters in M33 and LMC are
$\sim 10-50$ times smaller than the masses of their parent molecular
clouds \citep{fukui_kawamura10,portegies_etal10}. Given that $m_*$ is related to the mass of molecular gas in the cell via
the star formation recipe, Equation~\ref{eq:mstclmax} establishes a
relation between the masses of largest star cluster and mass of the
molecular gas in the cell. Depending on assumed star formation
efficiency, the value $\mu_{\rm max}\sim 0.1-1$ is reasonable. The minimum mass of the clumps can be taken to be $\Mclmin\approx 100\ \Msun$ \citep{lada_lada03}. In this work we fix the slope of the mass function to the above suggested value of $\beta=1.7$. For this slope, most of the mass of clumps is in most massive clumps. This will maximize radiation pressure feedback compared to mass functions that have slopes $\beta>2$. 

For a clump of mass $\Mcl$ and half-mass radius $R_{\rm h}$, we can define
gas surface density $\Scl=(M/2)/(\pi R_{\rm h}^2)$, velocity
dispersion $\sigma=\sqrt{0.4G\Mcl/R_{\rm h}}$, escape velocity $\vel_{\rm
e}=2\sigma$, and crossing time $t_{\rm c}=R_{\rm h}/\sigma$, all of which could be fully characterized in terms of mass if we
assume that clump mass and radius are related via power law:
\begin{equation}
R_{\rm h}=C_{\rm R}\Mcl^{\alpha}.
\end{equation}
The slope of this relation is constrained by observed slope of the average
$\Scl-\Mcl$ relation, $1-2\alpha$, and corresponding relation for young star clusters as indicated in Figure~\ref{fig:mr}. In this work we adopt the relation given by Equation~\ref{eq:mr}.

We obtain the luminosity $L(t)$ from {\small STARBURST99}, and define 
\begin{equation}
L(t)=L_1(t)\Mstcl,
\end{equation}
where $L_1(t)$ is bolometric luminosity per $\Msun$. $L_1$ is approximately constant at $\approx 3.8\times 10^{36}\rm\ ergs\,s^{-1}\,\Msun^{-1}$ for $t\lesssim 3\times 10^6$~yrs, and decreases roughly as $\sim t^{-1.25}$ at late times, see Figure~\ref{fig:sb99}. Using the above relations, the total momentum rate $\dot{p}_{\rm tot}$ in Equation~\ref{eq:ptot} may now be evaluated. The relation has two terms; let us consider them in turn. 

{\it The first term} is independent of surface density and, when integrated over the clump mass function, will simply give
\begin{equation}
\dot{p}_{\rm tot,1}=\eta_1\frac{L_1(t)}{c}m_*,
\end{equation}
where $t$ is the age of a given stellar particle of mass $m_*$ in cell under consideration. 
This contribution can simply be summed up for all young stellar particles in the cell with ages as old as it is deemed to be significant, typically for a few Myr.

{\it The second term} depends on the surface density. Before the parent molecular clumps are dispersed by their child star clusters, i.e. for time less than clump lifetime $t<\tcl$, radiation pressure operates on the surface density of the clumps,
\begin{equation}
\Scl=(1-\ecl)\frac{\Mcl}{2\pi R^2_{\rm h}},
\end{equation}
where $(1-\ecl)$ factor takes into account the fact that fraction $\ecl$ of clump mass was turned into stars, while factor of $0.5$ takes into account that $R_{\rm h}$ in the assumed mass-radius relation is the half-mass radius. 
Thus, for $t<\tcl$, using the equations above, and integrating over the clump mass function, we obtain
\begin{equation}
\label{eq:prad2a}
\dot{p}_{\rm tot,2a}=\frac{\eta_2\kappa_{\rm IR}}{2\pi C_R^2}\frac{(1-\ecl)(2-\beta)}{3-2\alpha-\beta}\left(\frac{\mu_{\rm max}}{\ecl}\right)^{1-2\alpha}\frac{1-(\Mclmin/\Mclmax)^{3-2\alpha-\beta}}{1-(\Mclmin/\Mclmax)^{2-\beta}}\frac{L_1(t)}{c}m_*^{2(1-\alpha)}.
\end{equation} 

For $t>\tcl$, the clump is dispersed and the radiation pressure will simply act on the surface density of
the cell, $\Sigma_{\rm gas,c}$, with a possible boosting by some clumping
factor to account for a clumpy nature of parent molecular cloud. The
latter can be introduced via $\eta_2$. The total momentum rate in this case will therefore be: 
\begin{equation}
\dot{p}_{\rm tot,2b}=\eta_2\kappa_{\rm IR}\Sigma_{\rm gas,c}m_*\frac{L_1(t)}{c}
\end{equation}
Summarizing, the total momentum rate is
\begin{equation}
\label{eq:pradfull}
\dot{p}_{\rm tot}=\left\{\begin{array}{l l} \dot{p}_{\rm tot,1}+\dot{p}_{\rm tot,2a} &\quad\mbox{if $t<\tcl$,}\\
 \dot{p}_{\rm tot,1}+\dot{p}_{\rm tot,2b}=(\eta_1+\eta_2\kappa_{\rm IR}\Sigma_{\rm gas,c})m_*\frac{L_1(t)}{c}&\quad\mbox{if $t>\tcl$,}\\
\end{array}\right.
\end{equation}

The clump lifetime $t_{\rm cl}$ is a highly uncertain factor, but can reasonably be assumed to be a fixed
multiple of the clump crossing time: $t_{\rm cl}=N_{\rm c}t_{c}$ with
$N_{\rm c}\sim 5-10$ \citep{palla_stahler00}, where crossing time $t_c$ is given by 
\begin{eqnarray}
t_c&\equiv&\frac{R_{\rm h}}{\sigma}=\frac{C_{\rm R}^{3/2}}{\sqrt{0.4G}}\Mcl^{(\alpha-1)/2},
\end{eqnarray}
where $R=C_R\Mcl^{\alpha}$ relation was used. For massive clusters, $C_{\rm R}\approx 2.5$~pc and $\alpha\approx 0$ and 
\begin{equation}
t_c=9.3\times 10^4\left(\frac{\Mstcl}{10^6\rm\,\Msun}\right)^{-1/2}\ \rm yrs.
\end{equation}
Thus, for $N_c\sim 5-10$, the life time of a clump is $0.5-1\times
10^6$~yrs. However, the
first stage of cluster life before clump dispersal is highly
uncertain \citep[e.g.,][]{portegies_etal10} and we do not know neither
the exact lifetime nor its scaling with cluster mass.

In the current work, we adopt the above model for radiation pressure feedback assuming $t_{\rm cl}=3\Myr$, $\epsilon_{\rm cl}=0.2$, $\mu_{\rm max}=1$, $\kappa_{\rm IR}=5\,(Z_\odot/Z)\,{\rm cm}^2\,{\rm g}^{-1}$ and $\eta_1=\eta_2=2$ for our fiducial model, unless noted otherwise. The metallicity scaling on the opacity is a crude way of accounting for the varying dust-to-gas ratios. The values of $\eta$ account for the fact that the actual measured momentum injected by a star particle is found to be reduced during advection of gas through computational mesh\footnote{This is due to two effects: smearing occurring at the grid level and momentum cancellations introduced by the nearest grid point approach. The latter occurs as a star particle discretely switches cells and reduce the momentum flux injected from a previous time-step. We have measured this effect to be on the order of $\sim 15-25\%$ per spatial dimension for translational velocities (relative to the grid) up to $\vel\sim1000\kmsec$.}.

\subsection{Caveats related to the stellar mass used for radiation pressure}
The particle mass $m_*$ entering the terms of Equation~\ref{eq:pradfull} does not necessarily need to be interpreted as the mass of each star particle. In fact, this is not preferred as the strength of radiation pressure depends on $m_*$ in a non-linear fashion. As the particle mass usually is a function or numerical resolution, and the effect of radiation pressure would weaken at higher resolution when the numerical scheme allows for the formation of lower mass star particles. When calculating the effect of radiation pressure we therefore adopt
\begin{equation}
m_* \rightarrow m_*=\sum_{i=1}^n m_{*,n}(t) \quad \mbox{for}\quad t\leq t_{\rm cl},
\end{equation}
i.e. $m_*$ is the binned mass of all $n$ star particles in a cell younger than some given age. As indicated in the above relation,we simply set this age to be the same as the clump life time $t_{\rm cl}$ defined above, for which we adopt the fiducial value of 3 Myr. This is consistent with estimates of the duration of embedded stage of young clusters \citep[e.g.,][]{portegies_etal10}.

Another caveat in our implementation of radiation pressure is that we only consider the young stars in one computational cell when estimating $\tau_{\rm IR}$. The actual cell has nothing to do with the physics of the problem, and is just a convenient implementation choice. At very high resolution, $\Delta x\sim$ few $\pc$, star formation will be spread over several cells in massive GMCs and the collective effect of radiation pressure will be underestimated.  This issue can be avoided by searching neighboring cell for young stars, or by using a star cluster finder, as suggested by \cite{Hopkins2011}. In the regime of $\Delta x\sim 50-100\pc$, adopted for the galactic disks in \S~\ref{sect:disks}, where cell sizes matches observed sizes of massive GMCs, this is not an issue.

\section{Implementation of non-thermal pressure}
\label{sect:nonthermalP}
As discussed in \S~\ref{sect:implementation}, feedback momentum can be injected directly to the computational grid via ``kicks.'' An alternative approach is to allow for the hydro scheme to generate momentum by appropriately pressurizing the local volume in which momentum is injected. In this approach, which we adopt in a sub-set of our simulations, we define the non-thermal pressure due to injected momentum as 
\begin{equation}  
P_{\rm nt}=\dot{p}/A 
\end{equation}
where the area $A$ is an arbitrary computational region, here chosen to be the surface area of computational cell ($A=6\Delta x^2$) containing a young star particle used to compute the momentum injection in the subgrid model described above. This pressure is added to the \emph{effective pressure}, 
\begin{equation}
\label{eq:peff}
P_{\rm eff}=P_{\rm therm}+P_{\rm nt},
\end{equation}
where $P_{\rm therm}$ is the thermal pressure. $P_{\rm eff}$ replaces $P_{\rm therm}$ in the sound speed definition when calculating the time step, and is otherwise only actively involved in the flux calculation (i.e., in the Godunov step). Here $P_{\rm eff}$ is consistently traced to the cell interfaces (here using the piecewise linear approximation) as a separate pressure variable, using its own TVD slopes. The left and right-hand states of $P_{\rm eff}$ then enter the Riemann solver, where it replaces the thermal pressure. Specifically, the momentum and energy equations that we aim to solve are 
\begin{equation}
\label{eq:chap2:eul2}
\frac{\partial}{\partial t}(\rho\vel)+\nab\cdot(\rho\vel\otimes\vel+P_{\rm eff})=-\rho\nab\phi \\
\end{equation}
and
\begin{equation}
\label{eq:chap2:eul3}
\frac{\partial}{\partial t}(\rho E)+\nab\cdot[\rho\vel(E+P_{\rm eff}/\rho)]=-\rho\vel\cdot\nab\phi, 
\end{equation}
where $E$ is the specific total energy, $\phi$ the gravitational potential, and $\otimes$ is the outer vector product. The effective pressure hence never enters into the specific total energy (as $P_{\rm eff}/\rho(\gamma-1)$) which is an important distinction to make in the MUSCL-scheme \citep{vanleer79} adopted in the {\small RAMSES} code \citep{teyssier02}, as it traditionally only stores one variable representing the total energy (the conservative variable), and pressure (the primitive variable) is derived from it after subtracting the kinetic energy. To universally replace $P_{\rm therm}$ by $P_{\rm eff}$ everywhere in the method would hence not be consistent with the equations we want to solve (as well as in  the cooling routines). For our implementation of radiation pressure, the effective pressure is not advected, but is updated every \emph{fine} time-step in the feedback routine, and stored as a separate variable.

In the case of the second feedback energy variable $E_{\rm fb}$, introduced in \S~\ref{sect:fbvar}, we calculate a non-thermal pressure $P_{\rm nt}=(\gamma-1)\rho E_{\rm fb}$, which enters the effective pressure as in Equation~\ref{eq:peff}. This quantity is then treated exactly as described above, although $E_{\rm fb}$ is passively advected with the flow, i.e. it obeys
\begin{equation}
\label{eq:chap2:eul4}
\frac{\partial}{\partial t}(\rho E_{\rm fb})+\nab\cdot(\rho\vel E_{\rm fb})=0.
\end{equation}

\end{document}